\global\def\draftcontrol{0}

   \def\versionno{ n=2 quenches }

\catcode`\@=11

\expandafter\ifx\csname draftcontrol\endcsname\relax\global\def\draftcontrol{0}
\fi

{\count255=\time\divide\count255 by 60
\xdef\hourmin{\number\count255}
\multiply\count255 by-60\advance\count255 by\time
\xdef\hourmin{\hourmin:\ifnum\count255<10 0\fi\the\count255}}
\def\draftdate{\number\month/\number\day/\number\year\ \ \ \hourmin }

\newcommand\makepapertitle{\par
  \begingroup
    \renewcommand\thefootnote{\@fnsymbol\c@footnote}%
    \def\@makefnmark{\rlap{\@textsuperscript{\normalfont\@thefnmark}}}%
    \long\def\@makefntext##1{\parindent 1em\noindent
            \hb@xt@1.8em{%
                \hss\@textsuperscript{\normalfont\@thefnmark}}##1}%
     \newpage
     \global\@topnum\z@   
     \@makepapertitle
     \thispagestyle{empty}\@thanks
  \endgroup
  \setcounter{footnote}{0}%
  \global\let\thanks\relax
  \global\let\makepapertitle\relax
  \global\let\@makepapertitle\relax
  \global\let\@thanks\@empty
  \global\let\@author\@empty
  \global\let\@date\@empty
  \global\let\@title\@empty
  \global\let\title\relax
  \global\let\author\relax
  \global\let\date\relax
  \global\let\and\relax
  \def\version{\let\version\@version\@gobble}
}
\def\@makepapertitle{%
  \newpage
   \ifnum\draftcontrol=1 {}
   \version\versionno
   \vskip 3em%
   \else
   \hfill\hbox to 3cm {\parbox{4cm}{\@pubnum}\hss}%
   \vskip 3em%
   \fi
   \begin{center}%
   \let \footnote \thanks
     {\LARGE {\@title}}%
     \vskip 1.5em%
     {\normalsize
       \lineskip .5em%
       \begin{tabular}[t]{c}%
         \@author
       \end{tabular}\par}%
     \vskip 1.5em%
     {\@bstract}%
     \end{center}%
     \vskip 1.5em
     \@date%
   \par
}

\gdef\@pubnum{}
\def\pubnum#1{%
  \gdef\@pubnum{#1}}

\gdef\@bstract{}
\def\Abstract#1{%
  \gdef\@bstract{%
   \parbox{\textwidth-0pc}{%
   \centerline{\bf Abstract}\penalty1000%
\kern.2cm%
\noindent
\renewcommand\baselinestretch{1.0}%
{#1}}}
}

\def\ps@paper{\let\@mkboth\@gobbletwo%
     \ifnum\draftcontrol=1
    \def\@oddfoot{\hbox to \textwidth{\tiny \versionno \hfil\tiny\draftdate}%
    \hskip -\textwidth \hbox to \textwidth{\hfil\rm\thepage\hfil}}%
     \else\def\@oddfoot{\hbox to \textwidth{\hfil\rm\thepage\hfil}}
     \fi
     \let\@evenfoot\@oddfoot
}

\def\body{\clearpage
          \pagestyle{paper}
    }

\def\@version#1{\ifnum\draftcontrol=1
\typeout{}\typeout{#1}\typeout{}
\vskip3mm\centerline{\hbox{\fbox{\normalsize{\tt DRAFT -- #1 -- }
                   {\draftdate}}}}\vskip3mm
\fi}
\let\version\@version
\long\def\eqlabel#1{\ifnum\draftcontrol=1
                    \tag@false  
                    \tag*{(\theequation) \hbox to -0.2cm{\hspace{0cm}\small{#1}\hss}}
                    \refstepcounter{equation}
                    \edef\@currentlabel{\theequation}
                    \ltx@label{#1}          
                    \else
                    \label{#1}
                    \fi
                    }
\let\st@bibitem\@bibitem
\let\st@lbibitem\@lbibitem
\ifnum\draftcontrol=1
  \def\@bibitem#1{%
    \st@bibitem{#1}\a@@label{#1}\ignorespaces}
  \def\@lbibitem[#1]#2{%
    \st@lbibitem[#1]{#2}\a@@label{#2}\ignorespaces}
  \def\a@@label#1{%
    \gdef\a@lab{\smash{\normalfont\small#1}}
    \ifvmode
      \if@inlabel
        \global\setbox\@labels\hbox{%
          \llap{\a@lab\let\a@lab\relax
                \kern\@totalleftmargin\kern\marginparsep}%
          \box\@labels}%
      \fi
    \fi}
\fi

\documentclass[12pt,letterpaper]{article}

\usepackage{amsmath,amssymb,array,calc,epsfig,rotating,bm}
\usepackage[sort]{cite}
\usepackage{graphicx}
\usepackage{psfrag,verbatim}

\ifnum\draftcontrol=1
\fi

\renewcommand\baselinestretch{1.25}
\renewcommand\section{\@startsection {section}{1}{\z@}%
                                   {-3.5ex \@plus -1ex \@minus -.2ex}%
                                   {2.3ex \@plus.2ex}%
                                   {\normalfont\large\bfseries}}
\renewcommand\subsection{\@startsection{subsection}{2}{\z@}%
                                   {-3.25ex\@plus -1ex \@minus -.2ex}%
                                   {1.5ex \@plus .2ex}%
                                   {\normalfont\normalsize\bfseries}}
\renewcommand\subsubsection{\@startsection{subsubsection}{3}{\z@}%
                                   {-3.25ex\@plus -1ex \@minus -.2ex}%
                                   {1.5ex \@plus .2ex}%
                                   {\normalfont\normalsize\it}}
\renewcommand\paragraph{\@startsection{paragraph}{4}{\z@}%
                                   {-3.25ex\@plus -1ex \@minus -.2ex}%
                                   {1.5ex \@plus .2ex}%
                                   {\normalfont\normalsize\bf}}

\numberwithin{equation}{section}

\def\revise#1       {\raisebox{-0em}{\rule{3pt}{1em}}%
                     \marginpar{\raisebox{.5em}{\vrule width3pt\
                     \vrule width0pt height 0pt depth0.5em
                     \hbox to 0cm{\hspace{0cm}{%
                     \parbox[t]{4em}{\raggedright\footnotesize{#1}}}\hss}}}}

\newcommand\nxt[1]  {\\\fnxt#1}
\newcommand{\ie}{{\it i.e.,}\ }
\newcommand{\eg}{{\it e.g.,}\ }
\newcommand{\mt}[1]{\textrm{\tiny #1}}
\newcommand{\beq}{\begin{equation}}
\newcommand{\eeq}{\end{equation}}
\newcommand{\beqa}{\begin{eqnarray}}
\newcommand{\eeqa}{\end{eqnarray}}
\newcommand{\labell}[1]{\eqlabel{#1}}
\newcommand{\eqlabell}[1]{\eqlabel{#1}}
\newcommand{\reef}[1]{\eqref{#1}}
\newcommand{\ssc}{\scriptscriptstyle}

\def\calc         {{\cal C}}

\def\cale         {{\cal E}}
\def\calf         {{\cal F}}

\def\call         {{\cal L}}
\def\calm         {{\cal M}}
\def\caln         {{\cal N}}
\def\calo         {{\cal O}}
\def\calp         {{\cal P}}

\def\calt         {{\cal T}}

\def\calv         {{\cal V}}

\def\del          {\partial}

\def\tr           {\mathop{\rm Tr}}

\def\sqr#1#2{{\vcenter{\vbox{\hrule height.#2pt
 \hbox{\vrule width.#2pt height#1pt \kern#1pt
 \vrule width.#2pt}\hrule height.#2pt}}}}

\newcommand{\fft}[2]{{\frac{#1}{#2}}}
\newcommand{\ft}[2]{{\textstyle{\frac{#1}{#2}}}}
\def\jsquare{\mathop{\mathchoice{\sqr{8}{32}}{\sqr{8}{32}}
{\sqr{6.3}{9}}{\sqr{4.5}{9}}}}

\def\a{\alpha}
\def\b{\beta}
\def\w{\omega}
\def\r{\rho}
\def\dd{\delta}
\def\e{\epsilon}

\def\g{\gamma}

\def\aa1{\phi}
\def\cc1{\psi}

\def\ta{\tilde{a}}
\def\l{\lambda}

\def\ra{\Longrightarrow}

\def\ha{\hat{a}}

\def\t{\tau}

\def\la{\langle}
\def\ra{\rangle}
\def\tp{\tilde{p}}
\def\ta{\tilde{a}}
\def\hp{\hat{\phi}}

\catcode`\@=12

\begin{document}


\title{\bf  Thermal quenches
in $\caln=2^*$ plasmas} \pubnum{UWO-TH-12/3}

\date{June 28, 2012}

\author{
Alex Buchel,$^{1,2}$ Luis Lehner$^{1,3}$ and Robert C. Myers$^1$\\[0.4cm]
\it $^1$\,Perimeter Institute for Theoretical Physics\\
\it Waterloo, Ontario N2J 2W9, Canada\\[0.2cm]
\it $^2$\,Department of Applied Mathematics,
University of Western Ontario\\
\it London, Ontario N6A 5B7, Canada\\[0.2cm]
\it $^3$\,Department of Physics, University of  Guelph\\
\it Guelph, Ontario N1G 2W1, Canada\\[0.2cm]
}

\Abstract{We exploit gauge/gravity duality to study `thermal quenches'
in a plasma of the strongly coupled $\caln=2^*$ gauge theory.
Specifically, we consider the response of an initial thermal
equilibrium state of the theory under variations of the bosonic or
fermionic mass, to leading order in ${m}/{T}\ll 1$.
When the masses are made to vary in time, novel new counterterms must
be introduced to renormalize the boundary theory. We consider
transitions the conformal super-Yang-Mills theory to the mass deformed
gauge theory and also the reverse transitions. By construction, these
transitions are controlled by a characteristic time scale $\calt$ and
we show how the response of the system depends on the ratio of this
time scale to the thermal time scale $1/T$. The response shows
interesting scaling behaviour both in the limit of fast quenches with
$T\calt\ll1$ and slow quenches with $T\calt\gg1$. In the limit that
$T\calt\to\infty$, we observe the expected adiabatic response. For fast
quenches, the relaxation to the final equilibrium is controlled by the
lowest quasinormal mode of the bulk scalar dual to the quenched
operator. For slow quenches, the system relaxes with a (nearly)
adiabatic response that is governed entirely by the late time profile
of the mass. We describe new renormalization scheme ambiguities in
defining gauge invariant observables for the theory with time dependant
couplings. }

\makepapertitle

\body

\version\versionno
\tableofcontents

\section{Introduction}

Consider quantum mechanics with a Hamiltonian which depends on some
external parameter $\l$,
\begin{equation}
H_\l=H_0 + \l\ \delta H\,.
\eqlabell{H}
\end{equation}
The dynamics of the system induced by variations in $\l$ is
well-understood, \eg, see \cite{LL}. In particular, consider beginning
with $\l=0$ and preparing the system in an energy eigenstate of the
Hamiltonian $H_0$. If the new coupling is turned on adiabatically, the
system continues in an eigenstate with a time-dependent energy which
simply traces the changes in $\l(t)$. In contrast, if the coupling is
abruptly turned on, \eg $\l=\l_0\,\theta(t)$, the system would evolve
forward in a complicated linear superposition of eigenstates of the new
Hamiltonian. While the description of adiabatically evolving couplings
is easily adapted to quantum field theory (QFT) \cite{qft}, the
description of the latter `quantum quenches' is less well understood in
the context. However, it has become the subject of a vigorous research
program motivated by the recent advances in cold atom experiments
\cite{more,cc1,cc2}.

Gauge/gravity duality \cite{m1} provides a remarkable framework for the
study of certain strongly coupled gauge theories. Although the most
applications of this correspondence have been directed at analyzing the
static properties of the boundary theories, there is no conceptual
obstacle in applying this holographic framework to time dependant
problems and in particular, to the study of quantum quenches
\cite{das4}.
In fact, early attention was given to the related question of
describing `thermalization' within this holographic framework
\cite{early} and motivated by connections with the strongly coupled
quark-gluon plasma, there has been a renewed interest in this subject
\cite{shiraz,vaidya,janik,shur}. However, given the complexities of the
bulk description of rapid changes in the boundary theory, numerical
relativity is increasingly being applied to study these `far from
equilibrium' situations
\cite{cy,cy2,garfinkle,Bantilan:2012vu,Heller:2012km,Heller:2012je,toby}.

In this paper we begin a study of quenches in the strongly couple
$\caln=2^*$ gauge theory \cite{pw,bpp,j} applying the techniques of
numerical relativity. Recall that $\caln=2^*$ gauge theory is obtained
as a deformation of the $\caln=4$ super-Yang-Mills (SYM), where a
$\caln=2$ hypermultiplet acquires a mass $m$. For technical reasons, we
will limit our present investigation to `thermal quenches,' where the
initial state is a thermal state, \ie the $\caln=2^*$ theory is
prepared in a microcanonical ensemble, and we work to leading order in
in a high temperature expansion with ${m}\ll
T$.\footnote{Thermodynamics of $\caln=2^*$ plasma was discussed in
\cite{n21,n22,n23,opera}.} As explained in \cite{n23}, in such a
thermal state, we can split the masses  of the bosonic and the
fermionic components of the massive hypermultiplet. Hence we
investigate two separate classes of thermal quenches with
\begin{equation}
\call_{SYM}+\l_\Delta(t)\ \calo_\Delta\,,
\eqlabell{con1}
\end{equation}
where we may have either the bosonic mass operator $\calo_2$ or the
fermionic mass operator $\calo_3$ (with dimensions $\Delta=2$ and 3,
respectively). Of course, the couplings which vary in time are then
simply the corresponding masses, \ie $\l_2=m_b^2$ and $\l_3=m_f$. In
particular, we will consider the following specific profiles for these
couplings
\begin{equation}
\l_\Delta(t)=\frac{1}{2}\ \l_{\Delta}^0\ \Big[\,1 \pm \tanh(t/\calt)\,\Big] \,.
\eqlabell{ratel}
\end{equation}
Note that the profile in eq.~\eqref{ratel} with the $+$ sign has
$\l_\Delta(t\to-\infty)=0$ and $\l_\Delta(t\to+\infty)=\l_\Delta^0$.
Hence it produces a transition from an initial thermal state in the
conformal SYM gauge theory to a final state of a massive gauge theory.
Meanwhile the profile with the $-$ sign then corresponds to the reverse
transition from the massive theory to the conformal theory.

We have introduced the time scale $\calt$ in eq.~\reef{ratel} to
control the rate of these transitions. This will allow us to consider
not only abrupt quenches (with $\calt\to0$) but also transitions made
over a finite time. In particular, we will also consider the adiabatic
limit where $\calt\to\infty$. We will find that the response to these
different quenches will depend on how $\calt$ compares to the thermal
time scale $1/T$. Hence it will be convenient in the following to
define the dimensionless ratio $\a\equiv\pi T_i\,\calt$, where $T_i$ is
the temperature of the initial state. Note that while our calculations
of these thermal quenches are perturbative in $m/T_i$, they are no
restrictions\footnote{One caveat to this statement will discussed in
section \ref{kronk}.} on the rate controlling ratio $T_i\calt$.

The holographic dual of the $\caln=2^*$ gauge theory consists of
Einstein gravity coupled to two massive scalar fields in five
dimensions, as will be described in detail in the next section. Of
course, within this framework, the initial thermal state is described
by an asymptotically anti-de Sitter (AdS) black hole and the quenches
are realized by varying the asymptotic of behaviour of the
corresponding scalar with the given profile \reef{ratel}. This
variation essentially excites the bulk scalar with a wave packet that
falls down into the horizon from the asymptotic boundary. Our numerical
simulations evaluate the evolution of these scalar excitations in
detail and in particular, allow us to determine the response of the
boundary theory, as measured by the expectation values of the stress
energy $\langle T_{ij} \rangle$ and the operator $\langle \calo_\Delta
\rangle$. Our main focus is to study the transitions from the massive
theory to the conformal theory, however, as we will explain, the
results for the reverse transitions readily follows from the previous
case, in the high temperature limit.

Before proceeding, let us summarize the results of our study:
\nxt New ultraviolet divergences appear into the quantum field theory
when the couplings are made to vary in time, as for the quenches
described above. Holography provides a well-defined approach to
regulating these new divergences and renormalizing the boundary theory,
even in the presence of such time varying couplings. In particular, the
necessary new counterterms are readily identified.
\nxt The response to fast quenches with $\a\ll 1$ exhibits an
interesting scaling behaviour in $\a$, similar to that noted previously
in \cite{shiraz,cy}. This scaling is apparent both in the immediate
response of the observables, \eg $\langle \calo_\Delta(t) \rangle$, and
in the comparison of the initial and final state, \eg the overall
change in the energy density. These scalings are particularly dramatic
in the case of the dimension three operator, where the immediate
response scales as $\a^{-2}\,\ln(1/\a)$ as $\a\to 0$. For both
operators, this scaling behaviour is divergent and naively suggests
that an abrupt quench with $\a=0$ is unphysical. However, our present
perturbative analysis is not sufficient to reliably confirm this
conclusion.
\nxt For very slow transitions with $\a\gg1$, the system essentially
maintains thermodynamic equilibrium throughout the process and the
response simply tracks the change in the coupling. The approach to the
adiabatic limit with $\a\to \infty$ also exhibits an interesting
scaling behaviour with $\a$. As expected the entropy production goes to
zero in this limit confirming that the adiabatic quenches are
reversible.
\nxt The relaxation of the system after a thermal quench of the
coupling is sensitive to whether the ratio of the characteristic
time-scale $\calt$ to the thermal scale $1/T_i$ is larger or smaller
than one, \ie whether $\a\gtrsim 1$ or $\a\lesssim 1$. For fast
quenches (with $\a\ll 1$), the relaxation is dominated by the lowest
quasinormal mode of the bulk scalar. This behaviour indicates that the
relaxation after fast quenches is universally controlled by the
characteristic thermal time scale $1/T_i$, as has been observed
previously in \cite{cy,Bantilan:2012vu,Heller:2012je}. In contrast, for
slow quenches (with $\a\gg 1$), the system relaxes with a (nearly)
adiabatic response and so the relaxation is governed entirely by the
late time profile of the coupling.
\nxt The holographic renormalization of the boundary theory with
time-dependent couplings introduces new scheme-dependent ambiguities.
This situation is rather reminiscent of the ambiguities that arise in
studying QFT in curved spacetime \cite{birrell}. All of these new
ambiguities only play a role while the couplings are varying and so
they are not important after the coupling has achieved its final value.
\vspace{0.5cm}

The remainder is organized as follows: In section \ref{describe}, we
review the dual gravitational description of strongly coupled
$\caln=2^*$ gauge theory, its holographic renormalization for arbitrary
sources, and the high-temperature equilibrium thermodynamics. In
section \ref{gravityX}, we derive the bulk equations of motion
describing the gravitational dual of the mass quenches \eqref{ratel}.
We also compute the expectation values of the stress-energy tensor and
the operator $\calo_\Delta$ throughout the quench. Our numerical
procedure is outlined in section \ref{numercl}. In section \ref{res},
we provide a detailed discussion of the results of our numerical
simulations. We conclude with further comments  and future directions
in section \ref{kronk}. Further details of the analysis, as well as the
description of the discretization of the evolution equations, are
discussed in appendix \ref{appd}.

\section{Gravitational description of $\caln=2^*$ gauge theory}
\label{describe}

In this paper, we will use holography to study quenches of the masses
for the $\caln=2^*$ gauge theory at strong coupling \cite{pw,bpp,j}.
However, to begin let us recall the field theoretic description of the
mass deformation of the $\caln=4$ SYM which yields the $\caln=2^*$
theory. The field content of the $\caln=4$ SYM theory includes the
gauge field $A_\mu$, four Majorana fermions $\psi_a$ and three complex
scalars $\phi_i$, where all of these fields are in the adjoint
representation. Now the SYM theory can be deformed by adding two
independent `mass' terms \cite{opera}
 \beq
\delta \call= -2\,\int d^4x\,\left[ \,m_b^2\,\calo_2
+m_f\,\calo_3\,\right]
 \labell{massterm}
 \eeq
where
 \beqa
\calo_2&=&\frac13 {\tr}\left(\, |\phi_1|^2 + |\phi_2|^2 - 2\,|\phi_3|^2
\,\right)\,,
 \labell{massb}\\
\calo_3&=& -{\tr}\left( i\,\psi_1\psi_2 -\sqrt{2}g_\mt{YM}\,\phi_3
[\phi_1,\phi_1^\dagger] +\sqrt{2}g_\mt{YM}\,\phi_3
[\phi_2^\dagger,\phi_2] + {\rm h.c.}\right)
 \labell{massf}\\
&&\qquad\qquad +\frac23 m_f\, {\tr}\left(\, |\phi_1|^2 + |\phi_2|^2 +
|\phi_3|^2\, \right)\,.
 \nonumber
 \eeqa
If one fixes $m_b=m_f$, the effect is to give masses to an $\caln=2$
hypermultiplet consisting of $\lambda_{1,2}$ and $\phi_{1,2}$, \ie the
resulting theory still preserves $\caln=2$ supersymmetry. Note that
beyond a fermion mass term, the dimension-3 operator contains trilinear
couplings between the hypermultiplet scalars and $\phi_3$, as well as a
mass term for all three scalars. Note that the latter is a
coupling-dependent correction induced at finite mass, \ie so that
$m_f\,\calo_3$ contains a contribution of order $m_f^2$. The presence
of these additional interactions is dictated by the supersymmetry
algebra and the latter mass terms can be distinguished from those in
$\calo_2$ by their transformation properties under the $SO(6)_R$
symmetry of the $\caln=4$ theory \cite{opera}. Further note that the
dimension-2 operator $\calo_2$ contains an unstable mass term for the
scalar $\phi_3$, which is in the $\caln=2$ vector multiplet, as well as
positive masses for the two scalars in the hypermultiplet. The two mass
contributions for $\phi_3$ only cancel with $m_b=m_f$, leaving this
scalar massless in the supersymmetric theory. In the following study of
mass quenches, we will vary $m_b$ and $m_f$ independently and so we are
not restricting our analysis to the supersymmetric theory. Our
assumption will be that the structure of the two independent operators
remains as described above. However, one should worry then that the
theory may be unstable,\footnote{A full description of these
instabilities (including their end-point) at strong coupling would
require extending the holographic framework described below and may
well require considering the full ten-dimensional string theory in the
bulk.} \ie when $m_b>m_f$, however, we will also be working at finite
temperature with $m_{b,f}/T\ll1$. In this case, `large' positive
thermal masses will be induced for all of the scalars, in particular
$\phi_3$, which will prevent any such instabilities from arising.

As we next describe, the holographic dual of these mass deformations is
well understood \cite{pw,bpp,j}. In the dual five-dimensional
supergravity, we can restrict our attention to a pair of scalars, which
we denote $\alpha$ and $\chi$. This construction and its `uplift' to
ten dimensions was originally elucidated by Pilch and Warner \cite{pw}.
In the full ten-dimensional type IIb supergravity, the two scalars are
Kaluza-Klein modes which deform the original $AdS_5\times S^5$ geometry
dual to the $\caln=4$ SYM theory. According to the general framework of
holographic RG flows \cite{m1}, the asymptotic boundary behaviour of
the supergravity scalars contains information about the couplings and
expectation values of the dual operators in the boundary theory. Here,
these are the mass parameters and the two operators presented in
eq.~\eqref{massterm}. Our nomenclature will be that the scalar $\a$ is
dual to the `{\it bosonic}' mass term $\calo_2$, in eq.~\reef{massb},
while $\chi$ is dual to the `{\it fermionic}' mass term $\calo_3$ in
eq.~\reef{massf}.

The appropriate terms in the five-dimensional supergravity action,
including the scalars $\alpha$ and $\chi$, can be written as
\begin{equation}
\begin{split}
I_5=&\,
\int_{\calm_5} d\xi^5 \sqrt{-g}\ \call_5=\frac{1}{4\pi G_5}\,
\int_{\calm_5} d\xi^5 \sqrt{-g}\left[\ft14 R-3 (\del\a)^2- (\del\chi)^2-
\calv(\a,\chi) \right]\,,
\end{split}
\eqlabell{action5o}
\end{equation}
where the potential takes the form
\begin{equation}
\calv(\a,\chi)=-e^{-4\alpha}-2e^{2\alpha}\cosh(2\chi)
+\frac1{4}\,e^{8\alpha}\sinh^2(2\chi)\,.
 \eqlabell{pp}
\end{equation}
Implicitly here, we set have the curvature radius of the
five-dimensional AdS vacuum to be unity, \ie $L=1$. With these
conventions, the five-dimensional Newton's constant becomes
\begin{equation}
G_5\equiv \frac{\pi}{2 N^2}\,.
\eqlabell{g5}
\end{equation}
The resulting Einstein equations can be written as
\begin{equation}
R_{\mu\nu}=12 \del_\mu \a \del_\nu\a+ 4 \del_\mu\chi\del_\nu\chi
+\frac{4}{3} g_{\mu\nu} \calv\,,
\eqlabell{ee}
\end{equation}
while the equations of motion for the scalars become
\begin{equation}
\jsquare\alpha=\fft16\fft{\del\calv}{\del\alpha}\qquad{\rm and}
\qquad
\jsquare\chi=\fft12\fft{\del\calv}{\del\chi}\,.
\eqlabell{scalar}
\end{equation}

As commented above, we are considering quenches of the mass parameters
at finite temperature and in the regime $m_{b,f}/T\ll1$. This means
that the dual holographic background will be an asymptotically AdS
black hole and we will work perturbatively in the amplitude of the bulk
scalars. In this case, the scalar sector of the effective action
\eqref{action5o} can be expanded up to quadratic order in $\a$ and
$\chi$. Furthermore, since the potential \reef{pp} does not contain a
quadratic term mixing the two scalars, each of them can be treated
independently, \ie we study quenches in
\begin{equation}
I_5=\frac{1}{16\pi G_5}\int d^5\xi \sqrt{-g} \left(R+12-\frac 12(\del\phi)^2
-\frac 12 m^2\phi^2+\calo(\phi^3)\right)\,,
\eqlabell{action5}
\end{equation}
with
\begin{equation}
m^2=\begin{cases}
-3\,,\ {\rm if}\ \phi=2\sqrt{2}\,\chi\,\ \Longleftrightarrow\
{\rm fermionic\ mass\ operator}\ \calo_3\,,\\
-4\,,\  {\rm if}\ \phi=2\sqrt{6}\,\a \,\ \Longleftrightarrow\
{\rm bosonic\ mass\ operator}\ \calo_2\,.
\end{cases}
\eqlabell{m23def}
\end{equation}

The quenches which we study will be homogeneous and isotropic in the
spatial directions on the boundary. Hence our ansatz for both the
background metric and the bulk scalar field depend only\footnote{ Some
limitations of this approximation are discussed in section 6. } on a
radial coordinate $r$ and a time $v$:
\begin{equation}
ds_5^2=-A(v,r)\ dv^2+\Sigma(v,r)^2\ d\vec{x}^2+2\,dr\, dv\,,
\qquad \phi=\phi(v,r)\,.
\eqlabell{anzats}
\end{equation}
The metric ansatz is adapted to infalling Eddington-Finkelstein (EF)
coordinates, which has proven useful in holographic investigations of
the dynamics of thermal systems \cite{cy,cy2,gfluid}. With this ansatz,
we will assume that the asymptotic AdS boundary occurs at $r\to\infty$
with
 \beq
 A(v,r)\to r^2\qquad{\rm and} \qquad \Sigma\to r\,.
 \labell{bound}
 \eeq

Given eq.~\reef{anzats}, we obtain the following equations of motion:
 \beqa
0&=&\frac 2A\ \del_r\!(\dot{\phi})+\frac{3\,\del_r\!\Sigma}{\Sigma A}\
\dot{\phi}+\frac{3\,\del_r\phi}{\Sigma A}\ \dot{\Sigma}-\frac{m^2}{A}\
\phi\,,
 \eqlabell{eoms3}\\
0&=&\Sigma\, \del_r\!(\dot{\Sigma})+2\dot{\Sigma}\,\del_r\!\Sigma
-2\Sigma^2+\frac{1}{12}m^2\phi^2\Sigma^2\,,
 \eqlabell{eoms1}\\
0&=&\del_r^2A-\frac{12}{\Sigma^2}\dot{\Sigma}\,\del_r\!\Sigma
+4+\dot{\phi}\,\del_r\phi-\frac 16 m^2\phi^2\,,
 \eqlabell{eoms2}\\
0&=&\ddot{\Sigma}-\frac 12 \del_r\!A\,\dot{\Sigma}+\frac 16 \Sigma\,
(\dot{\phi})^2\,,
 \eqlabell{eoms4}\\
0&=&\del_r^2\Sigma+\frac 16\Sigma\, (\del_r\phi)^2\,,
 \eqlabell{eoms5}
 \eeqa
where we have defined for any function $h(r,v)$,
\begin{equation}
\dot{h}\equiv \del_v h+\frac 12 A\, \del_r h\,.
\eqlabell{der}
\end{equation}
Note that eqs.~\reef{eoms4} and \eqref{eoms5} are constraint equations.
In particular, the radial derivative of eq.~\reef{eoms4} and the time
derivative of eq.~\reef{eoms5} are both implied by the first three
equations. Note also that the form of the metric \eqref{anzats} is invariant
under the residual diffeomorphism \cite{cy}
\begin{equation}
r\to r+f(v)\,,
\eqlabell{residual}
\end{equation}
for an arbitrary function $f(v)$ --- we use this as a check in our
analysis below.

Given eq.~\reef{bound}, it is straightforward to find the asymptotic
solution to eqs.~(\ref{eoms3}--\ref{eoms5}):\\
\nxt When $m^2=-3$,
 \beqa
\phi&=&\frac 1r\ p_0+\frac{1}{r^2}\ \left(-\frac12 p_0 a_1+p_0'\right)
+\frac{1}{r^3}\biggl(p_2-\left(\frac12 p_0''+\frac16 p_0^3\right) \ln
r\biggr) +\calo(r^{-4 }\ln r)\,,
 \nonumber\\
\Sigma&=&r+\frac 12 a_1+\calo(r^{-1})\,,
 \eqlabell{uvnonliner}\\
A&=&r^2+a_1 r+\frac14 a_1^2-\frac16 p_0^2 -\ a_1'+\frac
{1}{r^2}\biggl(a_4+\left(\frac16 p_0 p_0''+\frac{1}{36} p_0^4
-\frac16 (p_0')^2\right) \ln r\biggr)
 \nonumber\\
&&\qquad+\calo(r^{-3}\ln r)\,,
 \nonumber
 \eeqa
where $\{p_0,p_2,a_1,a_4\}$ are functions of the time $v$ and we have
defined $h'\equiv\del_v h$ for any function $h(v)$. In addition, on the
boundary, the constraint \eqref{eoms4} becomes
 \beqa
0&=&-2 a_4'+\frac23 (p_0')^2 a_1 +\frac{5}{27} p_0^3 p_0'-\frac23 p_0
a_1 p_0''+\frac23
 p_0' p_2-\frac23 p_0 p_2'-\frac19 p_0' p_0''+
\frac49 p_0 p_0'''
 \nonumber\\
&&\qquad-\frac23 p_0 a_1' p_0'+\frac13 a_1' p_0^2 a_1\,.
 \eqlabell{constnon}
 \eeqa
Note that if this constraint is imposed on the boundary (\ie at $r\to
\infty$) then eq.~\reef{eoms4} will be satisfied at all values of $r$
when the equations of motion (\ref{eoms3}--\ref{eoms2}) are satisfied.
Finally, we observe that under the residual diffeomorphism
\eqref{residual} one has
\begin{equation}
\left( \begin{array}{c}
p_0 \\
p_2 \\
a_1\\
a_4
\end{array} \right)
\to \left( \begin{array}{c}
p_0 \\
p_2 +p_0 (f^2+f a_1)-2 f p_0' \\
a_1+2 f\\
a_4
\end{array} \right) \,.
\eqlabell{resdim3}
\end{equation}
We emphasize that this transformation will allow us to choose a gauge
where $a_1=0$, which will do in section \ref{gravityX}. Further, we
note that the constraint \eqref{constnon} is invariant under the
transformation \eqref{resdim3}.\\
\nxt When $m^2=-4$,
 \beqa
\phi&=&\frac {1}{r^2}\ \left(p_0+p_0^l\ \ln r\right) +\calo(r^{-3 }\ln
r)\,,
 \nonumber\\
\Sigma&=&r+\frac 12 a_1+\calo(r^{-3}\ln^2 r)\,,
 \eqlabell{uvnonlinerdim2}\\
A&=&r^2+a_1 r+\frac14 a_1^2-a_1' +\frac
{1}{r^2}\biggl(a_4-\left(\frac{1}{54} (p_0^l)^2+\frac{2}{9} p_0 p_0^l
\right)\ \ln r-\frac 19 (p_0^l)^2\ \ln^2 r\biggr)
 \nonumber\\
&&\qquad+\calo(r^{-3}\ln r)\,, \nonumber
 \eeqa
where $\{p_0,p_0^l,a_1,a_4\}$ are functions of $v$. In this case, the
boundary constraint provided by eq.~\eqref{eoms4} is,
\begin{equation}
\begin{split}
0=&-2 a_4'-\frac{20}{81}p_0^l (p_0^l)'+\frac{8}{27} p_0 (p_0^l)'
-\frac{10}{27} p_0^l p_0'-\frac49 p_0 p_0'\,.
\end{split}
\eqlabell{constnon1}
\end{equation}
Under residual diffeomorphism  \eqref{residual} one has
\begin{equation}
\left( \begin{array}{c}
p_0^l \\
p_0 \\
a_1\\
a_4
\end{array} \right)
\to \left( \begin{array}{c}
p_0^l \\
p_0 \\
a_1+2 f\\
a_4
\end{array} \right) \,.
\eqlabell{resdim2}
\end{equation}
Hence again, we can use this diffeomorphism to set $a_1=0$. Further,
the constraint \eqref{constnon1} is now trivially invariant under the
transformation \eqref{resdim2}.

\subsection{Holographic renormalization} \label{holonorm8} 

The physical observables of the boundary theory are the correlation
functions of the gauge invariant operators. In the following analysis,
we will consider the simplest of these, namely, the one-point
functions. That is, we will examine the response of $\la T_{ij}\ra$ and
$\la\,\calo_\Delta\ra$ to quenches in the mass couplings, as given by
eq.~\eqref{ratel}. We will present our investigation of non-local
probes of these quenches (\eg two-point functions, Wilson lines and
entanglement entropy) elsewhere \cite{wip}.

The holographic computation of these one-point functions is most easily
implemented using Fefferman-Graham (FG) coordinates \cite{fefferman},
which we write here as
\begin{equation}
ds_5^2=\frac{d\r^2}{\r^2}+ \frac1{\r^2}\, G_{ij}(x^k,\r)\ dx^i dx^j  \ \,,
\eqlabell{mfg}
\end{equation}
and
\begin{equation}
\phi=\phi(x^k,\rho)\,.
\eqlabell{scalar3}
\end{equation}
The asymptotic AdS boundary now appears at $\r\to0$. As usual, to
produce finite results in the holographic calculations, one regulates
the bulk by introducing a cut-off surface at $\r=\e$, with some small
$\e$. However, the one-point functions extracted directly from the
action \eqref{action5} would still diverge as this cut-off is removed,
\ie as $\e\to 0$, and thus these observables need to be holographically
renormalized. The details of the holographic renormalization are
sensitive to the conformal dimension of the particular boundary
operator of interest, or correspondingly, to the mass $m^2$ of the bulk
scalar $\phi$ in eq.~\eqref{action5}. Hence we discuss the two cases
for $m^2$ separately.

\subsubsection{Renormalization of $m^2=-3$ bulk scalar}

First, we note that we are working with a flat boundary metric and
thus in the FG metric \reef{mfg},
\begin{equation}
\lim_{\r\to 0}G_{ij}=\eta_{ij}\,.
\eqlabell{gijas}
\end{equation}
Next we need to translate between the EF coordinates in
eq.~\reef{anzats} and the FG coordinates in eq.~\reef{mfg}. This
coordinate transformation is easily constructed perturbatively for
small $\r$ (large $r$) by comparing the asymptotic expansion of the
bulk solutions of eqs.~(\ref{eoms3}--\ref{eoms2}). Hence, using
eq.~\eqref{uvnonliner} for $m^2=-3$, we find
 \beqa
v&=&t-\r-\frac{1}{72} \r^3\ p_0(t)^2+\frac{1}{144} \r^4\ p_0(t)
p_0'(t)+\calo(\r^5\ln\r)\,,
 \nonumber\\
r&=&\frac{1}{\r}\biggl( 1-\frac 12 \r\ a_1(t)+\r^2 \left(\frac 12
a_1'(t)+\frac{1}{24} p_0(t)^2\right)
 \eqlabell{effg3}\\
&&\qquad +\r^3 \left(-\frac14 a_1''(t)-\frac{1}{18} p_0(t)
p_0'(t)\right)+\calo(\r^4\ln\r) \biggr)\,.
 \nonumber
 \eeqa
Note that as one approaches the boundary with $\r\to 0$, the EF time
$v$ and the FG or boundary time $t$ coincide, \ie
\begin{equation}
\lim_{r\to 0} v=t\,.
\eqlabell{EFisFG}
\end{equation}

Now the relevant boundary counterterms are given by \cite{n22,sk1,aby}:
 \beq
I_{ct}=I_{ct}^{divergent}+I_{ct}^{finite} \labell{scount0}
 \eeq
where
 \beqa
I_{ct}^{divergent}&=&\frac{1}{16\pi G_5}\int_{\del\calm_5,
\r=\e}\!\!\!\!\!\! d^4 x\ \sqrt{-\gamma}\biggl(6+\frac 12 \phi^2
+\frac{1}{12}\phi^4 \ln\r
 \eqlabell{scount}\\
&&\qquad\qquad\qquad\qquad\qquad+\frac{1}{2}\
\g^{ij}\del_i\phi\del_j\phi\ \ln \r+\frac {1}{12} R^\g \phi^2\
\ln\r\biggr)\,,
 \nonumber\\
I_{ct}^{finite}&=&\frac{1}{16\pi G_5}\int_{\del\calm_5,
\r=\e}\!\!\!\!\!\! d^4 x\ \sqrt{-\gamma}\biggl( \delta_1\ \phi^4+
\delta_2\ \g^{ij}\del_i\phi\del_j\phi+\delta_3\ R^\g \phi^2 \biggr)\,.
 \eqlabell{scountf}
 \eeqa
These expressions are written using
\begin{equation}
\g_{ij}\ dx^jdx^j \equiv \r^{-2}\ G_{ij}(x^k,\r)\ dx^i dx^j\,.
\eqlabell{defgamma}
\end{equation}
Implicitly then, the boundary action contains many potentially
divergent factors, \eg $\sqrt{-\g}=O(\e^{-4})$. Above, we have
separated the counterterms which diverge in the limit $\e\to 0$ from
the finite counterterms. Further $R^\g$ is the Ricci scalar constructed
from $\g_{ij}$ treated as a four-dimensional metric with an external
parameter $\r$. Note that even though the boundary metric \eqref{gijas}
is flat, the term proportional to $R^\g \phi^2$ contributes to the
holographic renormalization in the presence of space-time varying
sources \cite{aby}. As usual, the coefficients $\delta_i$ of the finite
counterterms \reef{scountf} are arbitrary constants, which are related
to ambiguities in the renormalization scheme.

Combining all of these expressions to compute the renormalized
one-point functions for
\begin{equation}
\la T_{ij}\ra\equiv \biggl(\cale,\calp,\calp,\calp\biggr)\,,
\eqlabell{defep}
\end{equation}
 and
$\la\calo_3\ra$, we find
\begin{equation}
\begin{split}
8\pi G_5\ \cale=&-\frac32 a_4-\frac{1}{12} (p_0')^2
-\frac12 p_0 a_1 p_0'+\frac18 p_0^2 a_1^2-\frac12 p_0 p_2
+\frac13 p_0 p_0''+\frac{7}{288} p_0^4\\
&+\cale^{ambiguity}\,,\\
8\pi G_5\ \calp=&-\frac12 a_4-\frac{1}{36} (p_0')^2+\frac16 p_0 a_1
p_0'-\frac{1}{24} p_0^2 a_1^2+\frac{1}{6} p_0 p_2-\frac{1}{18} p_0
p_0''+\frac{7}{864} p_0^4\\
&+\calp^{ambiguity}\,,\\
16\pi G_5\ \la\calo_3\ra=&\frac12 p_0''-\frac{1}{12} p_0^3-2 a_1
p_0'+\frac12 p_0 a_1^2-2 p_2+\calo_3^{ambiguity}\,.
\end{split}
\eqlabell{vev1}
\end{equation}
Here we employ the superscript `${ambiguity}$' to denote
renormalization scheme ambiguities introduced by the finite
counterterms \eqref{scountf}
\begin{equation}
\begin{split}
\cale^{ambiguity}=&\frac12 \delta_1 p_0^4+\frac12 \delta_2 (p_0')^2\,,\\
\calp^{ambiguity}=&-2 \delta_3 (p_0')^2-2 \delta_3 p_0 (p_0'')-\frac12
 \delta_1 p_0^4+\frac12 \delta_2 (p_0')^2\,,\\
\calo_3^{ambiguity}=&4 \delta_1 p_0^3+2 \delta_2 p_0''\,.
\end{split}
\eqlabell{ambiguities}
\end{equation}
As expected, the physical one-point functions in eq.~\eqref{vev1} are
invariant under the residual diffeomorphism \eqref{resdim3} of the
background metric \eqref{anzats}.

These one-point correlators are subject to various Ward identities
\cite{sk1}. In particular, one has the diffeomorphism Ward identity
\begin{equation}
\del^i \la\, T_{ij}\ra=-\la\calo_3\ra\ \del_j p_0\,.
\eqlabell{warddiffeo}
\end{equation}
Of course, when the mass parameter is held constant, this expression
reduces to the conservation of energy and momentum in the boundary
theory. In the present case, the configurations are independent of the
spatial coordinates and so only the $j=t$ component of
eq.~\reef{warddiffeo} is nontrivial, yielding
 \beq
\partial_t\,\cale = \la\calo_3\ra\ \del_t p_0\,.
 \labell{worky}
 \eeq
With a time-dependent $p_0$, the contribution on the right-hand side
describes the work done by varying the fermionic mass in the boundary
theory.\footnote{Later we will see that $p_0=\sqrt{2}\,m_f$.} We also
note that this equation is scheme independent, \ie
eqs.~\eqref{warddiffeo} and \reef{worky} hold independent of the values
of $\delta_i$. Finally we observe that the constraint \eqref{eoms4}
reduces to this identity on the asymptotic boundary $\r=0$, \ie
eqs.~\reef{constnon} and \reef{worky} are precisely the same.

We also have the conformal Ward identify
\begin{equation}
\la\, T_{i}^{\ i}\ra=-p_0\ \la\calo_3\ra
+\frac{1}{16\pi G_5}\left(\frac 12 p_0 \Box p_0-\frac{1}{12}p_0^4+\dd\,
\Box(p_0^2)\right)\,,
\eqlabell{wardconf}
\end{equation}
where
\begin{equation}
\dd=-\dd_2+6\,\dd_3\,.
\eqlabell{d3tune}
\end{equation}
Examining the right-hand side of this expression, the first term
represents the `classical' contribution from breaking conformal
invariance by the introduction of a dimensionful coupling $p_0$. Note
that this term carries an overall factor of $(\Delta-d)$, which happens
to be --1 in the present case. The remaining terms can be interpreted
as anomalous contributions. The quadratic term is a standard scalar
anomaly that was first studied in \cite{pet}. In a curved space
background, this term would take the form of a Paneitz operator:
$p_0(\Box- R^\g/6)p_0$ \cite{graham}. The quartic term was previously
observed in \cite{n22}. The final term is a scheme dependent total
derivative that is induced by the finite counterterms. While the
$\dd_2$ and $\dd_3$ counterterms in eq.~\eqref{scount} are not
conformally invariant, they still respect scale invariance and so can
only contribute with a total derivative in eq.~\eqref{wardconf}
\cite{scale}. Note that with $\dd_3=\dd_2/6$, these two counterterms
combine to form a conformally invariant combination and hence the total
derivative contribution above vanishes. Of course, the usual
curvature-squared contributions to the trace anomaly \cite{sken} do not
appear on the right-hand side of eq.~\eqref{wardconf} because we are
examining the boundary theory in flat space, \eg see eq.~\eqref{gijas}.

\subsubsection{Renormalization of $m^2=-4$ bulk scalar}

Constructing the transformation between EF and FG coordinates, using
\eqref{uvnonlinerdim2} for $m^2=-4$, we find
\begin{equation}
\begin{split}
v=&\ t-\r+\calo(\r^5\ln\r)\,,\\
r=&\ \frac{1}{\r}\biggl(
1-\frac12 \r\ a_1(t)+\frac 12 \r^2\ a_1'(t)-\frac 14 \r^3\ a_1'(t)
+\calo(\r^4\ln\r)
\biggr)\,.
\end{split}
\eqlabell{effg2}
\end{equation}
Note that as in the previous case, we have $v\to t$ on the asymptotic
boundary, \ie as $\r\to0$.

In this case, we write the relevant boundary counterterms as
\cite{n22,sk1}:
 \beq
I_{ct}=I_{ct}^{divergent}+I_{ct}^{finite}\,,
 \labell{scountdim2t}
 \eeq
with
 \beqa
I_{ct}^{divergent}&=&\frac{1}{16\pi G_5}\int_{\del\calm_5, \r=\e}
\!\!\! \!\!\! d^4 x\ \sqrt{-\gamma}\biggl(6+ \phi^2 +\frac{1}{2\ln\r}
\,\phi^2\biggr)\,,
 \labell{scountdim2}\\
I_{ct}^{finite}&=&\frac{1}{16\pi G_5}\int_{\del\calm_5, \r=\e} \!\!\!
\!\!\! d^4 x\ \sqrt{-\gamma}\biggl( \frac{\dd_1}{\ln^2\r}\
\phi^2+\frac{\dd_2}{\ln\r}\ R^\g \phi \biggr)\,.
 \eqlabell{scountdim2f}
 \eeqa
Again we have separated the counterterms which diverge in the limit
$\e\to 0$ from those which remain finite. Further recall that the
coefficients $\delta_i$ of the finite counterterms \reef{scountdim2f}
are arbitrary constants reflecting the ambiguities in the
renormalization scheme. Notice that there is a finite counterterm
involving the curvature term, \ie $\frac{1}{\ln\r}\ R^\g\phi$, even
though the divergent counterterm $R^\g\phi$ is absent. The absence of
such a divergent counterterm was established in \cite{n22}. Since the
bulk action \eqref{action5} is even in $\phi$, all divergences must be
even in $\phi$ as well. However, the candidate counterterm in question
is linear in $\phi$ and so does not occur.

Transforming the expressions in eq.~\eqref{uvnonlinerdim2} to FG form
and computing the renormalized one-point functions of $\la T_{ij}\ra$
and $\la\calo_2\ra$, we find
\begin{equation}
\begin{split}
8\pi G_5\ \cale=&-\frac{5}{18} p_0^l p_0-\frac32 a_4
-\frac{5}{54} (p_0^l)^2-\frac16 p_0^2+\cale^{ambiguity}\,,\\
8\pi G_5\ \calp=&\, \frac{13}{54} p_0^l p_0-\frac12 a_4+\frac{17}{324} (p_0^l)^2
-\frac{1}{18} p_0^2+\calp^{ambiguity}\,,\\
16\pi G_5\ \la\calo_2\ra=&-p_0+\calo_2^{ambiguity}\,,
\end{split}
\eqlabell{vev12}
\end{equation}
where
\begin{equation}
\begin{split}
\cale^{ambiguity}=&\frac12 \delta_1 (p_0^l)^2\,,\\
\calp^{ambiguity}=&-\frac 12 \delta_1 (p_0^l)^2+\delta_2 (p_0^l)''\,,\\
\calo_2^{ambiguity}=&2 \delta_1 p_0^l\,.
\end{split}
\eqlabell{ambiguities2}
\end{equation}
As expected, the physical one-point correlation functions in
\eqref{vev12} are invariant under the residual diffeomorphisms
\eqref{resdim2} of the background metric in eq.~\eqref{anzats}.

Again, the one-point correlators satisfy Ward identities. In particular,
the diffeomorphism Ward identity now becomes
\begin{equation}
\del^i \la \,T_{ij}\ra=-\la\calo_2\ra\ \del_j p_0^l\,.
\eqlabell{warddiffeo1}
\end{equation}
As in the previous case, this equation contains a single nontrivial
component here,\footnote{Note that we will see
$p_0^l=\sqrt{2/3}\,m_b^2$ below.}
 \beq
\del_t\,\cale = \la\calo_2\ra\ \del_t p_0^l\,.
 \labell{worky2}
 \eeq
As before, this expression is scheme independent, \ie it holds for
arbitrary $\delta_i$, and it is equivalent to the constraint
\eqref{constnon1} when the latter is taken to the asymptotic boundary.
The conformal Ward identify becomes
\begin{equation}
\la T_{i}^{\ i}\ra=-2 p_0^l\ \la\calo_2\ra
+\frac{1}{16\pi G_5}\ \left(\frac 12 (p_0^l)^2-6\delta_2\,
\Box(p_0^l)\right)\,,
\eqlabell{wardconf1}
\end{equation}
which has the same general form as described with the fermionic mass
operator in the previous subsection.

\subsection{High temperature equilibrium thermodynamics}
\label{equib}

Having established the equations of motion for the bulk theory
\reef{action5} of gravity coupled to a free scalar, as well as the
asymptotic expansions required to extract the one-point functions, we
apply these results here to examine time-independent or equilibrium
configurations. High temperature equilibrium thermodynamics of the
$\caln=2^*$ gauge theory has been extensively discussed in
\cite{n21,n22,n23}. In the following, we highlight some of the salient
results of these investigations. In particular, this allows us to
connect the bulk parameters to those in the $\caln=2^*$ gauge theory,
\eg as given in eqs.~\reef{map3} and \reef{bamm2}. Note that our
nomenclature here should be interpreted as follows: `high temperature'
implies that $m_{b,f}/T\ll1$ while `equilibrium' implies that the bulk
gravity solution will be time-independent.

Recall that in the high temperature limit, we are working
perturbatively in the amplitude of the bulk scalar. Hence the zero'th
order solution will be a planar AdS black hole with the bulk scalar
field set to zero. To leading order in the expansion in $m_{b,f}/T$, it
is sufficient to solve the linearized equations for the scalar $\phi$
in the black hole background. The scalar only backreacts on this
background geometry at the quadratic order. Thus given the ansatz
\reef{anzats}, we write
\begin{equation}
\begin{split}
&\phi(v,r)=\lambda\  \phi_1^e(r)+\calo(\l^3)\,,\\
&A(v,r)=r^2-\frac{\mu^4}{r^2}+\l^2\ \mu^2\ A_2^e(r)+\calo(\l^4)\,,\\
&\Sigma(v,r)=r+\l^2\ \mu\ \Sigma_2^e(r)+\calo(\l^4) \,,
\end{split}
\eqlabell{perturb}
\end{equation}
where we have introduced $\l$ as an expansion counting parameter.
Setting $\lambda=0$ leaves the AdS black hole, where $r=\mu$ is the
position of the event horizon. Note that we are using the superscript
${}^e$ above to denote quantities corresponding to the equilibrium (or
static) solution.

From eq.~\eqref{eoms3}, the linearized scalar equation becomes
\begin{equation}
0=\del^2_{rr}\phi_1^e+\frac{5r^4-\mu^4}{r(r^4-\mu^4)}\
\del_r \phi_1^e+\frac{m^2 r^2}{r^4-\mu^4}\ \phi_1^e\,.
\eqlabell{eq1}
\end{equation}
It is convenient to introduce a new radial coordinate
$\r$,\footnote{Note that this radial coordinate is distinct from the
the FG radius $\r$, which is related to the EF radius $r$ in
eqs.~\reef{effg3} and \reef{effg2}.}
\begin{equation}
\r=\frac \mu r\,,
\eqlabell{defrho}
\end{equation}
such that the asymptotic boundary ($r\to\infty$) occurs at $\rho\to0$
while the horizon ($r\to\mu$) is positioned at $\r\to1^-$ (to zero'th
order in the $\lambda$ expansion). In terms of $\rho$, eq.~\eqref{eq1}
takes the form
\begin{equation}
0=\del^2_{\r\r}\phi_1^e-\frac{3+\rho^4}{\rho(1-\rho^4)}\
\del_\r\phi_1^e-\frac{m^2}{\rho^2(1-\rho^4)}\ \phi_1^e\,.
\eqlabell{eq1rho}
\end{equation}
The solutions are readily expressed in terms of hypergeometric functions
and then in each case, they take the form:\\
\nxt {For} $m^2=-3$,
\begin{equation}
\phi_1^e=\pi^{-1/2}\ \Gamma\left(\frac 34\right)^2\ \r^3
\  _2 F_1\left(\frac 34,\frac34, 1, 1-\r^4\right)\,.
\eqlabell{solve3}
\end{equation}
This particular solution is chosen by demanding regularity at the
horizon. The overall normalization constant is chosen here in such a
way that
\begin{equation}
p_0=\l\ \mu\,,
 \labell{bamm}
\end{equation}
where $p_0$ is the leading coefficient in the asymptotic expansion of
the scalar given in eq.~\eqref{uvnonliner}. Given the above solution
\eqref{solve3}, we can also extract the next independent coefficient at
order $\rho^3$ ($\propto r^{-3}$) in the latter expansion as
\begin{equation}
p_2=-\l\ \mu^3\ \frac{\Gamma\left(\frac 34\right)^4}{\pi^2}\,.
\eqlabell{p2}
\end{equation}
Further given the linearized solution \eqref{solve3}, we can choose the
asymptotic solution of $A_2^e$  as
\begin{equation}
A_2^e(\r)=-\frac16+\calc\r^2+O(\r^3\ln\r)\,,
\eqlabell{a2edim3}
\end{equation}
where $\calc$ is an arbitrary constant. With this choice, we identify
in eq.~\eqref{uvnonliner}
\begin{equation}
a_1=0\,,\qquad a_4=-\mu^4+\l^2\mu^4\calc\,.
\eqlabell{dim3a1a4}
\end{equation}
As noted at eq.~\reef{resdim3}, $a_1=0$ corresponds to an implicit
gauge choice that was made in choosing the form of $A_2^e(\r)$ in
eq.~\reef{a2edim3} --- compare this expression to the general form in
eq.~\reef{uvnonliner}. Similarly, the constant $\calc$ can be chosen
arbitrarily with a gauge choice which fixes the position of the
horizon, \ie $r_{\ssc H}^4 = \mu^4\left(1-\l^2\calc +\calo(\l^4)
\right)$. In a discussion of the equilibrium thermodynamics, the most
convenient choice is to simply set $\calc=0$. However, we will see the
analogous constant cannot be avoided in the following studies since the
quenches naturally cause the horizon to grow.

Now using eq.~\eqref{vev1}, we can compute the equilibrium expectation
values for the energy density $\cale^e$, the pressure $\calp^e$ and the
fermionic mass operator $\langle\calo_3\rangle^e$:
\begin{equation}
\begin{split}
8\pi G_5\ \cale^e=&\ \frac32 \mu^4\left(1-\l^2\calc\right)+ \mu^4\l^2\
\frac{\Gamma\left(\frac 34\right)^4}{2\pi^2}+\calo(\l^4)\,,\\
8\pi G_5\ \calp^e=&\ \frac12 \mu^4\left(1-\l^2\calc\right)
-\mu^4\l^2\ \frac{\Gamma\left(\frac 34\right)^4}{6\
\pi^2}+\calo(\l^4)\,,\\
16\pi G_5\ \la\calo_3\ra^e=&\ \mu^3\l\  \frac{2\ \Gamma\left(\frac 34\right)^4}{\pi^2}
+\calo(\l^3)\,.
\end{split}
\eqlabell{vev1e}
\end{equation}
Note that to the order presented, there is no ambiguity in $\{\cale^e,
\calp^e, \la\calo_3\ra^e\}$. From eq.~\reef{ambiguities}, we can see
that for a time independent mass, the ambiguities first arise at order
$\l^3$ in $\la\calo_3\ra^e$, while they only arise at order $\l^4$ in
$\cale^e$ and $\calp^e$.

Next we compare eq.~\eqref{vev1e} with the energy density and the
pressure given in \cite{n23}
\begin{equation}
\begin{split}
\cale^e=&\frac 38 \pi^2 N^2 T^4 \left(1-\frac {2\
\Gamma\left(\frac 34\right)^4}{3\pi^4}\  \frac{m_f^2}{T^2}\right)\,,\\
\calp^e=&\frac 18 \pi^2 N^2 T^4 \left(1-\frac {2\
\Gamma\left(\frac 34\right)^4}{\pi^4}\  \frac{m_f^2}{T^2}\right)\,,
\end{split}
\eqlabell{equil3}
\end{equation}
for the high-temperature thermodynamics of $\caln=2^*$ gauge theory.
This comparison allows us to identify
\begin{equation}
\mu=\pi T\left(1+\left(\frac{\calc}{2\pi^2}-\frac{\Gamma\left(\frac 34\right)^4}
{3\pi^4}\right)\ \frac{m_f^2}{T^2}+O\left(\frac{m_f^4}{T^4}\right)\right)\,,
\qquad \l=\frac{\sqrt{2} m_f}{\pi T} \left(1
+O\left(\frac{m_f^2}{T^2}\right)\right)\,.
\eqlabell{map3}
\end{equation}
Combining these expressions with eq.~\reef{bamm} yields
 \beq
 p_0=\sqrt{2}\,m_f\,\left(1+O\left(\frac{m_f^2}{T^2}\right)\right)\,.
 \labell{bamm2}
 \eeq
Similarly, the equilibrium expectation value of the fermionic mass term
in eq.~\reef{vev1e} becomes
 \beq
\la\calo_3\ra^e=\frac{\sqrt{2}\,\Gamma\left(\frac 34\right)^4}{4\pi^2}
\,N^2\,m_f\,T^2\,\left(1+O\left(\frac{m_f^2}{T^2}\right)\right)\,.
 \labell{expO3}
 \eeq
{}
\nxt {For} $m^2=-4$, the linearized solution for the scalar field
becomes
\begin{equation}
\phi_1^e=\frac\pi4\ \r^2\  _2 F_1\left(\frac 12,\frac12, 1, 1-\r^4\right)\,.
\eqlabell{solve2}
\end{equation}
Again, this particular solution is chosen to be regular at the horizon.
Here the overall normalization constant is chosen in such a way that
\begin{equation}
p_0^l=\l\ \mu^2\,,
 \labell{bamm3}
\end{equation}
in the asymptotic expansion of the bulk scalar in
eq.~\eqref{uvnonlinerdim2}. Further, the second independent coefficient
in the latter expansion becomes
\begin{equation}
p_0=-\l\ \mu^2\ \ln (\mu/2)\,,
\eqlabell{p2dim2}
\end{equation}
given the solution \reef{solve2} above. Without loss of generality, we
write the asymptotic solution of $A_2^e$  as
\begin{equation}
A_2^e(\r)=\left(\calc+\left(\frac{1}{54}+\frac 29 \ln 2\right)\ln \r
-\frac 19 \ln^2\r\right)\r^2+\calo(\r^3\ln\r)\,,
\eqlabell{ae2edim3}
\end{equation}
where $\calc$ is an arbitrary constant. In this case comparing
eq.~\eqref{uvnonlinerdim2}, we identify
\begin{equation}
a_1=0\,,\qquad a_4=-\mu^4+\l^2\mu^4\left(\calc+\left(\frac{1}{54}+
\frac 29 \ln 2\right)\ln \mu -\frac 19 \ln^2\mu\right)\,.
\eqlabell{dim2a1a4}
\end{equation}
As discussed above, gauge choices are implicitly at work in setting
$a_1=0$ and fixing a final value for $\calc$.

Using eq.~\eqref{vev12}, we can now compute the equilibrium expectation
values for the energy density $\cale^e$, the pressure $\calp^e$ and the
bosonic mass operator $\langle\calo_2\rangle^e$:
\begin{equation}
\begin{split}
&8\pi G_5\ \cale^e=\frac32 \mu^4+ \l^2\ \mu^4\
\biggl(-\frac32\calc+ \frac 14\ln\mu
-\frac{5}{18}\ln2-\frac{5}{54}-\frac16\ln^22\biggr)+\cale^e_{ambiguity}
+\calo(\l^4)\,,\\
&8\pi G_5 \calp^e=\frac12 \mu^4+\l^2\ \mu^4\
\biggl(-\frac12\calc-\frac14\ln\mu+
\frac{13}{54}\ln 2+\frac{17}{324}-\frac{1}
{18}\ln^22
\biggr)
+\calp^e_{ambiguity}
+\calo(\l^4)\,,\\
&16\pi G_5\ \la\calo_2\ra^e=\l\ \mu^2\ \ln\frac \mu2+\calo_{ambiguity}^{e}
+\calo(\l^3)\,,
\end{split}
\eqlabell{vev1edim2}
\end{equation}
where
\begin{equation}
\begin{split}
\cale^e_{ambiguity}=&\ \frac 12 \dd_1 \l^2\ \mu^4\,,\\
\calp^e_{ambiguity}=&\, -\frac 12 \dd_1 \l^2\ \mu^4\,,\\
\calo_{ambiguity}^{e}=&\ 2\ \dd_1 \l\ \mu^2\,.
\end{split}
\eqlabell{renam2}
\end{equation}

Next we compare the energy density and pressure in
eq.~\eqref{vev1edim2} with the analogous expressions in \cite{n23}:
\begin{equation}
\begin{split}
\cale^e=&\frac 38 \pi^2 N^2 T^4 \left(1+\left(\ln\frac{\pi T}{\Lambda}
-1\right)
 \frac{m_b^4}{9\pi^4T^4}\right)\,,\\
\calp^e=&\frac 18 \pi^2 N^2 T^4 \left(1
-\ln\frac{\pi T}{\Lambda}\ \frac{m_b^4}{3\pi^4 T^4}\right)\,,
\end{split}
\eqlabell{equil2}
\end{equation}
where $\Lambda$ is an arbitrary scale in the theory. In this way, we
identify
\begin{equation}
\begin{split}
\mu=&\ \pi T \biggl(1+\frac{m_b^4}{972\,\pi^4 T^4} \left(162\,\calc+3\ln 2 +18\ln^22-17\right)
\biggr)\,,\\
\l=&\ \sqrt{\frac23}\ \frac{m_b^2}{\pi^2 T^2}\,,\qquad \dd_1=
-\frac 12\ln\frac\Lambda2\,.
\end{split}
\eqlabell{map2}
\end{equation}
Notice that the renormalization scheme ambiguity $\dd_1$ is identified
with the ambiguity of choosing the scale $\Lambda$ in
eq.~\eqref{equil2}. Combining these expressions with eq.~\reef{bamm}
yields
 \beq
 p_0^l=\sqrt{\frac23}\,m_b^2\,\left(1+O\left(\frac{m_b^4}{T^4}\right)\right)\,.
 \labell{bamm4}
 \eeq
Similarly, the equilibrium expectation value of the fermionic mass term
in eq.~\reef{vev1e} becomes
 \beq
\la\calo_2\ra^e=\sqrt{\frac23}\frac{N^2}{8\pi^2} \,m_b^2\,\ln\frac{\pi
T}{\Lambda}\,\left(1+O\left(\frac{m_b^4}{T^4}\right)\right)\,.
 \labell{expO2}
 \eeq

\section{Holographic mass quenches at high temperatures}
\label{gravityX}

Next we apply our results here to holographic quenches where the mass
parameters in the dual gauge theory are varied as in eq.~\reef{ratel}.
Let us sketch the general approach here for $m^2=-3$:\footnote{This
discussion extends to the case $m^2=-4$ in an obvious way.} Choosing a
specific time-dependent profile for $m_f$ corresponds to fixing the
asymptotic function $p_0$ according to eq.~\reef{bamm2}. Given this
boundary condition, we then numerically solve the scalar field equation
\reef{eoms3} which allows us to determine the subleading coefficient
$p_2$. Note that as in the previous section with the high temperature
approximation, we are working perturbatively in the amplitude of the
scalar and so this equation is solved to linear order in $\lambda$ with
the static black hole background (\ie setting $A=r^2-\mu/r^2$ and
$\Sigma=r$). Given both $p_0$ and $p_2$, we can evaluate the one-point
function $\la \calo_3\ra$ using eq.~\reef{vev1}.\footnote{We will make
the gauge choice $a_1=0$ to simplify this expression, as well as later
calculations.} Further, we can integrate the boundary constraint
\reef{constnon} to determine the metric coefficient $a_4$. Having
calculated the latter, we can evaluate the energy density $\cale$ and
pressure $\calp$ (at order $\lambda^2$) using the expressions in
eq.~\reef{vev1}. Note then that we are able to evaluate all of the
one-point functions to leading order in the high temperature
approximation without actually solving for the time-dependence of bulk
metric (\ie without solving for $A_2(v,r)$ and $\Sigma_2(v,r)$ below).
However, the latter will be required to evaluate non-local probes of
the quenches \cite{wip}.

As noted above, we are still investigating the time-dependent
configurations in the high-temperature approximation, $m_{f,b}/T\ll1$.
Hence it is still sufficient to first solve the linearized equation for
the scalar $\phi$ with the equilibrium black hole background and then
consider the backreation of the scalar on the geometry at quadratic
order. We thus organize the perturbative expansion of the ansatz
\reef{anzats} with
\begin{equation}
\begin{split}
&\phi(v,r)=\lambda\  \phi_1(v,r)+\calo(\l^3)\,,\\
&A(v,r)=r^2-\frac{\mu^4}{r^2}+\l^2\ \mu^2\ A_2(v,r)+\calo(\l^4)\,,\\
&\Sigma(v,r)=r+\l^2\ \mu\ \Sigma_2(v,r)+\calo(\l^4)\,,
\end{split}
\eqlabell{perturbd}
\end{equation}
which is a simple extension of eq.~\reef{perturb} to a time-dependent
situation. Then introducing  coordinates,
\begin{equation}
\r=\frac \mu r\qquad{\rm and}\qquad \t=\mu\ v\,,
\eqlabell{coordinate}
\end{equation}
the scalar field equation \reef{eoms3} becomes, to linear order in $\l$:
\begin{equation}
0=\del^2_{\t\r}\phi_1-\frac 12(1-\r^4)\ \del^2_{\r\r}\phi_1
-\frac {3}{2\r}\del_\t \phi_1+\frac{3+\r^4}{2\r}\ \del_\r\phi_1+\frac {m^2}{2\r^2}\
\phi_1\,.
\eqlabell{dec2}
\end{equation}
Of course, this equation reduces to eq.~\reef{eq1rho} with a
time-independent scalar. Further at order $\l^2$, the metric equations
\reef{eoms1} and \reef{eoms2} yield
\begin{equation}
\begin{split}
0=&\del_{\t\r}\Sigma_2-\frac 12 (1-\r^4)\ \del_{\r\r}^2\Sigma_2-\frac 2\r\ \del_\t\Sigma_2
+\frac 2\r\
\del_\r\Sigma_2 +\frac 12\ \del_\r A_2+\frac{(3-\r^4)}{\r^2}\ \Sigma_2-\frac 1\r\ A_2
\\
&-\frac{m^2}{12\r^3}\ \phi_1^2\,, \\
0=&\del_{\r\r}^2 A_2+\frac 2\r\ \del_\r A_2+\frac{12(1-\r^4)}{\r^2}\ \del_\r\Sigma_2
-\frac{12}{\r^2}\ \del_\t\Sigma_2+\frac{12(1-\r^4)}{\r^3}\ \Sigma_2-\frac{6}{\r^2}\ A_2
\\
&+\frac{1-\r^4}{2\r^2}\ \left(\del_\r\phi_1\right)^2-\frac {1}{\r^2}\del_\r\phi_1\del_\t\phi_1
-\frac{m^2}{6\r^4}\ \phi_1^2 \,,
\end{split}
\eqlabell{dec3}
\end{equation}
while the constraint equations \reef{eoms4} and \reef{eoms5} become
\begin{equation}
\begin{split}
0=&\del^2_{\t\t}\Sigma_2+ \frac14 (1-\r^4)^2\ \del_{\r\r}^2\Sigma_2-(1-\r^4)\ \del_{\t\r}^2\Sigma_2
-\frac{1+\r^4}{\r}\ \del_\t \Sigma_2+\frac{(1-\r^4)^2}{2\r}\ \del_\r\Sigma_2\\
&+\frac 12\
\del_\t A_2
-\frac {1-\r^4}{6\r}\ \del_\t\phi_1\del_\r\phi_1
+\frac {(1-\r^4)^2}{24\r}\ \left(\del_\r\phi_1\right)^2+\frac{1}{6\r}\ \left(\del_\t\phi_1\right)^2\,,
\\
0=&\del_{\r\r}^2\Sigma_2+\frac 2\r\ \del_\r\Sigma_2+\frac {1}{6\r}\ \left(\del_\r\phi_1\right)^2\,.
\end{split}
\eqlabell{dec4}
\end{equation}
It is now straightforward to find the asymptotic solutions (as $\r\to
0$) of eqs.~(\ref{dec2}--\ref{dec4}).
\\
\nxt When $m^2=-3$
\begin{equation}
\begin{split}
\phi_1=&p_{1,0}\ \rho+p_{1,0}'\ \rho^2+\left(p_{1,2}+\frac 12 p_{1,0}''\ \ln\rho\right)\ \rho^3+\calo(\r^4\ln\r)\,,\\
\Sigma_2=&-\frac{1}{12}p_{1,0}^2\ \rho-\frac{1}{9} p_{1,0}  p_{1,0}'\ \rho^2 +\calo(\r^3\ln\r)\,,\\
A_2=&-\frac 16 p_{1,0}^2+\left(a_{2,4}+\frac 16 \left((p_{1,0}')^2- p_{1,0} p_{1,0}''
\right) \ln\rho\right)\r^2+\calo(\r^3\ln\r)\,.
\end{split}
\eqlabell{uvperdim3}
\end{equation}
In addition, the first equation in eq.~\eqref{dec4} provides the
following constraint:
\begin{equation}
0=a_{2,4}'+\frac 13 \left(p_{1,0}\ p_{1,2}'- p_{1,0}'\ p_{1,2}\right)+\frac{1
}{18} p_{1,0}'\ p_{1,0}''-\frac 29 p_{1,0}\ p_{1,0}'''\,.
\eqlabell{uvperdim3c}
\end{equation}
In eqs.~\eqref{uvperdim3} and \eqref{uvperdim3c}, we have
$p_{1,0}=p_{1,0}(\t)$, $p_{1,2}=p_{1,2}(\tau)$ and
$a_{2,4}=a_{2,4}(\t)$, while the prime ${}'$ denotes differentiation
with respect to $\t$.

Comparing eqs.~\eqref{uvnonliner} and \eqref{uvperdim3},  we identify
\begin{equation}
\begin{split}
&p_0=\l\mu\ p_{1,0}\,,\qquad p_2=\l\mu^3\ \left(p_{1,2}+p_{1,0}''\ \frac 12 \ln \mu \right)\,,\\
&a_1=0\,,\qquad a_4=-\mu^4+\l^2 \mu^4\
\left(a_{2,4}+\frac 16 \left((p_{1,0}')^2- p_{1,0} p_{1,0}''
\right)\ln\mu\right)\,.
\end{split}
\eqlabell{iddim3}
\end{equation}
Note that $a_1=0$ was an implicit gauge choice made in writing
$\Sigma_2$ and $A_2$ in eq.~\reef{uvperdim3}. Further we note that the
boundary constraint \reef{uvperdim3c} is precisely the constraint
\reef{constnon} evaluated at order $\l^2$.

As argued above, it is enough to solve numerically only the scalar field
equation \eqref{dec2} to compute $p_2$ for a specified profile $p_0$.
Then we obtain $a_{2,4}$ by directly integrating the constraint
\eqref{uvperdim3c},
\begin{equation}
a_{2,4}=\calc_3
-\frac 13 p_{1,0}\ p_{1,2}-\frac{5}{36}(p_{1,0}')^2+\frac 29 p_{1,0}\ p_{1,0}''+\frac 23
\int_{-\infty}^\tau \!\!\!ds\ p_{1,2}(s)\, p_{1,0}'(s) \,,
\eqlabell{a24dim3}
\end{equation}
where $\calc_3$ is an arbitrary integration constant which we will fix
to a convenient value below. Now we have determined both $p_{1,2}$ and
$a_{2,4}$ for a specific profile $p_{1,0}$ (as well as having set
$a_1=0$). Thus, as observed above, we have sufficient data to evaluate
the one-point correlation functions of $T_{ij}$ and $\calo_3$ using the
expressions in eq.~\eqref{vev1} for a mass quench to $\calo(\l^2)$.
\\
\nxt When $m^2=-4$
\begin{equation}
\begin{split}
\phi_1=&-\left(p_{1,0}+p_{1,0}^l\ln\r \right) \rho^2+\calo(\r^3\ln\r)\,,\\
\Sigma_2=&\calo(\r^3\ln^2\r)\,,\\
A_2=&\left(a_{2,4}+\left(\frac {1}{54} (p_{1,0}^l)^2-\frac 29 p_{1,0} p_{1,0}^l
\right) \ln\rho-\frac 19 (p_{1,0}^l)^2\ln^2\r\right)\r^2+\calo(\r^3\ln^2\r)\,.
\end{split}
\eqlabell{uvperdim2}
\end{equation}
The first equation in eq.~\eqref{dec4} now yields the following
boundary constraint:
\begin{equation}
0=a_{2,4}'-\frac{5}{27}p_{1,0}^l\ p_{1,0}'+\frac{4}{27}(p_0^l)'\ p_{1,0}
+\frac{10}{81} p_{1,0}^l\ (p_{1,0}^l)'+\frac{2}{9} p_{1,0}\ p_{1,0}'\,.
\eqlabell{uvperdim2c}
\end{equation}
As before, the boundary profiles are functions of $\t$, \ie
$p_{1,0}^l=p_{1,0}^l(\t)$, $p_{1,0}=p_{1,0}(\tau)$ and
$a_{2,4}=a_{2,4}(\t)$ and the prime denotes derivative with respect to
$\t$.

Comparing eqs.~\eqref{uvnonlinerdim2} and \eqref{uvperdim2}, we
identify
\begin{equation}
\begin{split}
&p_0^l=\l\mu^2\ p_{1,0}^l\,,\qquad p_0=-\l\mu^2 (p_{1,0}+p_{1,0}^l\ln\mu)\,,\\
&a_1=0\,,\qquad a_4=-\mu^4+\l^2 \mu^4\
\left(a_{2,4}+
\left(\frac {1}{54} (p_{1,0}^l)^2-\frac 29 p_{1,0} p_{1,0}^l
\right) \ln\mu-\frac 19 (p_{1,0}^l)^2\ln^2\mu
\right)\,.
\end{split}
\eqlabell{iddim2}
\end{equation}
Again, note that it suffices to solve numerically only
eq.~\eqref{dec2}, which allows us to extract $p_{1,0}$ for a given mass
profile specified by $p_{1,0}^l$ and then compute the one-point
function $\la\calo_2\ra$ using eq.~\eqref{vev12}. These expressions
also allow us to evaluate $\la T_{ij}\ra$ after we determine $a_{2,4}$
by integrating eq.~\eqref{uvperdim2c}:
 \beq
a_{2,4}= \calc_2 -\frac{5}{81}(p_{1,0}^l)^2-\frac 19
p_{1,0}^2+\frac{5}{27}p_{1,0}\ p_{1,0}^l -\frac 13 \int_{-\infty}^\t
\!\!\!ds\ p_{1,0}(s)\, (p_{1,0}^l(s))'\,,
 \eqlabell{a24dim2}
 \eeq
where $\calc_2$ is another integration constant, for which we make a
convenient choice below.

\subsection{Physical observables and ambiguities} \label{result}

As already discussed, the physical observables which we track during the
quenches are the one-point correlation functions of the stress-energy
tensor $T_{ij}$ and the operator dual to the bulk scalar $\phi$  which
induces the quench, \ie $\calo_3$ operator for $m^2=-3$ and $\calo_2$
for $m^2=-4$.  Further we will consider two types of mass quenches:
First, we start at $t\to -\infty$ with the CFT (\ie $m_{f,b}=0$) in a
thermal equilibrium state at temperature $T_i$, and after the quench as
$t\to +\infty$, it will equilibrate to a thermal state of the massive
theory with a final temperature $T_f$ but still in the regime where
$m_{f,b}\ll T_f$. The second scenario will be mass quenches where we
start at $t\to -\infty$ with the massive theory in thermal equilibrium
with $m_{f,b}\ll T_i$ and the quench will take the mass in the boundary
theory to zero. Hence as $t\to +\infty$, it equilibrates to a new
thermal state now of the CFT with temperature $T_f$.

\subsubsection{$m^2=-3$} \label{subm3}

From substituting the expressions in eq.~\eqref{iddim3} into
eq.~\eqref{vev1}, we find
\begin{equation}
\begin{split}
8\pi G_5\ \cale=&\frac32 \mu^4-\mu^4\l^2 \
\biggl(\frac32 a_{2,4}+\frac 14 \ln\mu\ (p_{1,0}')^2+\frac{1}{12}p_{1,0}'^2+\frac 12 p_{1,0} p_{1,2}-
\frac 13 p_{1,0}'' p_{1,0}\biggr) \\
&+\cale^{ambiguity}+\calo(\l^4)\,,\\
8\pi G_5 \calp=&\frac 12 \mu^4-\l^2 \mu^4 \
\biggl(\frac12 a_{2,4}+ \ln\mu\ \left(-\frac16 p_{1,0}'' p_{1,0}+\frac{1}{12} (p_{1,0}')^2\right)
+\frac{1}{36}
(p_{1,0}')^2\\
&-\frac 16 p_{1,0} p_{1,2}+\frac{1}{18} p_{1,0}'' p_{1,0}\biggr)
+\calp^{ambiguity}+\calo(\l^4)\,,\\
16\pi G_5\ \la\calo_3\ra=&\frac 12 \mu^3 \l\ (p_{1,0}''-4 p_{1,2}-2 \ln\mu\ p_{1,0}'')
+\calo_3^{ambiguity}+\calo(\l^3)\,,
\end{split}
\eqlabell{vev1l2}
\end{equation}
where:
\begin{equation}
\begin{split}
\cale^{ambiguity}=&\frac12 \dd_2\ \mu^4 \l^2\ (p_{1,0}')^2\,, \\
\calp^{ambiguity}=&\mu^4 \l^2 \left(\frac 12 \dd_2\ (p_{1,0}')^2-2\dd_3\ ( (p_{1,0}')^2+ p_{1,0}'' p_{1,0}) \right)\,,\\
\calo_3^{ambiguity}=&2 \dd_2\ \mu^3 \l\ p_{1,0}'' \,.
\end{split}
\eqlabell{ambiguitiesl2}
\end{equation}
{}\\
As described above, we consider two classes of quenches:\\
\nxt First, we start with a CFT as $t\to -\infty$ ($\t\to -\infty$) and quench to a massive theory
as $t\to +\infty$ ($\t\to +\infty$).
This quench is implemented as
\begin{equation}
\lim_{\t\to -\infty}p_{1,0}=0\,,\qquad \lim_{\t\to +\infty}p_{1,0}=1\,,\qquad \lim_{\t\to \pm \infty} p_{1,0}'=0\,.
\eqlabell{cftmass3}
\end{equation}
Further, the asymptotic values of $p_{1,2}$ match those in the
appropriate thermodynamic equilibrium,
\begin{equation}
\lim_{\t\to -\infty}p_{1,2}=0\,,\qquad \lim_{\t\to +\infty}p_{1,2}=-\frac{\Gamma\left(\frac34\right)^4}{\pi^2}\,.
\eqlabell{cftmass3a}
\end{equation}
In particular, the latter is given by comparing to eqs.~\reef{bamm} and
\reef{p2}. Once the profile $p_{1,0}(\t)$ is specified, the full
function $p_{1,2}(\t)$ will be determined by numerically integrating
the scalar field equation \reef{dec2}. The metric function $a_{2,4}$ is
then determined by eq.~\eqref{a24dim3} but we must first fix the
integration constant $\calc_3$. It turns out that a convenient choice
is simply $\calc_3=0$, which sets $a_{2,4}(\tau=-\infty)=0$ for the
present class of quenches. Matching the $\t\to \pm\infty$ limits of
$T_{ij} $ with the corresponding equilibrium results, we identify to
leading order
\begin{equation}
\mu=\pi T_i\,,\qquad \l=\frac{\sqrt{2}m_f^0}{\pi T_i}\,,\qquad m_f(\t)=m_f^0\ p_{1,0}(\t)\,.
\eqlabell{identity}
\end{equation}
Further we define two scales, $\Lambda_2$ and $\Lambda_3$, to be
associated with the ambiguities in the renormalization scheme,
\begin{equation}
\dd_2=\frac 12 \ln \Lambda_2\,,\qquad \dd_3=\frac{1}{12}\ln\ \Lambda_3\,.
\eqlabell{dddim31}
\end{equation}
Now we can rewrite eq.~\eqref{vev1l2} in terms of variables of the
boundary gauge theory:
\begin{equation}
\begin{split}
\cale=&\frac 38 \pi^2 N^2 T^4_i\biggl(1-
\biggl(2a_{2,4}+\frac 13 (p_{1,0}')^2\ \ln\frac{\pi T_i}{\Lambda_2}
+\frac 19 (p_{1,0}')^2-\frac 49 p_{1,0}p_{1,0}''\\
&+\frac 23 p_{1,0}p_{1,2}
 \biggr)\frac{(m_f^0)^2}{\pi^2T_i^2}
+\calo\left(\frac{(m_f^0)^4}{T_i^4}\right)
\biggr) \,,\\
\calp=&\frac 18 \pi^2 N^2 T_i^4\biggl(1-\biggl(
2a_{2,4}+\frac 19 (p_{1,0}')^2-\frac 23 p_{1,0}p_{1,2}+\frac 29 p_{1,0}p_{1,0}''
\\
&-\frac 23 \left(p_{1,0}p_{1,0}''+(p_{1,0}')^2\right)\ \ln\frac{\pi T_i}{\Lambda_3}
+ (p_{1,0}')^2\ \ln\frac{\pi T_i}{\Lambda_2}
 \biggr)\frac{(m_f^0)^2}{\pi^2 T_i^2}
+\calo\left(\frac{(m_f^0)^4}{T_i^4}\right)
\biggr)\,,\\
\calo_3=&-\frac{1}{2\sqrt{2}}N^2 T_i^2 m_f^0\biggl(
p_{1,2}-\frac 14 p_{1,0}''+\frac 12 \ln\frac{\pi T_i}{\Lambda_2}\ p_{1,0}''
+\calo\left(\frac{(m_f^0)^2}{T_i^2}\right)
\biggr) \,.
\end{split}
\eqlabell{vev1qft1}
\end{equation}
Notice that at the order that we are calculating these quantities, the
scheme-dependent ambiguities (the arbitrariness in the choice of
$\Lambda_2$ and $\Lambda_3$) arise only during the evolution. That is,
asymptotically $p_{1,0}'=0=p_{1,0}''$ and hence the terms involving
$\ln\Lambda_{1,2}$ above vanish in the initial and final equilibrium
configurations. Of course, this is in agreement with our observation in
the previous section that there are no renormalization ambiguities in
the one-point functions \eqref{vev1e} at equilibrium.

Finally, we compute the final temperature  and the final energy density/pressure after the quench.
Denoting
\begin{equation}
a_{2,4}^\infty=\lim_{\t\to +\infty} a_{2,4}(\t)=\frac{\Gamma\left(\frac34\right)^4}{3\pi^2}+\frac 23
\int_{-\infty}^{\infty} ds\ p_{1,2}(s)\ p_{1,0}'(s)\,,
\eqlabell{aindef}
\end{equation}
and matching the $\t\to +\infty$ limit of the above expressions
\eqref{vev1qft1} with the equilibrium values in eq.~\eqref{equil3}, we
find
\begin{equation}
\begin{split}
\frac{T_f}{T_i}=&\ 1+\left(\frac{\Gamma\left(\frac34\right)^4}{3\pi^2}
-\frac 12 a_{2,4}^\infty\right)\frac{(m_f^0)^2}{\pi^2 T_i^2}
+\calo\left(\frac{(m_f^0)^4}{T_i^4}\right)\,,\\
\frac{\cale_f}{\cale_i}=&\ 1+\left(\frac{2\Gamma\left(\frac34\right)^4}{3\pi^2}
-2 a_{2,4}^\infty\right)\frac{(m_f^0)^2}{\pi^2 T_i^2}
+\calo\left(\frac{(m_f^0)^4}{T_i^4}\right)\,,\\
\frac{\calp_f}{\calp_i}=&\ 1-\left(\frac{2\Gamma\left(\frac34\right)^4}{3\pi^2}
+2 a_{2,4}^\infty\right)\frac{(m_f^0)^2}{\pi^2 T_i^2}
+\calo\left(\frac{(m_f^0)^4}{T_i^4}\right)\,.
\end{split}
\eqlabell{ftdim3}
\end{equation}
It is also interesting to consider the entropy of the gauge theory. For
either of the asymptotic equilibrium states, we can use the
thermodynamic formula: $S=(\cale +\calp)/T$. Further, we have
$\cale_i=3\,\calp_i$ for the initial equilibrium of the CFT  and so we
find
 \beqa
\frac{S_f}{S_i}&=& \frac{T_i}{T_f}\left(\frac34\,
\frac{\cale_f}{\cale_i}+\frac14\,\frac{\calp_f}{\calp_i}\right)
\nonumber\\
&=& 1-\frac{3}{2} a_{2,4}^\infty\ \frac{(m_f^0)^2}{\pi^2 T_i^2}
+\calo\left(\frac{(m_f^0)^4}{T_i^4}\right)\,. \eqlabell{sfsi}
 \eeqa
This last expression makes clear that the constant $a_{2,4}^\infty$,
defined in eq.~\reef{aindef}, directly parameterizes the entropy
production of the mass quench. In our simulations, we find that
$a_{2,4}^\infty$ is always negative --- see section \ref{halo3} for
details. This result is, of course, in agreement with the expectation
that the entropy density should always increase.

Let us consider an adiabatic transition in the mass where
\begin{equation}
|p_{1,0}'|\ll 1
\eqlabell{ad1}
\end{equation}
everywhere. We expect (and later confirm numerically) that
\begin{equation}
p_{1,2}(\t)\approx p_{1,2}(\t=+\infty)\ p_{1,0}(\t)=
-\frac{\Gamma\left(\frac 34\right)^4}{\pi^2}\ p_{1,0}(\t)\,.
\eqlabell{ad2}
\end{equation}
That is, for a very slow transition, the system essentially maintains
thermodynamic equilibrium throughout the process and $p_{1,2}(\t)$
simply tracks $p_{1,0}(\t)$. Thus, from eq.~\eqref{aindef}, we have
\begin{equation}
a_{2,4}^\infty\approx \frac{\Gamma\left(\frac 34\right)^4}{3\pi^2}
\left(1-2\int_{-\infty}^\infty ds\
p_{1,0}(s)\ p_{1,0}'(s)\right)=0\,.
\eqlabell{ad3}
\end{equation}
Given this result, eq.~\reef{sfsi} indicates that the entropy density
is constant for slow changes of masses. Of course, from
eq.~\eqref{ftdim3}, we see that these adiabatic processes where the
mass is raised still produce an increase in temperature and energy
density but a decrease in pressure. This behaviour can be intuitively
understood with a quasi-particle picture of a system undergoing such an
adiabatic transition.\footnote{Of course, a quasi-particle picture is
not valid for strongly coupled $\caln=2^*$ plasma.} In such an
adiabatic process, the occupation number of energy levels is unchanged
but increasing mass increases the energy of any given level. Hence the
net energy density would increase and further, the occupation number of
the states after such a quench would correspond to a higher effective
temperature.

Notice that since we expect that $a_{2,4}^\infty$ is always negative,
eq.~\reef{ftdim3} indicates that the temperature and energy density
always increase, even for rapid mass quenches. However, the change in
the pressure can be either a decrease or an increase. We can expect
that slow transitions result in the pressure decreasing while rapid
quenches, with a large entropy production, should result in an increase
in the pressure.
{}\\
\nxt For the second type of quenches, we start with a massive theory as
$t\to -\infty$ ($\t\to -\infty$) and quench to a CFT as $t\to +\infty$
($\t\to +\infty$). This quench is implemented as\footnote{We use an upper symbol
 $^\sim$ to distinguish quenches of this type.}
\begin{equation}
\lim_{\t\to -\infty}\tp_{1,0}=1\,,\qquad
\lim_{\t\to +\infty}\tp_{1,0}=0\,,\qquad \lim_{\t\to \pm \infty} \tp_{1,0}'=0\,.
\eqlabell{cftmass3cft}
\end{equation}
In this case, the asymptotic values of $p_{1,2}$ become
\begin{equation}
\lim_{\t\to -\infty}\tp_{1,2}=-\frac{\Gamma\left(\frac34\right)^4}{\pi^2}
\,,\qquad \lim_{\t\to +\infty}\tp_{1,2}=0\,.
\eqlabell{cftmass3x}
\end{equation}
The metric function $a_{2,4}$ is again determined by
eq.~\eqref{a24dim3}, however, we have the freedom to choose a new value
for the integration constant $\calc_3$. A convenient choice in this
case turns out to be
\begin{equation}
\calc_3=\frac 13 \big[p_{1,0}\ p_{1,2}\big]_{\tau=-\infty}
=-\frac{\Gamma\left(\frac34\right)^4}{3\pi^2}
\,,
\eqlabell{Cdim3acft}
\end{equation}
which again ensures that $\ta_{2,4}(\t=-\infty)=0$. Now comparing to
eq.~\eqref{equil3}, we match the $\t\to - \infty$ limit of $T_{ij}$ to
the corresponding equilibrium results by identifying
\begin{equation}
\begin{split}
&\mu=\pi T_i\left(1-\frac{\Gamma\left(\frac34\right)^4}{3\pi^4}\frac{(m_f^0)^2}{T_i^2}
+\calo\left(\frac{(m_f^0)^4}{T_i^4}\right)\right)\,,\qquad
\l=\frac{\sqrt{2}m_f^0}{\pi T_i}\,,\\
&m_f(\t)=m_f^0\ \tp_{1,0}(\t)\,,
\end{split}
\eqlabell{identitycft}
\end{equation}
to leading order. Further we introduce the same scales, $\Lambda_2$
and $\Lambda_3$, as in eq.~\reef{dddim31} to define the renormalization
scheme. Then we rewrite eq.~\eqref{vev1l2} in terms of field theory
parameters:
\begin{equation}
\begin{split}
\cale=&\frac 38 \pi^2 N^2 T^4_i\biggl(1-
\biggl(2\ta_{2,4}+\frac{4\Gamma\left(\frac 34\right)^4}{3\pi^2}
+\frac 13 (\tp_{1,0}')^2\ \ln\frac{\pi T_i}{\Lambda_2}
+\frac 19 (\tp_{1,0}')^2-\frac 49 \tp_{1,0}\tp_{1,0}''\\
&+\frac 23 \tp_{1,0}\tp_{1,2}
 \biggr)\frac{(m_f^0)^2}{\pi^2T_i^2}
+\calo\left(\frac{(m_f^0)^4}{T_i^4}\right)
\biggr)\,, \\
\calp=&\frac 18 \pi^2 N^2 T_i^4\biggl(1-\biggl(
2\ta_{2,4}+\frac{4\Gamma\left(\frac 34\right)^4}{3\pi^2}+\frac 19
(\tp_{1,0}')^2-\frac 23 \tp_{1,0}\tp_{1,2}+\frac 29 \tp_{1,0}\tp_{1,0}''
\\
&-\frac 23 \left(\tp_{1,0}\tp_{1,0}''+(\tp_{1,0}')^2\right)\ \ln\frac{\pi T_i}{\Lambda_3}
+ (\tp_{1,0}')^2\ \ln\frac{\pi T_i}{\Lambda_2}
 \biggr)\frac{(m_f^0)^2}{\pi^2 T_i^2}
+\calo\left(\frac{(m_f^0)^4}{T_i^4}\right)
\biggr)\,,\\
\calo_3=&-\frac{1}{2\sqrt{2}}N^2 T_i^2 m_f^0\biggl(
\tp_{1,2}-\frac 14 \tp_{1,0}''+\frac 12 \ln\frac{\pi T_i}{\Lambda_2}\ \tp_{1,0}''
+\calo\left(\frac{(m_f^0)^2}{T_i^2}\right)
\biggr) \,.
\end{split}
\eqlabell{vev1qft1cft}
\end{equation}
Once again, the arbitrariness in the choice of $\Lambda_2$ and
$\Lambda_3$ only plays a role during the time-dependent portion of the
quench --- as already pointed out in eq.~\eqref{vev1e}, there are no
scheme dependent ambiguities at equilibrium in quenches of the
fermionic mass term.

Finally, we compute the final temperature  and the final energy
density/pressure after the quench. Denoting
\begin{equation}
\ta_{2,4}^\infty=\lim_{\t\to +\infty} \ta_{2,4}(\t)=
-\frac{\Gamma\left(\frac34\right)^4}{3\pi^2}+\frac 23
\int_{-\infty}^{\infty} ds\ \tp_{1,2}(s)\ \tp_{1,0}^{\,\prime}(s)\,,
\eqlabell{aindefcft}
\end{equation}
and matching the $\t\to +\infty$ limit of \eqref{vev1qft1} with
\eqref{equil3} we find
\begin{equation}
\begin{split}
\frac{T_f}{T_i}=&\ 1-\left(\frac{\Gamma\left(\frac34\right)^4}{3\pi^2}
+\frac 12 \ta_{2,4}^\infty\right)\frac{(m_f^0)^2}{\pi^2 T_i^2}
+\calo\left(\frac{(m_f^0)^4}{T_i^4}\right)\,,\\
\frac{\cale_f}{\cale_i}=&\ 1-\left(\frac{2\Gamma\left(\frac34\right)^4}{3\pi^2}
+2 \ta_{2,4}^\infty\right)\frac{(m_f^0)^2}{\pi^2 T_i^2}
+\calo\left(\frac{(m_f^0)^4}{T_i^4}\right)\,,\\
\frac{\calp_f}{\calp_i}=&\ 1+\left(\frac{2\Gamma\left(\frac34\right)^4}{3\pi^2}
-2 \ta_{2,4}^\infty\right)\frac{(m_f^0)^2}{\pi^2 T_i^2}
+\calo\left(\frac{(m_f^0)^4}{T_i^4}\right)\,.
\end{split}
\eqlabell{ftdim3cft}
\end{equation}
Turning to the entropy, it is straightforward to show that
\begin{equation}
\frac{S_f}{S_i}=1-\frac{3}{2} \ta_{2,4}^\infty\ \frac{(m_f^0)^2}{\pi^2 T_i^2}
+\calo\left(\frac{(m_f^0)^4}{T_i^4}\right)\,.
\eqlabell{sfsicft}
\end{equation}
So again the entropy production of these quenches is parameterized by
$\ta_{2,4}^\infty$. Further all of the general comments made above
about the transitions from the CFT to the massive theory can be
extended to analogous statements about the present class of quenches.
In particular, $\ta_{2,4}^\infty$ vanishes for adiabatic transitions
where $|\tp_{1,0}^{\,\prime}|\ll1$ everywhere.

Notice that if $p_{1,0}(\t)$
describes a quantum quench from a CFT to a massive theory, then
\begin{equation}
\tilde{p}_{1,0}(\t)\equiv 1-p_{1,0}(\t)
\eqlabell{reversedim31}
\end{equation}
describes a quench from a massive theory to a CFT. Because we are
working perturbatively in the amplitude of the scalar field, in this
`reverse' quench, we will also find
\begin{equation}
\tilde{p}_{1,2}(\t)=p_{2,0}(\t=+\infty)-p_{2,0}(\t)
=-\frac{\Gamma\left(\frac 34\right)^4}{\pi^2}-p_{2,0}(\t)\,.
\eqlabell{reversedim32}
\end{equation}
Further, comparing eqs.~\eqref{aindef} and \eqref{aindefcft}, we find
\begin{equation}
\ta_{2,4}^\infty=a_{2,4}^\infty\,.
\eqlabell{a24t}
\end{equation}
Hence the entropy production of a quench from the CFT to the massive
theory is precisely the same as in the reverse quench from the massive
theory to the CFT. Again, this result occurs because we are working
perturbatively in the high temperature approximation.

\subsubsection{$m^2=-4$}
Substituting eq.~\eqref{vev12} into eq.~\eqref{iddim2}, we find
\begin{equation}
\begin{split}
8\pi G_5\ \cale=&\frac 32 \mu^4-\mu^4\l^2
\left(\frac 32 a_{2,4}-\frac{5}{18} p_{1,0} p_{1,0}^l-\frac14 (p_{1,0}^l)^2 \ln\mu+\frac{5}{54} (p_{1,0}^l)^2
+\frac 16 p_{1,0}^2\right)\\
&+\cale^{ambiguity}+\calo(\l^4)\,,\\
8\pi G_5 \calp=&\frac12 \mu^4-\mu^4\l^2 \left(\frac 12 a_{2,4}+\frac{13}
{54} p_{1,0} p_{1,0}^l+\frac 14 (p_{1,0}^l)^2 \ln\mu-\frac{17}{324} (p_{1,0}^l)^2+\frac{1}{18} p_{1,0}^2
\right) \\&
+\calp^{ambiguity}+\calo(\l^4)\,,\\
16\pi G_5\ \la\calo_2\ra=&\mu^2\l\ (p_{1,0}+p_{1,0}^l \ln\mu)
+\calo_3^{ambiguity}+\calo(\l^3)\,,
\end{split}
\eqlabell{vev12l2}
\end{equation}
where:
\begin{equation}
\begin{split}
\cale^{ambiguity}=&\frac12 \dd_1\ \mu^4 \l^2\ (p_{1,0}^l)^2 \,,\\
\calp^{ambiguity}=&-\frac12 \dd_1\ \mu^4 \l^2\ (p_{1,0}^l)^2+\dd_2\ \mu^2\l\ (p_{1,0}^l)'' \,,\\
\calo_2^{ambiguity}=&2 \dd_1\ \mu^2 \l\ p_{1,0}^l \,.
\end{split}
\eqlabell{ambiguitiesl2dim2}
\end{equation}
As before, we  consider two classes of quenches:\\
\nxt First, we start with a CFT as $t\to -\infty$ ($\t\to -\infty$) and
quench to a massive theory as $t\to +\infty$ ($\t\to +\infty$). This
quench is implemented as
\begin{equation}
\lim_{\t\to -\infty}p_{1,0}^l=0\,,\qquad \lim_{\t\to +\infty}p_{1,0}^l=1\,,
\qquad \lim_{\t\to \pm \infty} (p_{1,0}^l)'=0\,.
\eqlabell{cftmass2}
\end{equation}
In this case, the asymptotic values of $p_{1,0}$ are
\begin{equation}
\lim_{\t\to -\infty}p_{1,0}=0
\,,\qquad \lim_{\t\to +\infty}p_{1,0}=-\ln2\,.
\eqlabell{mass2xx}
\end{equation}
The metric function $a_{2,4}$ is determined by eq.~\eqref{a24dim2} but
we must now fix the integration constant $\calc_2$. With some
foresight, we choose
 \beq
\calc_2=\left[-\frac{1}{54}p_{1,0}\
p_{1,0}^l+\frac{5}{81}(p_{1,0}^l)^2+\frac 19
p_{1,0}^2\right]_{\t=+\infty}=\frac{1}{54}\ln2+\frac{5}{81}+\frac{1}{9}\ln^22\,.
 \labell{C2cftx}
 \eeq
Note that with this choice, $a_{2,4}$ does not vanish for either
$\t\to\pm\infty$. However, we will see that this choice simplifies our
description of the entropy production below. Now matching the $\t\to
\pm \infty$ limits of $\cale$ and $\calp$ with the corresponding
equilibrium results, we identify to leading order
\begin{equation}
\mu=\pi T_i\left(1+\frac16\,\calc_2\frac{(m_b^0)^4}{\pi^4 T_i^4}\right)\,,
\qquad \l=\sqrt{\frac23}\ \frac{(m_b^0)^2}{\pi^2 T_i^2}\,,
\qquad m_b^2(\t)=(m_b^0)^2\ p_{1,0}^l(\t)\,.
\eqlabell{identitydim2}
\end{equation}
Next we introduce two scales, $\Lambda_1$ and $\Lambda_2$, that are
related to the ambiguities in the renormalization scheme with
\begin{equation}
\dd_1=-\frac 12 \ln \frac{\Lambda_1}{2}\,,\qquad \dd_2=\Lambda_2^2\,.
\eqlabell{dddim21}
\end{equation}
Then we can rewrite eq.~\eqref{vev1l2} using parameters in the dual
field theory:
\begin{eqnarray}
\cale&=&\frac 38 \pi^2 N^2 T^4_i\biggl(1-
\biggl(\frac 23 a_{2,4}-\frac23\calc_2-\frac{10}{81} p_{1,0}^l p_{1,0}
-\frac 19 (p_{1,0}^l)^2 \ln\frac{2\pi T_i }{\Lambda_1}
\nonumber\\
&&\qquad\qquad\qquad
+\frac{10}{243} (p_{1,0}^l)^2+\frac{2}{27} p_{1,0}^2
 \biggr)\frac{(m_b^0)^4}{\pi^4T_i^4}
+\calo\left(\frac{(m_b^0)^8}{T_i^8}\right)
\biggr) \,,
\eqlabell{vev12qft1}\\
\calp&=&\frac 18 \pi^2 N^2 T_i^4\biggl(1-\sqrt{\frac{8}{3}}\ \frac{\Lambda_2^2(m_b^0)^2}{\pi^4 T_i^4} (p_{1,0}^l)''
-\biggl(\frac 23 a_{2,4}-\frac23\calc_2+
\frac{26}{81} p_{1,0}^l p_{1,0}
\nonumber\\
&&\qquad\qquad
+\frac 13 (p_{1,0}^l)^2 \ln\frac{2 \pi T_i}{\Lambda_1}
-\frac{17}{243} (p_{1,0}^l)^2
+\frac{2}{27} p_{1,0}^2
 \biggr)\frac{(m_b^0)^4}{\pi^4 T_i^4}
+\calo\left(\frac{(m_b^0)^8}{T_i^8}\right)
\biggr)\,,
\nonumber\\
\calo_2&=&\frac{\sqrt{6}N^2}{24\pi^2}\ (m_b^0)^2\ \biggl(
p_{1,0}+p_{1,0}^l \ln\frac{2\pi T_i}{\Lambda_1}
+\calo\left(\frac{(m_b^0)^4}{T_i^4}\right)
\biggr)\,.
\nonumber
\end{eqnarray}
Note that in our perturbative expansion in powers of $(m_b^0)^2/T_i^2$,
the pressure is formally dominated by the term proportional to
$\Lambda_2^2$. However, this scheme dependent term vanishes in an
equilibrium configuration where $(p_{1,0}^l)''=0$.

Now we compute the final temperature  and the final energy
density/pressure after the quench. Denoting
\begin{equation}
a_{2,4}^\infty=\lim_{\t\to +\infty} a_{2,4}(\t)=-\frac16\ln 2-\frac 13
\int_{-\infty}^{\infty} ds\ p_{1,0}(s)\ (p_{1,0}^l)'(s)\,,
\eqlabell{aindef2}
\end{equation}
and matching the $\t\to +\infty$ limit of the above results
\eqref{vev12qft1} with the equilibrium values \eqref{equil2}, we find
\begin{equation}
\begin{split}
\frac{T_f}{T_i}=&\ 1+\left(\frac{1}{36}-\frac16 a_{2,4}^\infty \right)
\frac{(m_b^0)^4}{\pi^4 T_i^4}
+\calo\left(\frac{(m_b^0)^8}{T_i^8}\right)\,,\\
\frac{\cale_f}{\cale_i}=&\ 1+\left(\frac 19 \ln\frac{\pi T_i}{\Lambda_1}-\frac23 a_{2,4}^\infty
\right)
\frac{(m_b^0)^4}{\pi^4 T_i^4}
+\calo\left(\frac{(m_b^0)^8}{T_i^8}\right)\,,\\
\frac{\calp_f}{\calp_i}=&\ 1+\left(\frac{1}{9}-\frac 13 \ln\frac{\pi T_i}{\Lambda_1}
-\frac 23
a_{2,4}^\infty \right)
\frac{(m_b^0)^4}{\pi^4 T_i^4}
+\calo\left(\frac{(m_b^0)^8}{T_i^8}\right)\,.
\end{split}
\eqlabell{ftdim2}
\end{equation}
Turning to the entropy, since this quench begins with the CFT, we may
apply the same expression as given in eq.~\reef{sfsi} to find
 \beqa
\frac{S_f}{S_i}&=& \frac{T_i}{T_f}\left(\frac34\,
\frac{\cale_f}{\cale_i}+\frac14\,\frac{\calp_f}{\calp_i}\right)
\nonumber\\
&=& 1-\frac{1}{2} a_{2,4}^\infty\, \frac{(m_b^0)^4}{\pi^4 T_i^4}
+\calo\left(\frac{(m_b^0)^8}{T_i^8}\right)\,. \eqlabell{sfsi2}
 \eeqa
Hence with the judicious choice of the integration constant made in
eq.~\reef{C2cftx}, we find that $a_{2,4}^\infty$ directly parameterizes
the entropy production in these quenches of the bosonic mass. As
before, our simulations of the bosonic mass quenches seem to indicate
that $a_{2,4}^\infty$ is always negative
--- see section \ref{halo2} for details. Of course, this matches the
intuition that the entropy density must always increase.

If we consider an adiabatic transition where
\begin{equation}
|(p_{1,0}^l)'|\ll 1\,
\eqlabell{ad1x2}
\end{equation}
we expect (and later confirm numerically) that
\begin{equation}
p_{1,0}(\t)\approx p_{1,0}(\t=+\infty)\ p_{1,0}^l(\t)=
-\ln2\ p_{1,0}^l(\t)\,.
\eqlabell{ad2x2}
\end{equation}
Again, this reflects the expectation that for a very slow transition,
the system essentially maintains thermodynamic equilibrium throughout
the process and $p_{1,0}(\t)$ simply tracks $p_{1,0}^l(\t)$. Thus, from
eq.~\eqref{aindef}, we have
\begin{equation}
a_{2,4}^\infty\approx -\frac16\ln 2
\left(1-2\int_{-\infty}^\infty ds\
p_{1,0}^l(s)\ (p_{1,0}^l)'(s)\right)=0\,.
\eqlabell{ad3x2}
\end{equation}
Hence we again find that the entropy density is constant for adiabatic
changes of the mass. Notice that in this case, from eq.~\eqref{ftdim2},
these adiabatic processes always produce an increase in temperature,
however, the sign of change in the energy density and the pressure
depends on the choice of the renormalization scale $\Lambda_1$, which
is needed to describe thermodynamics in the massive theory, \eg see
eq.~\reef{equil2}. Even with $a_{2,4}^\infty$ always being negative,
similar statements still apply for rapid mass quenches.
{}\\
\nxt The second type of quench, which we consider, starts with the
massive theory as $t\to -\infty$ ($\t\to -\infty$) and makes a
transition to the CFT as $t\to +\infty$ ($\t\to +\infty$). These
quenches are implemented with\footnote{Here again, we use $ ^\sim$ to
distinguish this second class of quenches.}
\begin{equation}
\lim_{\t\to -\infty}\tp_{1,0}^l=1\,,\qquad \lim_{\t\to +\infty}\tp_{1,0}^l=0\,,
\qquad \lim_{\t\to \pm \infty} (\tp_{1,0}^l)'=0\,.
\eqlabell{cftmass2cft}
\end{equation}
In this case, the asymptotic values of $p_{1,0}$ become
\begin{equation}
\lim_{\t\to -\infty}\tp_{1,0}=-\ln2
\,,\qquad \lim_{\t\to +\infty}\tp_{1,0}=0\,.
\eqlabell{mass2xxcft}
\end{equation}
The metric function $a_{2,4}$ is determined by eq.~\eqref{a24dim2} and
we must now fix the integration constant $\calc_2$. Motivated by the
desire for a simple expression for the entropy production, we choose
 \beq
\calc_2=\left[-\frac{1}{6}\tp_{1,0}\
\tp_{1,0}^l\right]_{\t=-\infty}=\frac{1}{6}\ln2\,.
 \labell{C2cftxnew}
 \eeq
This choice of $\calc_2$ again means that $a_{2,4}$ is nonvanishing for
both $\t\to\pm\infty$. Now matching the $\t\to - \infty$ limit of
$T_{ij} $ with the corresponding equilibrium results \eqref{equil2}, we
identify to leading order
\begin{equation}
\mu=\pi T_i \biggl(1-\frac1{36}\frac{(m_b^0)^4}{\pi^4 T_i^4}
\biggr)\,,\qquad
\l=\sqrt{\frac23}\ \frac{(m_b^0)^2}{\pi^2 T_i^2}\,,\qquad m_b^2(\t)=(m_b^0)^2\ \tp_{1,0}^l(\t)\,.
\eqlabell{identitycftdim2}
\end{equation}
Finally, we introduce two renormalization scales, $\Lambda_1$ and
$\Lambda_2$, as in eq.~\reef{dddim21}. Then we can rewrite
eq.~\eqref{vev12l2} in terms of field theory parameters:
\begin{eqnarray}
\cale&=&\frac 38 \pi^2 N^2 T^4_i\biggl(1-
\biggl(\frac 23 \ta_{2,4}-\frac 23 \calc_2+\frac{1}{9}+\frac 19 \ln2
-\frac{10}{81} \tp_{1,0}^l \tp_{1,0}-\frac 19 (\tp_{1,0}^l)^2 \ln\frac{2\pi T_i }{\Lambda_1}
\nonumber\\
&&\qquad\qquad\qquad\qquad
+\frac{10}{243} (\tp_{1,0}^l)^2+\frac{2}{27} \tp_{1,0}^2
 \biggr)\frac{(m_b^0)^4}{\pi^4T_i^4}
+\calo\left(\frac{(m_b^0)^8}{T_i^8}\right)
\biggr) \,,
 \eqlabell{vev12qft1cft}\\
\calp&=&\frac 18 \pi^2 N^2 T_i^4\biggl(1-\sqrt{\frac{8}{3}}\ \frac{\Lambda_2^2(m_b^0)^2}{\pi^4 T_i^4} (\tp_{1,0}^l)''
-\biggl(\frac 23 \ta_{2,4}-\frac 23 \calc_2+\frac{1}{9}+\frac 19\ln2
\nonumber\\
&&+\frac{26}{81} \tp_{1,0}^l \tp_{1,0}
+\frac 13 (\tp_{1,0}^l)^2 \ln\frac{2 \pi T_i}{\Lambda_1}
-\frac{17}{243} (\tp_{1,0}^l)^2
+\frac{2}{27} \tp_{1,0}^2
 \biggr)\frac{(m_b^0)^4}{\pi^4 T_i^4}
+\calo\left(\frac{(m_b^0)^8}{T_b^8}\right)
\biggr)\,,
\nonumber\\
\calo_2&=&\frac{\sqrt{6}N^2}{24\pi^2}\ (m_b^0)^2\ \biggl(
\tp_{1,0}+\tp_{1,0}^l \ln\frac{2\pi T_i}{\Lambda_1}
+\calo\left(\frac{(m_b^0)^4}{T_i^24}\right)
\biggr) \,.
\nonumber
\end{eqnarray}

Finally, we compute the final temperature  and the final energy
density/pressure after the quench. Denoting
\begin{equation}
\ta_{2,4}^\infty=\lim_{\t\to +\infty} \ta_{2,4}(\t)=\frac16\ln 2-\frac 13
\int_{-\infty}^{\infty} ds\ p_{1,0}(s)\ (p_{1,0}^l)'(s)\,,
\eqlabell{aindef2cft}
\end{equation}
and matching the $\t\to \infty$ limit of eq.~\eqref{vev12qft1} with
eq.~\eqref{equil2}, we find
\begin{equation}
\begin{split}
\frac{T_f}{T_i}=&\ 1-\left(\frac{1}{36}+\frac16 \ta_{2,4}^\infty \right)
\frac{(m_b^0)^4}{\pi^4 T_i^4}
+\calo\left(\frac{(m_b^0)^8}{T_i^8}\right)\,,\\
\frac{\cale_f}{\cale_i}=&\ 1-\left(\frac 19 \ln\frac{\pi T_i}{\Lambda_1}
+\frac23 \ta_{2,4}^\infty\right)
\frac{(m_b^0)^4}{\pi^4 T_i^4}
+\calo\left(\frac{(m_b^0)^8}{T_i^8}\right)\,,\\
\frac{\calp_f}{\calp_i}=&\ 1-\left(\frac{1}{9} -\frac 13 \ln\frac{\pi T_i}{\Lambda_1}+\frac 23
 \ta_{2,4}^\infty\right)
\frac{(m_b^0)^4}{\pi^4 T_i^4}+\calo\left(\frac{(m_b^0)^8}{T_i^8}\right)\,.
\end{split}
\eqlabell{ftdim2cft}
\end{equation}
Given these expressions, it is straightforward to shown that the change
in the entropy is given by
 \beq
\frac{S_f}{S_i} = 1-\frac{1}{2} \ta_{2,4}^\infty\,
\frac{(m_b^0)^4}{\pi^4 T_i^4}
+\calo\left(\frac{(m_b^0)^8}{T_i^8}\right)\,. \eqlabell{sfsi2cft}
 \eeq
Hence we again find that $\ta_{2,4}^\infty$ gives a direct measure of
the entropy production in these quenches. Of course, this simple result
\reef{sfsi2cft} relies on the judicious choice of $\calc_2$ in
eq.~\reef{C2cftxnew}. Furthermore all of the general comments made
above about the transitions from the CFT to the massive theory can be
extended to analogous statements about the present class of quenches.

Again, we observe that if $p_{1,0}^l(\t)$ describes a quench from a CFT
to a massive theory, then
\begin{equation}
\tilde{p}_{1,0}^l(\t)\equiv 1-p_{1,0}^l(\t)
\eqlabell{reversedim21}
\end{equation}
will describe a quench from a massive theory to a CFT. Because our
calculations are perturbative in the amplitude of the scalar field, we
will also have
\begin{equation}
\tilde{p}_{1,0}(\t)=p_{1,0}(\t=+\infty)-p_{1,0}(\t)=-\ln 2-p_{2,0}(\t)
\eqlabell{reversedim22}
\end{equation}
for this `reverse' quench. Thus, comparing eqs.~\eqref{aindef2} and
\eqref{aindef2cft}, we again find
\begin{equation}
\ta_{2,4}^\infty=a_{2,4}^\infty\,.
\eqlabell{a24t2}
\end{equation}
Hence the entropy production of a quench from the CFT to the massive
theory is precisely the same as in the reverse quench from the massive
theory to the CFT. Again, this result occurs because we are working
perturbatively in the high temperature approximation.

\section{Numerical procedure} \label{numercl}

As described at the beginning of section 3, the focus of our
numerical calculations is the linearized scalar wave equation
\reef{dec2}. The non-normalizable mode of the scalar specifies the
time-dependent profile of the corresponding mass parameter in the dual
gauge theory. Solving the scalar equation allows us to extract the
normalizable mode, with which we can evaluate the one-point function of
the corresponding operator. With this information, it is also
straightforward to integrate the corresponding boundary constraint,
eq.~\reef{a24dim3} or \reef{a24dim2}, which then allows us to calculate
the energy density and pressure.

For the sake of clarity, we describe the numerical procedure for the
fermionic and bosonic operators, \ie $m^2=-3$ and $m^2=-4$, separately.
Below and in the following section, we also focus on the quenches which
go from the CFT to the massive theory. As described in the previous
section, the reverse quenches going from the massive theory to the CFT
are then easily obtained using eqs.~\eqref{reversedim31} and
\eqref{reversedim32} for $m^2=-3$ or eqs.~\eqref{reversedim21} and
\eqref{reversedim22} for $m^2=-4$.

\subsection{$m^2=-3$}

We study mass quenches generated by
\begin{equation}
p_{1,0}=\frac 12+\frac 12\tanh\frac{\t}{\a}=
\frac12+\frac 12\tanh\frac{t}{\calt}\,,
\eqlabell{mq3}
\end{equation}
where
 \beq
\calt=\frac{\a}{\mu} =\frac{\a}{\pi
T_i}\left(1+\calo\left(\frac{(m_f^0)^2}{T_i^2}\right)\right)\,.
 \labell{split3}
 \eeq
Here we have used the fact that at the boundary $t=v=\frac {\t}{\mu}$
--- see eq.~(\ref{EFisFG}). By varying the constant $\a$, we are able to
study quenches that are much faster or slower than the characteristic
thermal time-scale in the corresponding plasma, \ie $1/T_i$.

Solving the linearized equation \eqref{dec2} for $\phi_1$, we are then
able extract the normalizable coefficient $p_{1,2}(\t)$, which appears
as shown in eq.~\eqref{uvperdim3}. In practice, it is more convenient
to integrate the evolution of $\hp_1(\t,\r)$, which we define as
\begin{equation}
\phi_1(\t,\r)\equiv \r\ p_{1,0}+\r^2\ \del_\t p_{1,0}+\frac 12\r^3
\ \ln\r\ \del^2_{\t\t}p_{1,0}+\r\ \hp_1(\t,\r)  \,.
\eqlabell{defhp}
\end{equation}
Comparing this definition with eq.~\eqref{uvperdim3}, we see that
asymptotically as $\r\to0$,
\begin{equation}
\hp_1=p_{1,2}\ \r^2+\calo(\r^3\ln\r)\,.
\eqlabell{hpass}
\end{equation}
Hence organizing the scalar in this way provides for a natural way to
enforce the desired boundary condition for $p_{1,0}$ and to deal with a
regular field in the numerical implementation. The scalar wave equation
\eqref{dec2} now becomes
\begin{equation}
0=\del^2_{\t\r}\hp_1- \frac{1-\r^4}{2}\ \del^2_{\r\r} \hp_1-\frac{1}{2\r}\ \del_\t \hp_1
+\frac{1+3\r^4}{2\r}\ \del_\r \hp_1+\frac{\r^2}{2}\ \hp_1+J_0\,,
\eqlabell{hpdim3}
\end{equation}
with
\begin{equation}
J_0=\frac 14 \r (2+3 \ln\r)\ \del^3_{\t\t\t}p_{1,0}+\frac34 \r^4 (2+3 \ln\r)\
\del^2_{\t\t}p_{1,0}+2 \r^3\ \del_\t p_{1,0}+\frac 12 \r^2\ p_{1,0}\,.
\eqlabell{defaj3}
\end{equation}
From eq.~\reef{hpass}, the required boundary condition is $\hp_1=0$.

\subsection{$m^2=-4$}
We study mass quenches generated by
\begin{equation}
p_{1,0}^l=\frac 12+\frac 12\tanh\frac{\t}{\a}=
\frac12+\frac 12\tanh\frac{t}{\calt}\,,
\eqlabell{mq2}
\end{equation}
where
 \beq
\calt=\frac{\a}{\mu} =\frac{\a}{\pi
T_i}\left(1+\calo\left(\frac{(m_b^0)^4}{T_i^4}\right)\right)\,.
 \labell{Xct4}
 \eeq
We are again using $t=v=\frac {\t}{\mu}$ at the boundary and we will
vary $\a$ to study quenches that are much faster or slower than the
characteristic thermal time-scale, \ie $1/T_i$.

Solving the linearized equation \eqref{dec2} for $\phi_1$, we extract
the normalizable coefficient $p_{1,0}(\t)$, which appears as shown in
eq.~\eqref{uvperdim2}. In this case, it is convenient to introduce
$\hp_1(\t,\r)$ defined as
\begin{equation}
\phi_1(\t,\r)=\r^2\ln\r \ p_{1,0}^l+\r^3\ln\r\ \del_\t p_{1,0}^l+\r\ \hp_1(\t,\r)\,.
\eqlabell{defhpdim2}
\end{equation}
Comparing this definition with eq.~\eqref{uvperdim2}, we see that
asymptotically
\begin{equation}
\hp_1=p_{1,0}\ \r+\calo(\r)\,.
\eqlabell{hpassdim2}
\end{equation}
Writing the scalar equation \eqref{dec2} in terms of $\hp_1$, we have
\begin{equation}
0=\del^2_{\t\r}\hp_1- \frac{1-\r^4}{2}\ \del^2_{\r\r} \hp_1
-\frac{1}{2\r}\ \del_\t \hp_1 +\frac{1+3\r^4}{2\r}\ \del_\r \hp_1
-\frac{1-\r^4}{2\r^2}\ \hp_1+J_0\,,
\eqlabell{hpdim2}
\end{equation}
with
\begin{equation}
J_0=\frac 12 \r (2+3 \ln\r)\ \del^2_{\t\t}p_{1,0}^l+\frac32 \r^4 (2+3 \ln\r)\
\del_{\t}p_{1,0}^l+2 \r^3(1+\ln\r)\ p_{1,0}^l\,.
\eqlabell{defaj2}
\end{equation}
From eq.~\reef{hpassdim2}, the boundary condition is $\hp_1=0$ for
$\r\to 0$.

\begin{figure}[t]
\begin{center}
\psfrag{t}{{$\t$}}
\psfrag{pthree}{{$p_{1,2}(\t)$}}
\psfrag{ptwo}{{$p_{1,0}(\t)$}}
  \includegraphics[width=3in]{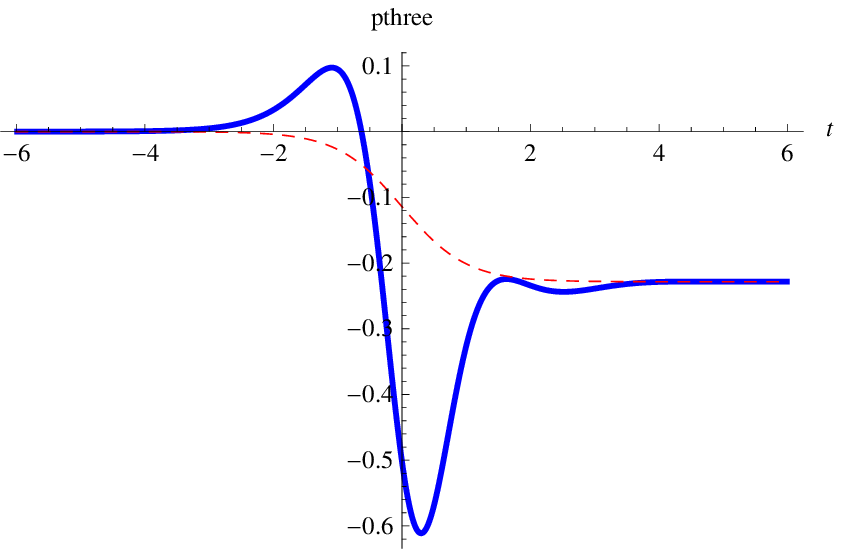}
  \includegraphics[width=3in]{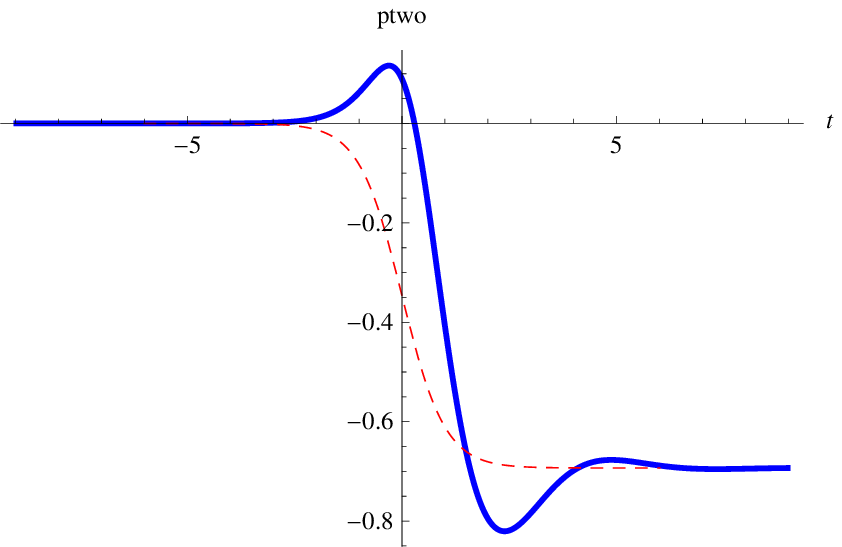}
\end{center}
  \caption{
(Colour online) Evolution of the normalizable component $p_{1,2}$ (left panel) and $p_{1,0}$
(right panel) during the quenches in eqs.~\eqref{mq3} and \eqref{mq2}, respectively, with $\a=1$.
The dashed red lines represent the adiabatic response given by eqs.~\eqref{adresponse}
and \eqref{adresponsedim2}, respectively.
}\label{figure1}
\end{figure}

Eqs.~(\ref{hpdim3}) and (\ref{hpdim2}) were discretized employing
second order finite difference approximations as discussed in Appendix
\ref{appd}. The normalizable response, \ie $p_{1,2}$ for $m^2=-3$ and
$p_{1,0}$ for $m^2=-4$, is then obtained by a straightforward fit from
the numerical solutions obtained. Examples of the computed values
versus time for the case $\alpha=1$ are shown in fig.~\ref{figure1},
together with the adiabatic responses from eqs.~\reef{ad2} and
\reef{ad2x2} for guidance:
\begin{eqnarray}
\big[p_{1,2}(\t)\big]_{adiabatic} &=& -\frac{\Gamma\left(\frac 34\right)^4}{\pi^2}
\ p_{1,0}(\t)\,,
 \eqlabell{adresponse}\\
\big[p_{1,0}(\t)\big]_{adiabatic} &=& -\ln 2\ p_{1,0}^l(\t)\,.
\eqlabell{adresponsedim2}
\end{eqnarray}

\section{Results}\label{res}

In this section we present the results for a range of different
quenches with both the fermionic and bosonic operators. We discuss each
case separately.

\subsection{Quenches with $\calo_3$} \label{halo3}

\begin{figure}[t]
\begin{center}
  \includegraphics[width=2.3in,angle=-90]{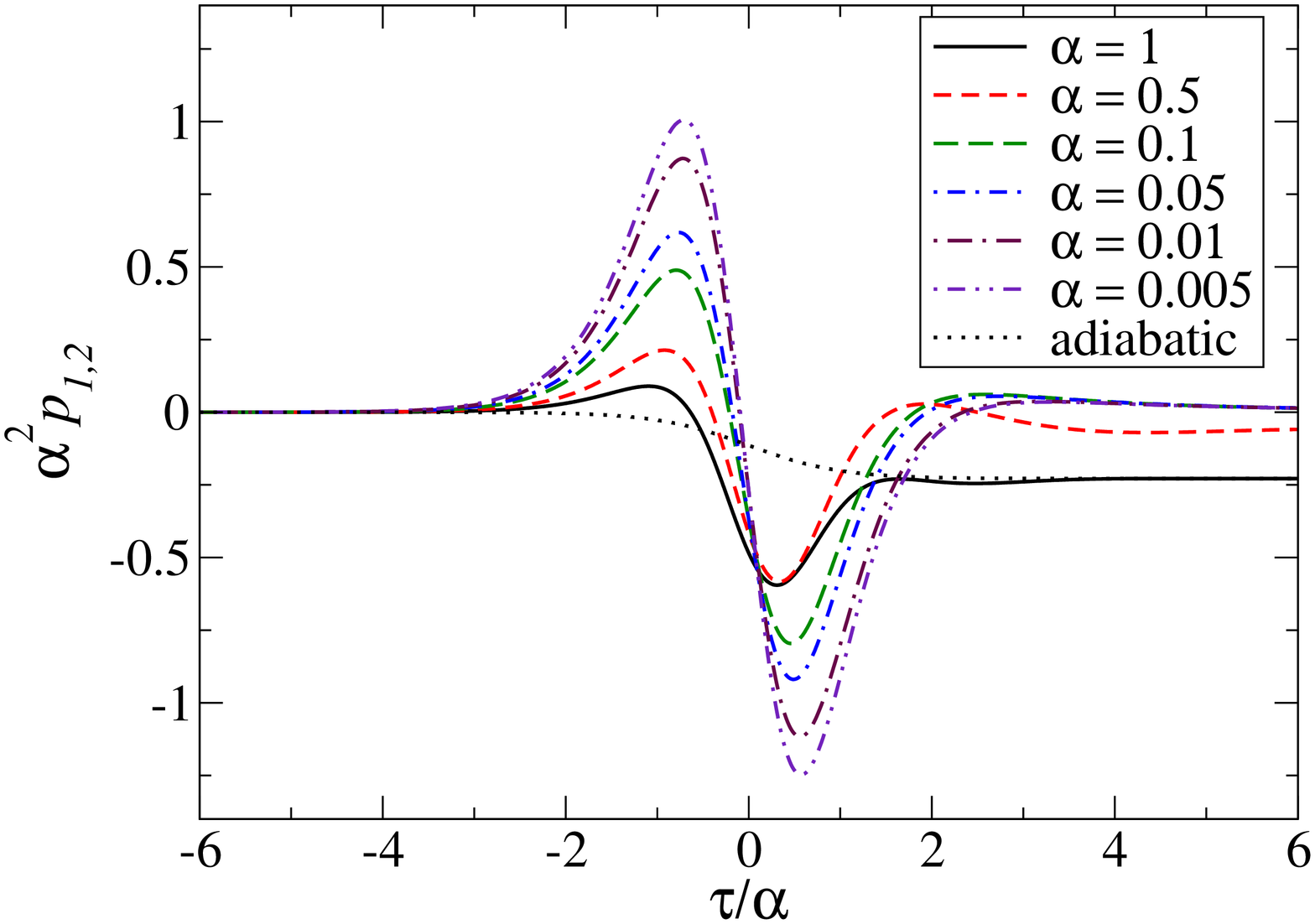}
  \includegraphics[width=2.3in,angle=-90]{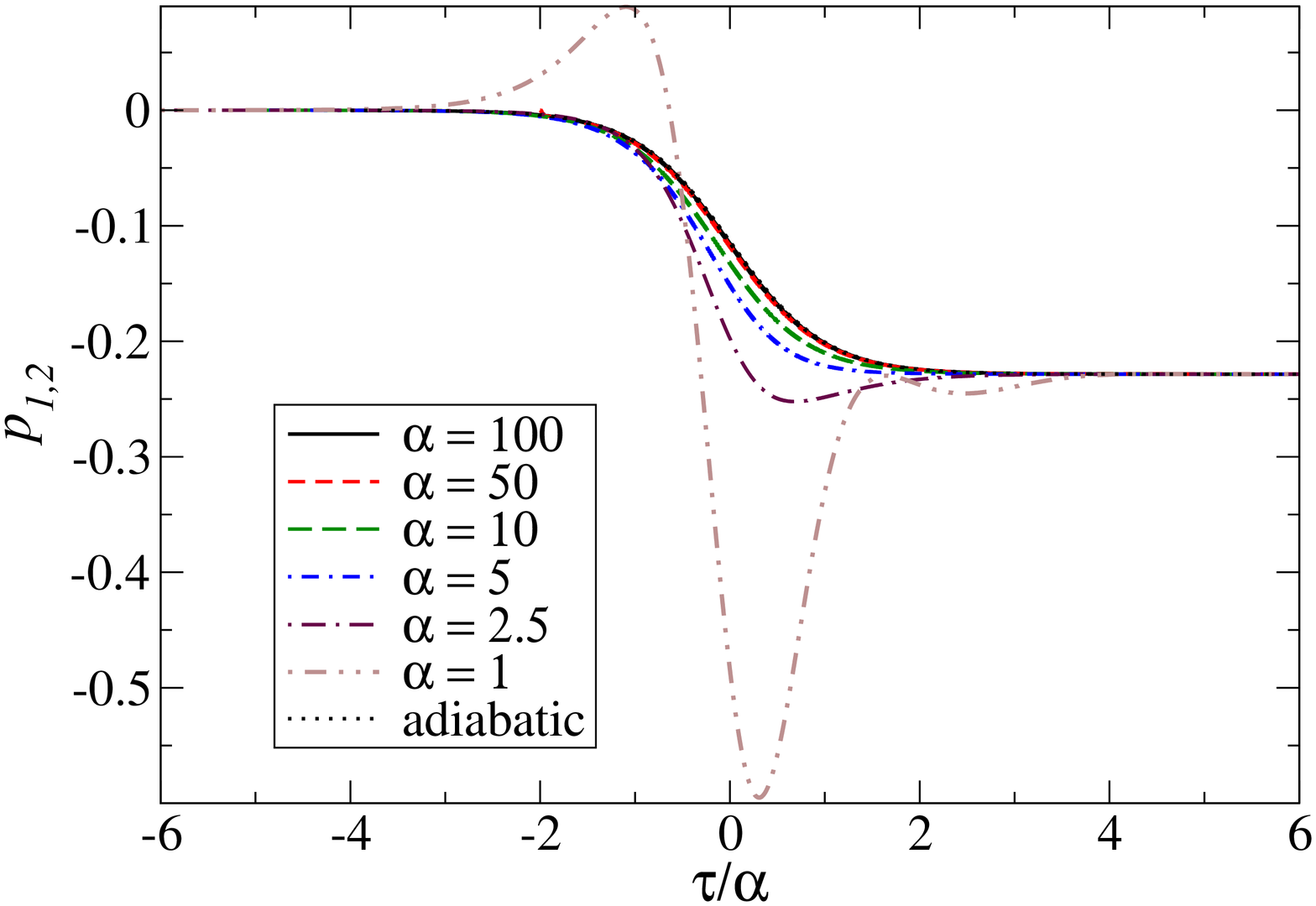}
\end{center}
  \caption{(Colour online)
The curves on the left plot represent the evolution of the
$\a$-rescaled normalizable component, $\a^2\ p_{1,2}$, as a function of
$\frac{\t}{\a}$ during the quench \eqref{mq3} with different values of
$\a$. The
curves on the right plot represent the evolution of the normalizable
component, $p_{1,2}$, as a function of $\frac{\t}{\a}$ during the
quench \eqref{mq3} for the representative values of $\a$.
}\label{figure2}
\end{figure}
%
%
\begin{figure}[t]
\psfrag{a}{{$\ln\a$}}
\psfrag{f}{{$\calf(\a)$}}
\begin{center}
  \includegraphics[width=4in]{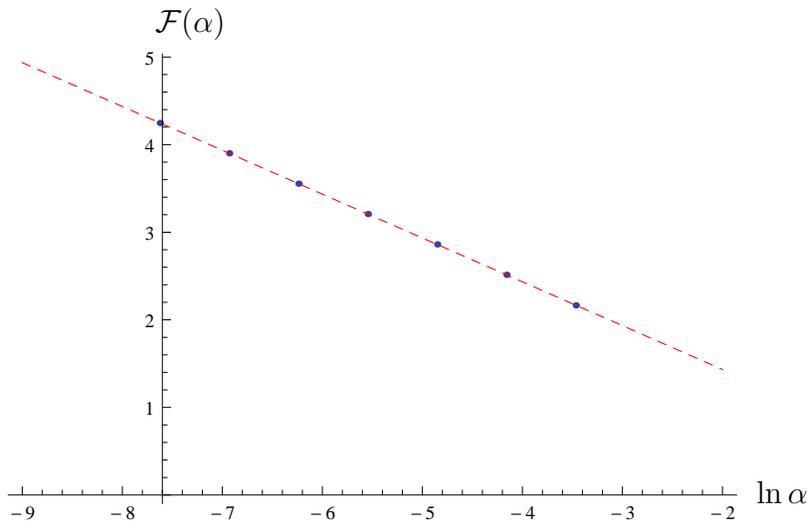}
\end{center}
  \caption{(Colour online) $\calf(\a)$ quantifies the response $p_{1,2}$
for abrupt quenches, \ie as $\a\to 0$ --- see the definition in
eq.~\eqref{deff}. The blue dots correspond to $\calf(\a_n)$ for
$\a_n=2^{-n}$ with $n=5,\ldots,11$. The dashed red line represents the
linear fit to these points given in eq.~\reef{fitfa}. }\label{figure2a}
\end{figure}
%
%
\begin{figure}[t]
\begin{center}
  \includegraphics[width=3.in,angle=-90]{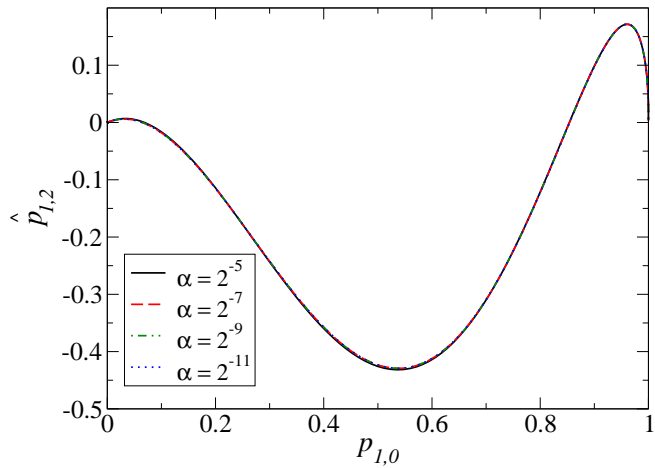}
\end{center}
  \caption{(Colour online) Universality of the subtracted response
$\hat{p}_{1,2}$ (defined in eq.~\eqref{p12s}) for abrupt quenches.
The different curves are virtually indistinguishable from each other.
}\label{figure2b}
\end{figure}
%
%
\begin{figure}[t]
\begin{center}
\psfrag{a}{{$\ln\a$}}
\psfrag{m}{{$\ln(-a_{2,4}^{\infty})$}}
  \includegraphics[width=3.5in]{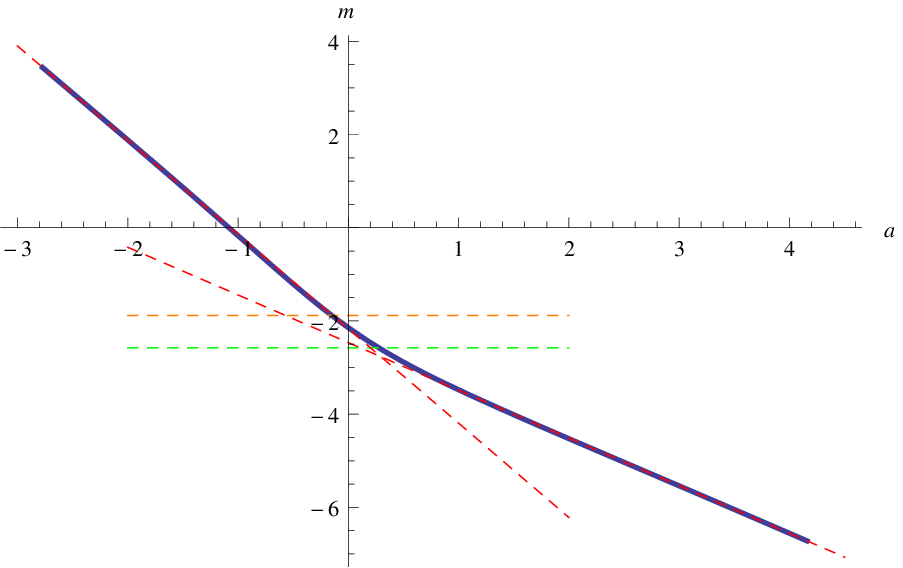}
\end{center}
  \caption{(Colour online)
Log-log plot of coefficient $(-a_{2,4}^\infty)$ as a function of $\a$.
The dashed red lines represent the linear fits to the data (blue curve)
for `fast' quenches with small $\alpha$ ($\ln\a \to -\infty$) and for
`slow' quenches with large $\a$ ($\ln\a \to +\infty$). The dashed green
and orange horizontal lines indicate the thresholds given in
eqs.~\reef{largeenergy} and \reef{larget}, respectively. For values of
$(-a_{2,4}^\infty)$ above the dashed green line, both classes of
quenches produce a final energy density which exceeds the initial
energy density. For values of $(-a_{2,4}^\infty)$ above dashed orange
line, the final temperature is always larger than the initial
temperature for either type of quench. See the discussion in the main
text.}\label{figure2t}
\end{figure}

The source $p_{1,0}$ is given by eq.~\eqref{mq3} and hence $\a$
indicates the the characteristic time scale of the quench. The response
of the normalizable component $p_{1,2}$ depends crucially on whether
the transition is `fast' with $\a<1$ or `slow' with $\a>1$. Our results
for both cases are presented in fig.~\ref{figure2}. The curves in the
left panel present the (rescaled) profile of $p_{1,2}$ as function of
$\frac \t\a$ for fast quenches. As indicated, here we have adopted
values $\a=\lbrace1,0.5,0.1,0.05,0.01,0.005 \rbrace$. The right panel
we shows the evolution of $p_{1,2}$ as a function of $\frac\t\a$ for
slow quenches with $\a=\lbrace1,2.5,5,10,50,100 \rbrace$.

Consider first the fast quenches in the left panel of
fig.~\ref{figure2}. The response of these quenches deviate very far
from the adiabatic response \eqref{adresponse}, which is represented by
the red dashed line (with no rescaling). Indeed for the adiabatic
response $|p_{1,2}|<0.3$ while the plots show that the maximum of the
response $p_{1,2}$ scales faster than $\frac {1}{\a^2}$ for small $\a$.
There are three distinct phases in the evolution of $p_{1,2}$. In the
infinite past the system starts at equilibrium with $p_{1,0}=0$ and
hence $p_{1,2}(-\infty)=0$ for all quenches. In the infinite future,
the system ends at equilibrium in the massive theory with $p_{1,0}=1$
and hence $p_{1,2}(+\infty)= -\Gamma\left(\frac 34\right)^4/\pi^2$ for
all quenches.\footnote{Note that the curves in fig.~\ref{figure2}
approach different values at large $\tau$ because we have rescaled the
response with a factor of $\a^2$ in this plot.} The component of
$p_{1,2}$ which scales as $1/\a^2$ is first excited whenever the source
$p_{1,0}$ noticeably deviates from $0$, \ie whenever $p_{1,0}>0.01$
which corresponds to $\frac\t\a>-2.3$. In fig.~\ref{figure2}, $\a^2
p_{1,2}$ also appears to rapidly approach the asymptotic
value\footnote{As we discuss below, the curves fig.~\ref{figure2} do
not allow us to resolve in the details  the approach to the final
equilibrium.} for large $\t$ when the source is sufficiently close to
$1$, \ie whenever $(1-p_{1,0})<0.01$ and correspondingly $\frac\t\a>
2.3$. Indeed, most of the variation in $\a^2 p_{1,2}$ occurs
approximately within the interval $\frac{\t}{\a}\in (-4,4)$. In this
sense, the $\a^2$-rescaled response {\it follows} the characteristic
time-scale of the source \eqref{mq3}. That is, faster variations in the
source $p_{1,0}$ (smaller values of $\a$) result in both the faster
excitation and the faster equilibration of $\a^2 p_{1,2}$.

The right panel in fig.~\ref{figure2}  presents the evolution of the
normalizable component $p_{1,2}$ for slow quenches, \ie for $\a>1$.
Note that we do not rescale $p_{1,2}$ in this plot. Here we see for
progressively larger values of $\a$, the profiles come closer to
resembling the adiabatic response given by eq.~\eqref{adresponse}
--- the latter is represented by a barely visible red dashed line.

While it is not surprising that in the limit $\a\to \infty$ the response
$p_{1,2}$ follows a universal adiabatic profile \eqref{adresponse}, it
is remarkable that the response $p_{1,2}$ is also universal for abrupt
quenches, \ie in the limit $\a\to 0$. In this regard, we first observe
that in fig.~\ref{figure2}, the profile $\a^2p_{1,2}$ has a relatively
simple form for very small $\a$ and further that this simple form is
remarkably similar to $\a^2p_{1,0}''=-\sinh(\t/\a)/\cosh^3(\t/\a)$.
This observation motivated us to define $\calf(\a)$ as the constant
which minimizes the following:
\begin{eqnarray}
\| p_{1,2}-\calf(\a)\ p_{1,0}''  \|
&\equiv&\int_0^1 d(p_{1,0})\ \left(p_{1,2}-\calf(\a)\ p_{1,0}''\right)^2
\nonumber\\
&=&\int_{-\infty}^{\infty}d\tau\ p'_{1,0}\ \left(p_{1,2}-\calf(\a)\ p_{1,0}''\right)^2\,.
\labell{deff}
\end{eqnarray}
The first expression defining the norm above was chosen since it seems
natural if $\calf(\a)$ is to describe the universal character of the
response. From the second expression, we see that since
$p'_{1,0}=1/(2\a \cosh(\t/\a)^2$, this measure weights most heavily the
region in the vicinity of $\tau=0$, where the peaks in $p_{1,2}$ occur.
Fig.~\ref{figure2a} shows $\calf(\a_n)$ for $\a_n=2^{-n}$ with
$n=5,\ldots,11$. As also shown in the figure, these points are very
well fit with a simple linear expression which takes the form:
\begin{equation}
\calf\big|_{fit}\simeq 0.43 - 0.50\ \ln\a\,.
\labell{fitfa}
\end{equation}
This simple result indicates that the leading contribution in response
$p_{1,2}$ actually grows as approximately $\alpha^{-2}\ln(1/\sqrt{\a})$
as $\a\to0$, \ie faster than $\alpha^{-2}$. Now let us define the
`subtracted' response $\hat{p}_{1,2}$ as follows
\begin{equation}
\hat{p}_{1,2}\equiv \a^2\ \left(p_{1,2} -\calf(\a)\ p_{1,0}''\right)\,,
\labell{p12s}
\end{equation}
which is then plotted as a function of $p_{1,0}$ in
fig.~\ref{figure2b}. Remarkably, all four curves in this plot are
almost indistinguishable from each other and so we see that
$\hat{p}_{1,2}$ also has a simple universal behaviour in the limit
$\a\to 0$.

Eqs.~\eqref{ftdim3} and \reef{sfsi} show the characteristics of the
final equilibrium state resulting from the quench, \ie the temperature
$T_f$, the energy density $\cale_f$, the pressure $\calp_f$ and entropy
density $S_f$, relative to the initial state parameters. All of these
ratios depend on coefficient $a_{2,4}^\infty$, given in
eq.~\reef{aindef}. As noted in the discussion in section \ref{subm3},
in the adiabatic limit, $\a\to \infty$, $a_{2,4}^\infty$ vanishes and
so the entropy density is constant. Further for these adiabatic
transitions from the thermal state of the CFT plasma to the
mass-deformed thermal state result in an increase of the temperature
$T_f>T_i$ and  the energy density ${\cale_f}>{\cale_i}$, while the
pressure is decreased $\calp_f<{\calp_i}$. In fact, in all of our
simulations with finite $\a$, we found that $a_{2,4}^\infty<0$ and
further $|a_{2,4}^\infty|$ grows as $\a$ becomes smaller --- see below.
Of course, in agreement with one's intuition then, eq.~\reef{sfsi}
indicates that the entropy density increases in a generic quench and in
fact, the entropy production grows as the quenches become faster.
Further, according to eq.~\eqref{ftdim3}, more rapid quenches result in
greater `heating', \ie the increase in the temperature and the energy
density grows as $\a$ becomes smaller.

We will consider these results more quantitatively below but first let
us consider the properties of quenches which make a transition from a
thermal state in the mass-deformed gauge theory to a thermal CFT state.
As described in section \ref{subm3}, we can make a direct translation
of any of the previous results to these `reverse' quenches using
eqs.~\reef{reversedim31} and \reef{reversedim32}. For these reverse
transitions, the characteristics of the final equilibrium relative to
the initial state are given by eqs.~\eqref{ftdim3cft} and
\reef{sfsicft} and these ratios are controlled by $\ta_{2,4}^\infty$,
given in eq.~\reef{aindefcft}. However, using the previous translation,
one finds that, as presented in eq.~\reef{a24t}:
$\ta_{2,4}^\infty=a_{2,4}^\infty$. In the adiabatic limit, these
transition from the mass-deformed theory to the CFT state result in
`cooling' of the system. That is, $\ta_{2,4}^\infty$ vanishes for
$\a\to\infty$ and while the entropy density is constant, the
temperature and energy density decrease. However, for very rapid
transitions, $\ta_{2,4}^\infty$ becomes large and negative and so
eq.~\eqref{ftdim3cft} shows that the quenches are again `heating' the
plasma with both $T_f>T_i$ and ${\cale_f}>{\cale_i}$. In fact, we see
that the thresholds for increasing the temperature and energy density
are slightly different. Specifically, from \eqref{ftdim3cft},
$\cale_f>\cale_i$ provided
\begin{equation}
-\ta_{2,4}^\infty> \frac{\Gamma\left(\frac 34\right)^4}{3\pi^2}\,,
\eqlabell{largeenergy}
\end{equation}
while $T_f>T_i$ requires
\begin{equation}
-\ta_{2,4}^\infty> \frac{2\Gamma\left(\frac 34\right)^4}{3\pi^2}\,.
\eqlabell{larget}
\end{equation}
As shown in fig.~\ref{figure2t}, both of these thresholds occur close
to $\a=1$:
\begin{equation}
\begin{split}
&\cale_f>\cale_i\qquad \Longleftrightarrow\qquad \a\lesssim 1.32\,, \\
&T_f>T_i\qquad \Longleftrightarrow\qquad \a\lesssim 0.86\,.\\
\end{split}
\eqlabel{thresholds}
\end{equation}

Fig.~\ref{figure2t} presents a log-log plot of the coefficient
$a_{2,4}^\infty$ as a function of $\a$. The behaviour of the plot
reveals simple behaviours in both the `slow' and `fast' regimes with
$\a\gg 1$ and $\a\ll 1$, respectively. In fig.~\ref{figure2t}, the
dashed red lines show the linear fits to the data for these ranges of
$\a$. In particular, we find
\begin{equation}
\begin{split}
{\rm slow}:\qquad \ln(-a_{2,4}^\infty)\big|_{fit}\simeq
-2.465-1.02\ \ln\a\,,\qquad \a\gg1 \,,\\
{\rm fast}:\qquad \ln(-a_{2,4}^\infty)\big|_{fit}\simeq
-2.170-2.02\ \ln\a\,,\qquad \a\ll1 \,.\\
\end{split}
\eqlabell{sffits}
\end{equation}
Hence we find that $a_{2,4}^\infty$ vanishes as $\a\to\infty$ with
$a_{2,4}^\infty\propto1/\a$. However, we also find that
$a_{2,4}^\infty$ is divergent for $\a\to0$ with
$a_{2,4}^\infty\propto1/\a^2$. Let us further observe that the constant
term in the `fast' fit seems to match:
$\ln\!\left[\Gamma\left(\frac34\right)^4/(2\pi^2)\right]\simeq
-2.169\cdots$. Hence our numerical results seems to indicate that the
leading behaviour as $\a\to0$ is given by
 \beq
a_{2,4}^\infty\simeq-\frac{\Gamma\left(\frac34\right)^4}{2\pi^2}
\,\frac1{\a^2}+\cdots\,,
 \labell{guess}
 \eeq
which, of course, calls for an analytic derivation. Unfortunately, at
present, we can provide no insight into such a derivation. However, we
note that the scaling for slow transitions can, in principle, be
deduced analytically from the evolution of $\phi_1$ with $\a\gg 1$
--- see the discussion below at eq.~\eqref{slow1}.

Using eq.~\eqref{ftdim3}, we can translate the asymptotic behaviour in
eq.~\eqref{sffits} into
\begin{equation}
\frac{\Delta \cale}{\cale_i}\equiv \frac{\cale_f-\cale_f^{adiabatic}}{\cale_i}
\propto
\begin{cases}
&\frac{1}{\a}\ \frac{(m_f^0)^2}{T_i^2}\quad{\rm for}\ \a\gg 1\,,\\
&\frac{1}{\a^2}\ \frac{(m_f^0)^2}{T_i^2}\quad{\rm for}\ \a\ll 1\,,
\end{cases}
\eqlabell{tfits}
\end{equation}
where $\cale_f^{adiabatic}$ is the final energy density for an
adiabatic transition, \ie
\begin{equation}
\cale_f^{adiabatic}=\cale_i\ \left(1+\frac{2\,\Gamma\left(\frac34\right)^4}{3\pi^2}
\ \frac{(m_f^0)^2}{\pi^2 T_i^2}
+\calo\left(\frac{(m_f^0)^4}{T_i^4}\right)\right)\,.
\eqlabell{tfadiab}
\end{equation}
Note that the same expression \reef{tfits} also applies for $\Delta
\cale/\cale_i$ in the `reverse' quenches describing transitions from
the mass-deformed gauge theory to the CFT. Further we can similarly
deduce the asymptotic behaviour for the relative change in the
temperature and the pressure using eq.~\eqref{ftdim3} and the entropy
using eq.~\reef{sfsi}. Using eq.~\reef{guess}, we have a more precise
expression for the limit $\a\to0$,
 \beq
\frac{\Delta
\cale}{\cale_i}\simeq\frac{\Gamma\left(\frac34\right)^4}{\pi^4}\,\frac{(m_f^0)^2}{
T_i^2} \,\frac1{\a^2}+\cdots\,.
 \labell{guess2}
 \eeq
In particular then, we note that a discontinuous quench with $\a=0$
would seem to produce physical divergences --- see further discussion
in section \ref{kronk}.

However, let us observe that previously we found that the leading
scaling of amplitude of the response was $\alpha^{-2}\ln(1/\a)$ --- see
discussion around eq.~\reef{fitfa}. Hence one might have expected an
even stronger divergence in eq.~\reef{guess2}, which `only' scales like
$\alpha^{-2}$. This behaviour can be understood as follows: The
divergence in eq.~\reef{guess2} simply reflects the divergence in
$a^\infty_{2,4}$ given in eq.~\reef{guess}. Now in eq.~\reef{aindef}
which defines the latter, the integral dominates for small $\a$ and so
from eq.~\reef{p12s}, we have
 \beqa
a^\infty_{2,4}&\simeq&\frac23\int_{-\infty}^\infty ds \left(
\calf(\a)\,p''_{1,0}(s)- \frac1{\a^2}\,\hat{p}_{1,2}(s)\right)
p'_{1,0}(s)
 \nonumber\\
 &=&\frac2{3\,\a^2}\int_{-\infty}^\infty ds\ \hat{p}_{1,2}(s)\,
p'_{1,0}(s)\,.
 \labell{hack1}
 \eeqa
That is, the specific form of the leading response, \ie being
proportional to $p''_{1,0}$, leads to a total derivative which yields a
vanishing contribution above. Hence, only the `subtracted' response
\reef{p12s} contributes in the integral to produce the $\a^{-2}$
scaling in eq.~\reef{guess2}.

%
%
\begin{figure}[t]
\begin{center}
\psfrag{tl}{{$\frac \t\a$}}
\psfrag{a}{{$\frac {\t_{ex}}{\a}$}}
\psfrag{b}{{$\frac {\t_{eq}}{\a}$}}
\psfrag{dp}{{$\delta_{neq}$}}
  \includegraphics[width=3.5in]{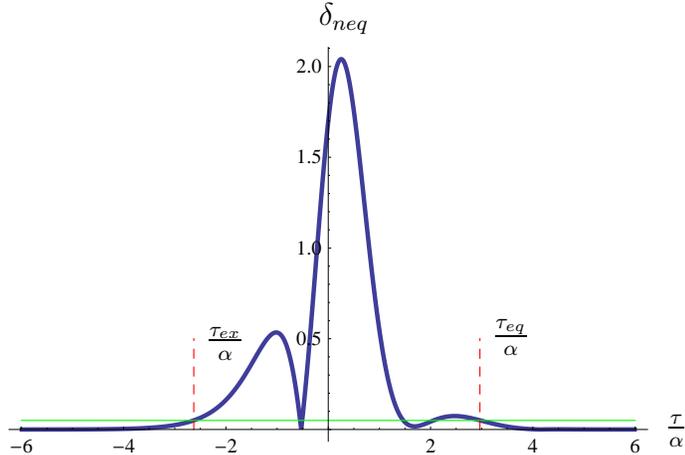}
\end{center}
  \caption{(Colour online) Extraction of the excitation/equilibration rates
for $\a=1$ quench. The horizontal green line is the threshold
for excitation/equilibration which we define to be $5\%$ away from local
equilibrium as determined by $\delta_{neq}$, as given defined in
eq.~\eqref{defddn}. The dashed red lines indicate the earliest and
latest times of crossing this threshold, which we denote as $\t_{ex}$
(for excitation time) and $\t_{eq}$ (for equilibration time), respectively.
}\label{relaxdim3p}
\end{figure}
%
%
\begin{figure}[t]
\begin{center}
\psfrag{la}{{$\ln\a$}}
\psfrag{t}{{$|\t_{ex}|/\a$}}
\psfrag{lt}{{$\ln \frac{\t_{eq}}{\a}$}}
  \includegraphics[width=2.5in]{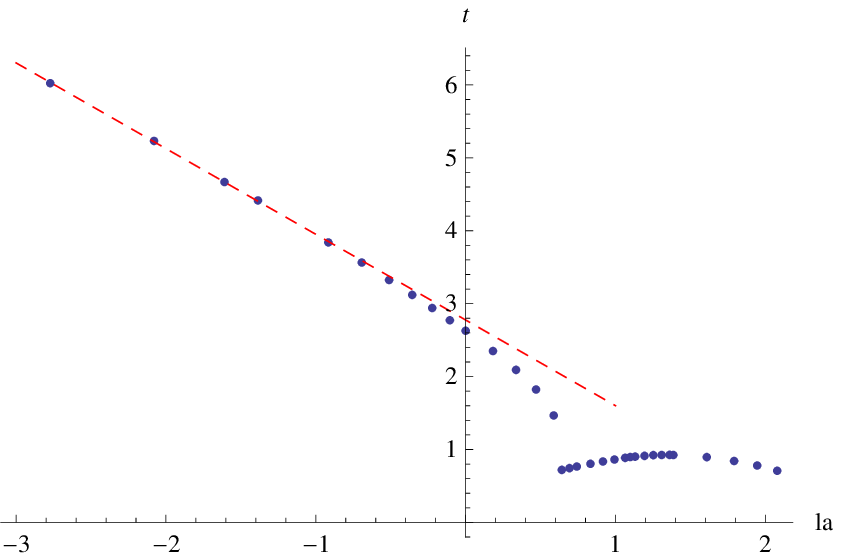}
  \includegraphics[width=2.5in]{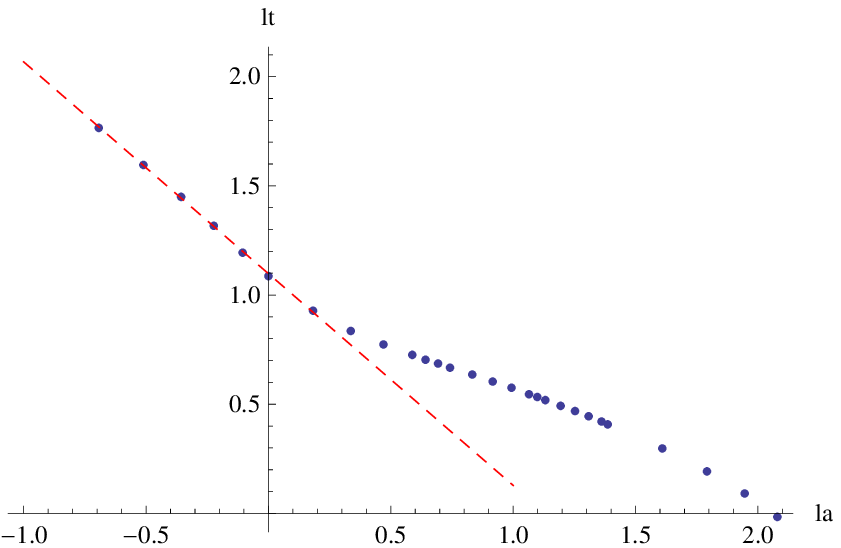}
\end{center}
  \caption{(Colour online) Excitation time $\t_{ex}$ (left panel) and
equilibration time $\t_{eq}$ (right panel) as a function of $\a$. The
threshold is fixed to be $\e=0.05$ --- see eqs.~\eqref{excitationt} and
\eqref{equilibrationt}. In both cases, a simple linear fit describes the
behaviour for small $\a$, as shown with the red dashed lines.}\label{exeqdim3}
\end{figure}
%
%
We conclude this section with  more quantitative analysis of the
excitation and the equilibration time scales in the evolution of  the
normalizable component $p_{1,2}(\t)$. As described above, for an
adiabatic transition, the system essentially maintains thermodynamic
equilibrium throughout the process and $p_{1,2}(\t)$ simply tracks
$p_{1,0}(\t)$, as indicated in eq.~\reef{ad2} or \eqref{adresponse}.
Hence, we define a measure of the deviation from the local equilibrium
as follows
\begin{equation}
\dd_{neq}(\t)\equiv \left|\frac{p_{1,2}(\t)-\left[p_{1,2}(\t)
\right]_{adiabatic}}{p_{1,2}^{equilibrium}}\right|\,,
\eqlabell{defddn}
\end{equation}
where $\left[p_{1,2}(\t)\right]_{adiabatic}$ is given in
eq.~\reef{adresponse} and
\begin{equation}
p_{1,2}^{equilibrium}=\lim_{\t\to+\infty}p_{1,2}(\t)
= -\frac{\Gamma\left(\frac
34\right)^4}{\pi^2}\,.
\eqlabell{defp12eqiui}
\end{equation}
Fig.~\ref{relaxdim3p} shows a typical plot of $\dd_{neq}$. We further
define the excitation time $\t_{ex}$ as the earliest time in the quench
at which $\dd_{neq}$ when the latter quantity exceeds a predefined
threshold value, which we choose to be $\e=0.05$, \ie
\begin{equation}
\dd_{neq}\  <\ \e=0.05\,,\qquad {\rm for\ all}\ \ \t\ <\ \t_{ex}\,.
\eqlabell{excitationt}
\end{equation}
While the choice of the threshold here is somewhat arbitrary, we do not
expect that it will greatly effect the following observations.
Similarly, we define the equilibration time $\t_{eq}$ as the latest
time at which $\dd_{neq}$ exceeds the same threshold value, \ie
\begin{equation}
\dd_{neq}\  <\ \e=0.05\,,\qquad {\rm for\ all}\ \ \t\ >\ \t_{eq}\,.
\eqlabell{equilibrationt}
\end{equation}
All of these quantities are identified in the sample plot shown in
fig.~\ref{relaxdim3p}. In fig.~\ref{exeqdim3}, we show $\t_{ex}$ and
$\t_{eq}$ for quenches across a wide range of $\a$.

Examining fig.~\ref{relaxdim3p}, one sees that $\dd_{neq}$ has several
the zeros (at finite $\t/\a$) which reflect the fact that $p_{1,2}$ is
oscillating about $[p_{1,2}]_{adiabatic}$ --- later we will see that
these oscillations are described by quasinormal modes of the scalar.
The amplitude of oscillations become smaller as $\alpha$ grows and in
fact, for $\a$ sufficiently large, $\dd_{neq}$ will be below any fixed
threshold $\e$ throughout the quench. While we did not try to pinpoint
the precise value of $\a$ at which this occurs, we did establish that
this behaviour sets in before $\a=16$ with $\e=.05$. Further, we see in
fig.~\ref{exeqdim3} that $\t_{ex}$ and $\t_{eq}$ behave irregularly for
large $\a$. In particular, $\t_{ex}$ is clearly discontinuous for
$\a\sim1.8$ and there is a similar discontinuity in $\t_{eq}$, albeit
less profound. Hence, as defined above, these time scales are not
useful diagnostics for slow quenches.

On the other hand, both $\t_{ex}$ and $\t_{eq}$ exhibit simple
behaviours for small $\a$, as shown with the linear fits in
fig.~\ref{exeqdim3}. In the left panel, the linear fit for the
excitation time is
\begin{equation}
\left.\frac{|\t_{ex}|}{\a}\right|_{fit}\simeq 2.8-1.2\ \ln\a\,.
\eqlabell{exfit}
\end{equation}
A heuristic understanding of this behaviour is as follows: In the
discussion around eqs.~(\ref{deff}--\ref{p12s}), we demonstrated that
the leading response takes the form
\begin{equation}
p_{1,2}(\t_{ex})\ \simeq\ \calf(\a)\, p_{1,0}''(\t_{ex})\ \simeq
\  -\frac{\ln\a }{\a^2}\ e^{2\t_{ex}/\a}\,, \qquad
\frac{\t_{ex}}{\a}\to -\infty\,.
\eqlabel{p12fast}
\end{equation}
where the latter uses the approximate form of $p_{1,0}''$ for large
negative $\tau$. Hence if we set a fixed threshold for the response at
early times, we should expect that
\begin{equation}
p_{1,2}(\t_{ex})\ \sim\  {\rm const}\qquad \Longrightarrow\ \qquad
\frac{|\t_{ex}|}{\a}\ \sim\ -\ln\a+\calo\left(\ln(-\ln(\a))\right)  \,,
\eqlabell{scalingex}
\end{equation}
which matches the result given in eq.~\reef{exfit}. Using
eqs.~\eqref{mq3} and \reef{split3}, this behaviour can be translated to
an excitation time in terms of the original boundary time
\begin{equation}
t_{ex}\equiv \calt\ \frac{\t_{ex}}{\a}\ \simeq\ \calt\
\ln\left(1/T_i\calt \right)\qquad {\rm for}\ \ T_i\calt\ll1\,.
\eqlabell{texT}
\end{equation}

The scaling of the equilibration time at small $\a$ is different. For
the right panel on fig.~\ref{exeqdim3}, the linear fit was
\begin{equation}
\left.\ln\frac{\t_{eq}}{\a}\right|_{fit}
\simeq 1.1- 1.0\ \ln\a\,.
\eqlabell{eqfit}
\end{equation}
Hence we have for the equilibration time in fast quenches
\begin{equation}
\frac{\t_{eq}}{\a}\ \sim\ \frac 1\a\,,
\eqlabell{eqscaling}
\end{equation}
or, in terms of the original boundary time,
\begin{equation}
t_{eq}\equiv \calt\ \frac{\t_{eq}}{\a}\ \sim\ \frac{1}{T_i}\,.
\eqlabell{teqT}
\end{equation}
That is, irrespective of how quickly the system is driven away from the
initial equilibrium, the return to the final equilibrium is determined
universally by the typical thermal time-scale.

\begin{figure}[t]
\begin{center}
  \includegraphics[width=3.in,angle=-90]{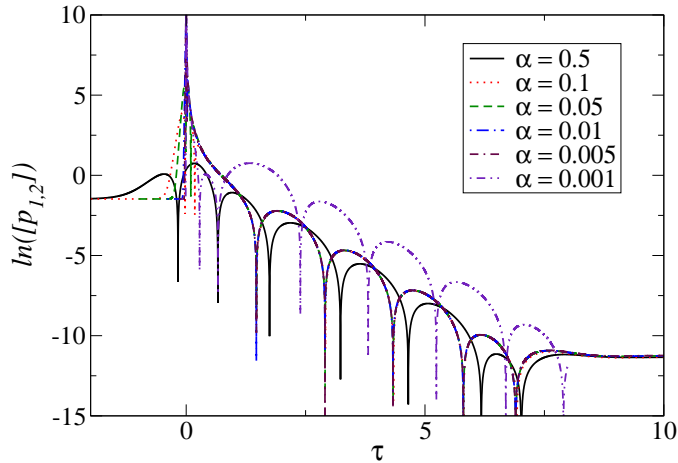}
\end{center}
  \caption{(Colour online) Behaviour of the response coefficient versus time for different
fast quenches. Note that on the vertical axis, we have defined
$[p_{1,2}] \equiv p_{1,2} +\Gamma(3/4)^4
\pi^{-2}$. In all cases, the same quasinormal mode governs the dynamics very soon
after the quench. }\label{figureQNM_3}
\end{figure}

This behaviour seems closely connected to the fact that the late-time
response is controlled by the quasinormal modes of the scalar field
\cite{genera,StarinetsNunez}. The quasinormal oscillations are well
illustrated in fig.~\ref{figureQNM_3}, which shows the late-time
evolution of $p_{1,2}$ for a range of small values of $\alpha$. For all
of these fast quenches, a single mode clearly determines the decay of
$p_{1,2}$ after a very short time. A rough fit of the oscillation
periods and the slopes indicates that in all cases, the decay is
governed by a quasinormal mode with $\omega/(2\pi T_i) \simeq (1.095 +
i\, 0.87)$. This value is consistent with the expected frequency
$(1.099 + i\, 0.879)$, found in \cite{StarinetsNunez}. Beyond this
fundamental mode, the first overtone can also be extracted for about
one oscillation period and a rough fit yields real and imaginary parts
of its frequency which agree with the expected values to within
approximately 25$\%$.

\subsection{Quenches with $\calo_2$} \label{halo2}

\begin{figure}[t]
\begin{center}
  \includegraphics[width=2.2in,angle=-90]{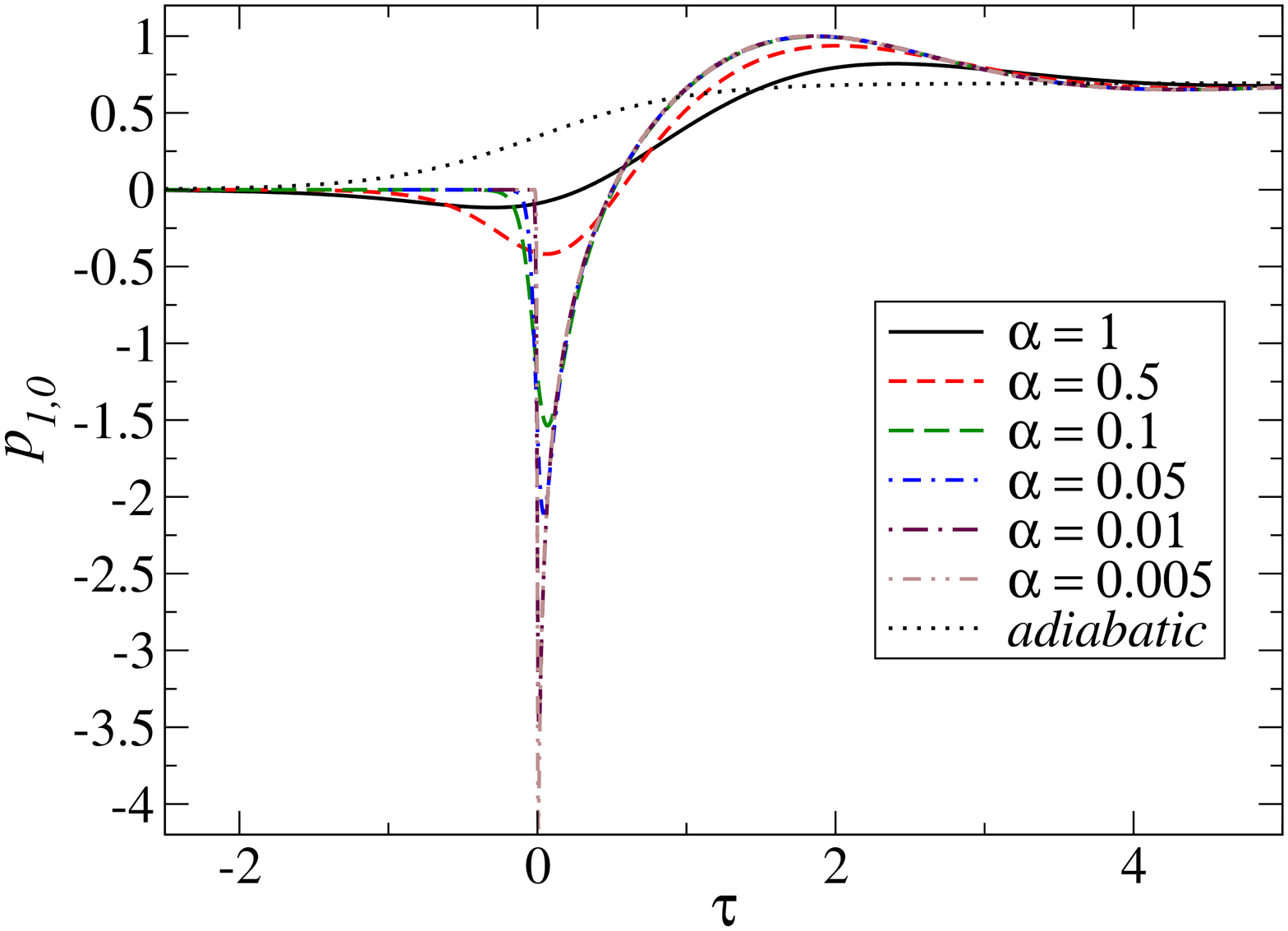}
  \includegraphics[width=2.2in,angle=-90]{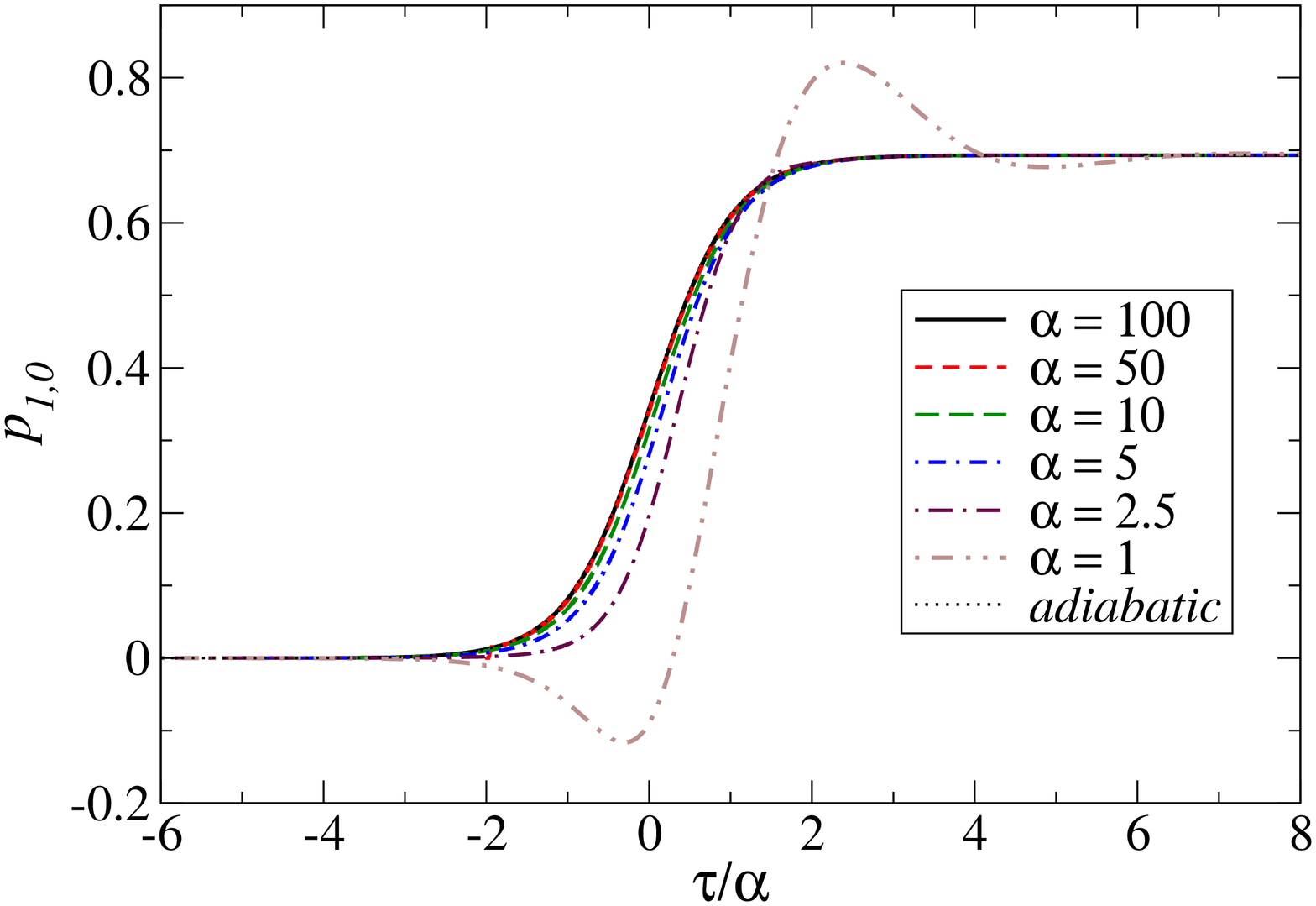}
\end{center}
  \caption{(Colour online) Normalizable component $p_{1,0}$ as a function of
${\t}$ for fast and slow quenches is shown in the left and the right
panels, respectively.
}\label{figure4}
\end{figure}

\begin{figure}[t]
\begin{center}
\psfrag{a}{{$\ln\a$}}
\psfrag{f}{{$\calf(\a)$}}
 \includegraphics[width=4in]{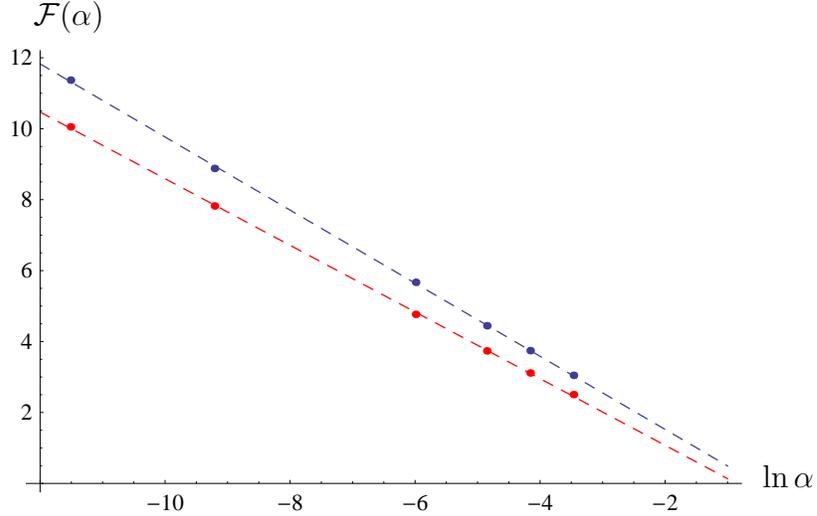}
\end{center}
 \caption{(Colour online) $\calf(\a)$ quantifies the response $p_{1,0}$ for
abrupt quenches, \ie as $\a\to 0$. The red dots correspond to
$\calf_0(\a_n)$ (defined in eq.~\eqref{univdim2}) for $\a_n =\{
{1}/{32},$ ${1}/{64},$ ${1}/{128},$  $0.0025 ,$ $0.0001,$ $0.00001\}$.
The dashed red line represents the linear fit to these points given in
eq.~\eqref{univdim22}. The blue dots correspond to $\calf_1(\a_n)$
(defined in eq.~\eqref{deff2}) for the same $\a_n$. The dashed blue
line represents the linear fit to the latter points given in
eq.~\eqref{univdim22z}.}\label{univcdim2a}
\end{figure}

\begin{figure}[t]
\begin{center}
\psfrag{x}{{$p_{1,0}^l$}}
\psfrag{y}{{${p}_{1,0}/\calf(\a)$}}
  \includegraphics[width=2.5in]{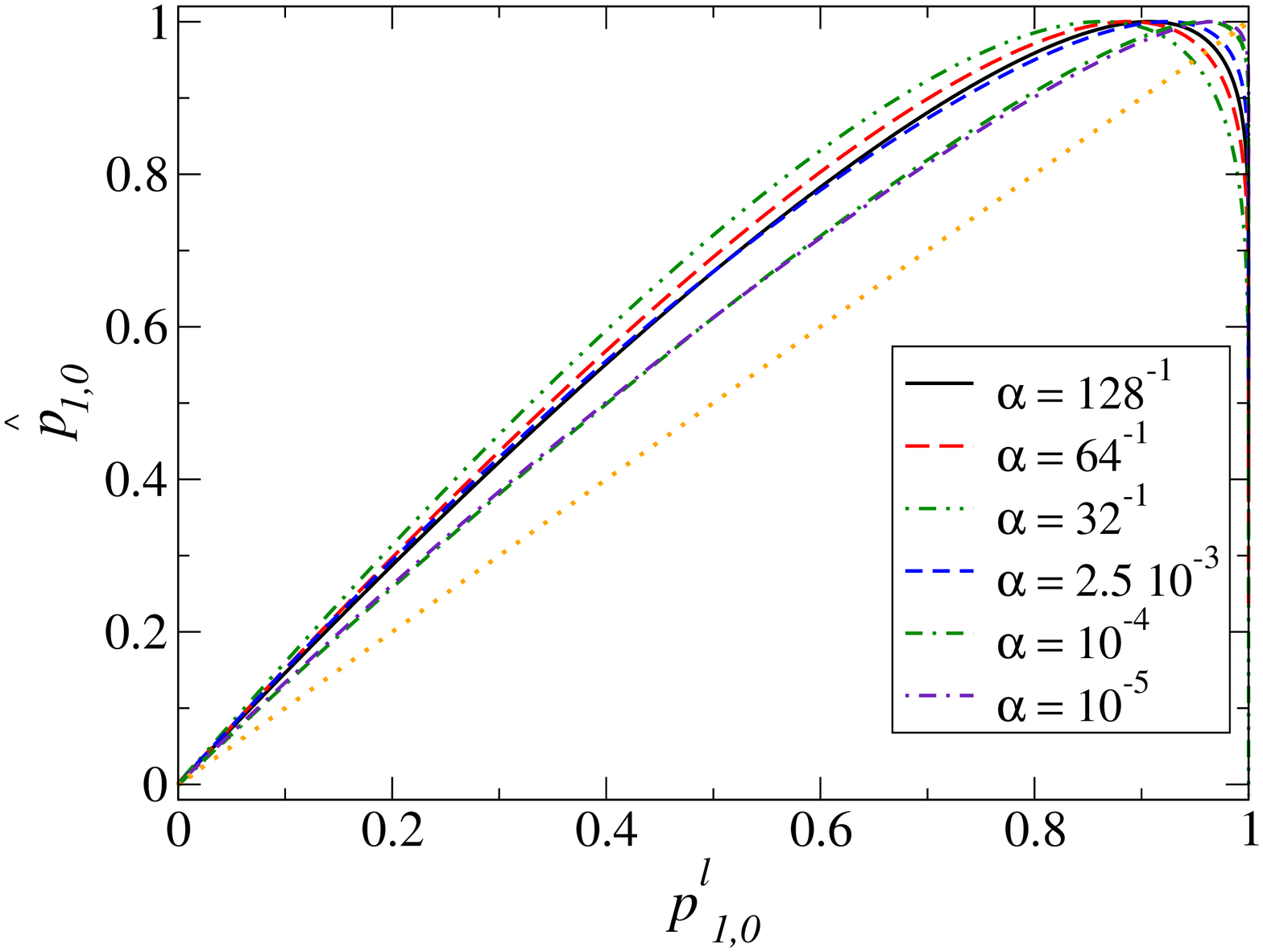}
  \includegraphics[width=2.5in]{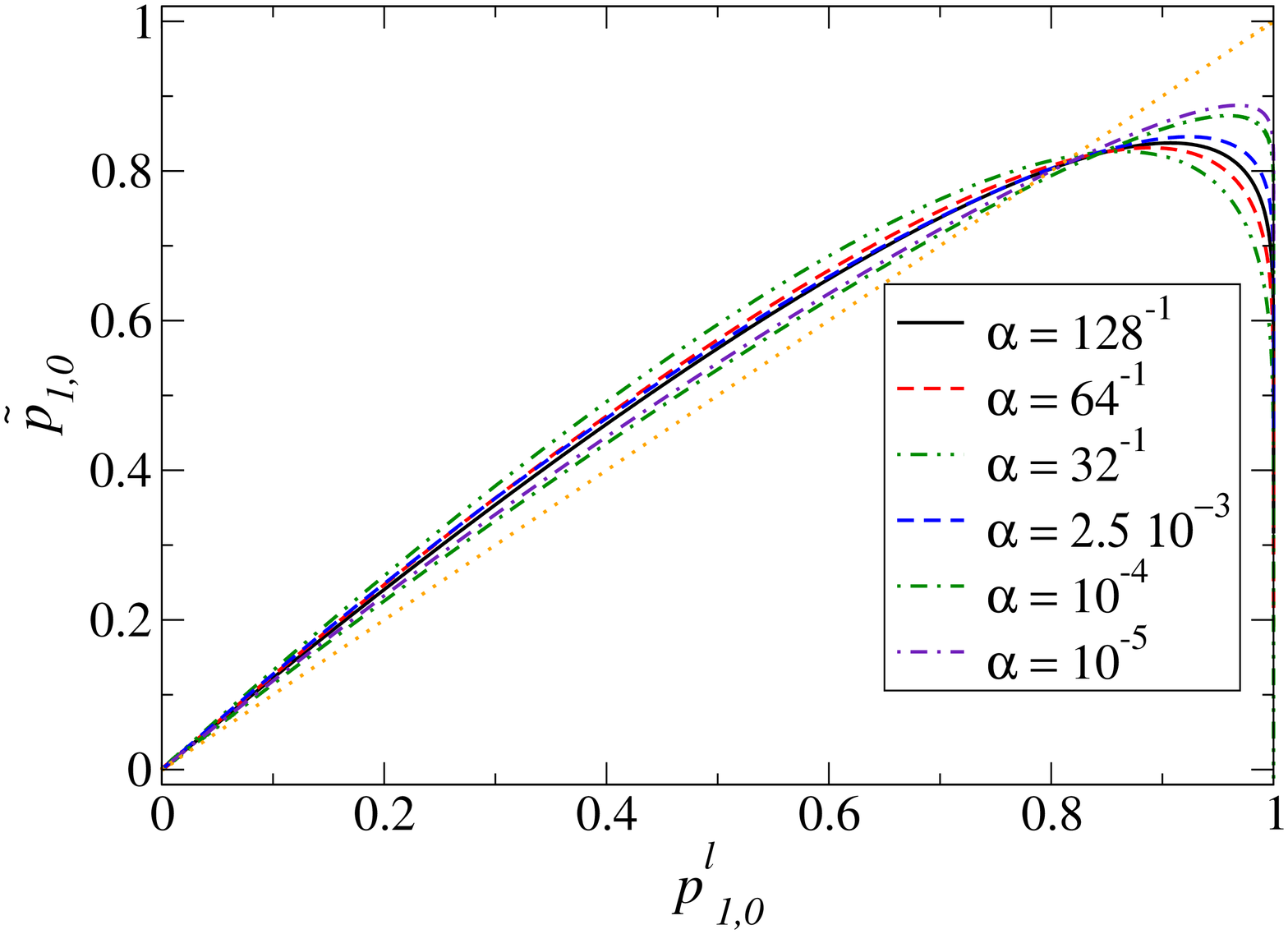}
\end{center}
  \caption{(Colour online) Limiting behaviour of the rescaled response for abrupt
quenches with $\hat{p}_{1,0}\equiv- {p}_{1,0}/\calf_0(\a)$ in the left
panel and $\tilde{p}_{1,0}\equiv- {p}_{1,0}/\calf_1(\a)$ in the right
panel. In both cases, the curves suggest a slow convergence towards
straight line where the rescaled response simply follows the source,
\eg $\hat{p}_{1,0}=p_{1,0}^l$ on the left, as $\a\to0$.
}\label{univcdim2b}
\end{figure}

\begin{figure}[t]
\begin{center}
\psfrag{la}{{$\ln\a$}}
\psfrag{lm}{{$\ln(-a_{2,4}^{\infty})$}}
\psfrag{m}{$-a_{2,4}^{\infty}$}
  \includegraphics[width=2.5in]{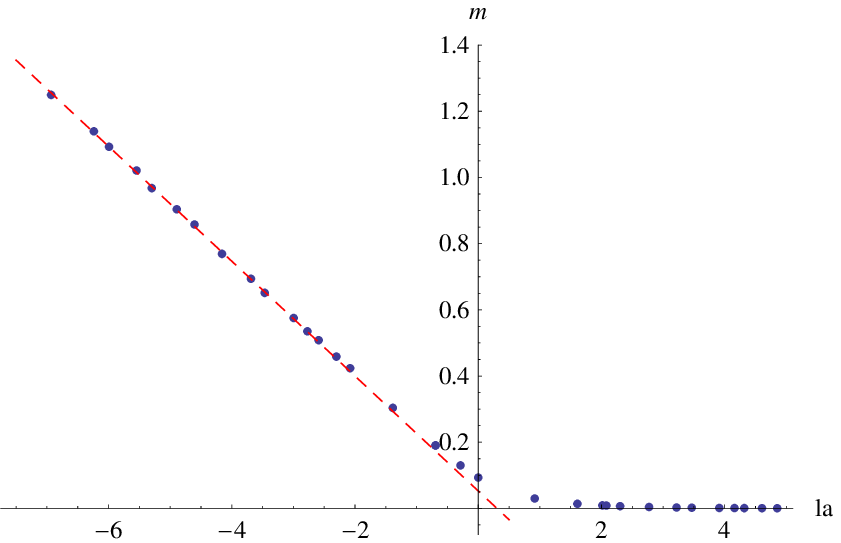}
  \includegraphics[width=2.5in]{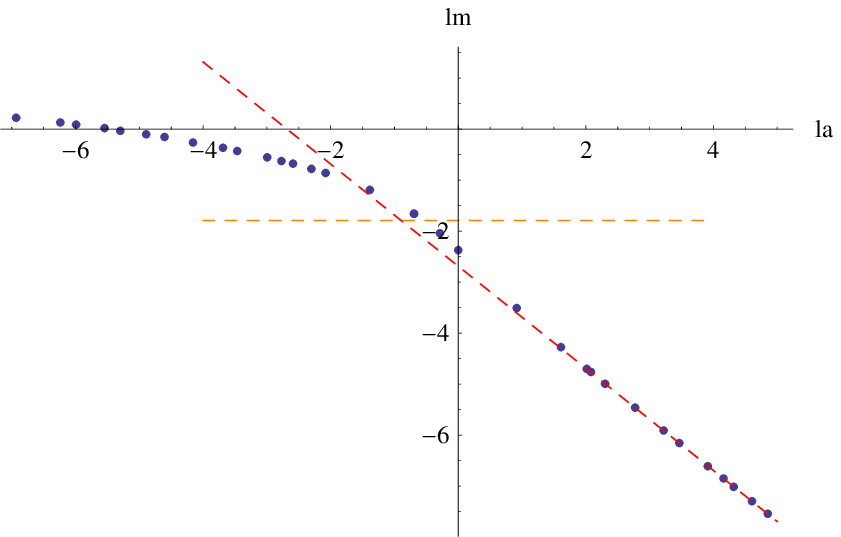}
\end{center}
  \caption{(Colour online) Coefficient $a_{2,4}^{\infty}$ as a function
of $\a$. The dashed red lines represent the linear fits to the data
(blue points) for fast (left panel) and slow (right panel) quenches.
Values of $(-a_{2,4}^\infty)$ above dashed orange line in the right panel
always result in quenches where the final temperature exceeds the
initial temperature. }\label{figure6}
\end{figure}

Our discussion of the results for quenching the bosonic mass operator
will be shorter, as it parallels the previous discussion of quenches of
the fermionic mass operator. However, we will see that the physical
characteristic of these types of quenches differ in many respects. For
the present quenches, the profile of the source $p_{1,0}^l$ is given by
eq.~\eqref{mq2}. Again, the characteristics of response in the
normalizable component $p_{1,0}$ depend on whether the transition is
`fast' with $\a<1$ or `slow' with $\a>1$. Our results for both cases
are presented in fig.~\ref{figure4}. The curves in the left and right
panels present $p_{1,0}$ as function of $\frac \t\a$ for fast and slow
quenches, respectively. We present fast quenches with
$\a=\lbrace1,0.5,0.1,0.05,0.01,0.005 \rbrace$ and slow quenches with
$\a=\lbrace1,2.5,5,10,50,100 \rbrace$.

First we consider the slow quenches, \ie with $\a>1$ in the right
panel. As found before with quenches of $\calo_3$ operator,
progressively larger values of $\a$ result in profiles that more
closely resemble the adiabatic response given here by
eq.~\eqref{adresponsedim2}. The latter profile `appears' in the plot as
a black dotted line, but for the most part, it is covered by the
profiles for $\a=50$ and 100 which track this adiabatic response.

Now we turn to the fast quenches in the left panel. The response shown
there is very far from the adiabatic curve, which corresponds to
eq.~\eqref{adresponsedim2}. Although the maximum amplitude of the
response grows as $\a$ becomes smaller and the quenches become faster,
comparing to fig.~\ref{figure2}, we see that the present increase is
not as nearly as dramatic as found for quenches of $\calo_3$ in the
previous section. Notice that neither the response $p_{1,0}$ nor the
time $\t$ is scaled for fast quenches in fig.~\ref{figure4}, in
contrast to the corresponding plot in fig.~\ref{figure2}. With this
slower grow, there is no distinctive profile that emerges here for
$p_{1,0}$ in this limit. Hence to determine if the response is scaling
with $\a$, we first simply consider the peak value of this normalizable
component, \ie
\begin{equation}
\calf_0(\a)
=\max_{\t\in(-\infty,+\infty)}|p_{1,0}|\,.
\eqlabel{univdim2}
\end{equation}
The left panel in fig.~\ref{univcdim2a} shows $\calf_0(\a_n)$ for the a
select set of values, $\a_n=\{ {1}/{32},$ ${1}/{64},$  ${1}/{128},$
$0.0025 ,$ $0.0001,$ $0.00001\}$. As the red dashed line in the figure
shows, these points are quite well fit with a simple linear expression
which takes the form
\begin{equation}
\calf_0\big|_{fit}\simeq -0.80 - 0.94\ \ln\a\,.
\eqlabel{univdim22}
\end{equation}
The rescaled response, $\hat{p}_{1,0}\equiv- {p}_{1,0}/\calf_0(\a)$, is
then shown in the left panel of fig.~\ref{univcdim2b} as a function of
${p}^l_{1,0}$. It seems that these curves are slowly converging to a
straight line such that in the limit $\a\to 0$, the profile of this
rescaled response is simply given by $\hat{p}_{1,0}=p_{1,0}^l$, which
is represented by a dotted orange line in the figure. That is, in the
limit of very fast quenches, the rescaled response simply tracks the
source. We must clarify this claim, which seems to suggest that profile
of $p_{1,0}$ should approach a simple step function as $\a\to0$.
Clearly, the latter is at odds with the profiles shown in
fig.~\ref{figure4}. However, note that in fig.~\ref{univcdim2b}, we are
plotting rescaled response as a function of ${p}^l_{1,0}$ and with the
profile in eq.~\reef{mq2}, this source varies across 90\% of its range
within $-3\a/2\,\le \tau\le\, 3\a/2$. Hence in this plot, we are really
focussing the behaviour of $p_{1,0}$ in a very narrow range around
$\t=0$ when $\a$ is small.\footnote{To see more clearly that as
$\a\to0$, $p_{1,0}$ is approaching this step function in time, one can
plot ${p}_{1,0}$ as a function of $\tau/\alpha$ -- rather than just
$tau$, as was done in fig.~\ref{figure4}. However, as is evident from
fig.~\ref{univcdim2b}, the response still falls off rapidly for
$\tau/\alpha>3/2$ for the values of $\alpha$ shown there.} Further,
this observation motivated us to define $\calf_1(\a)$ as the constant
which minimizes the following norm:
\begin{eqnarray}
\| p_{1,0}-\calf_1(\a)\ p_{1,0}^l  \|
&\equiv&\int_0^1 d(p_{1,0}^l)\ \left(p_{1,0}-\calf_1(\a)\ p_{1,0}^l\right)^2
\nonumber\\
&=&\int_{-\infty}^{\infty}d\tau\ (p_{1,0}^l)'\ \left(
p_{1,0}-\calf_1(\a)\ p_{1,0}^l\right)^2\,,
\labell{deff2}
\end{eqnarray}
in analogy with eq.~\reef{deff} for the dimension-3 operator. As
before, the factor $(p^l_{1,0})'=(2\a \cosh(\t/\a)^2)^{-1}$ in the last
expression weights most heavily the integration in a narrow range
around $\tau=0$ for small $\a$. The right panel in
fig.~\ref{univcdim2a} shows $\calf_1(\a_n)$ for the the same values
$\a_n$ used above. The blue dashed line in the figure indicates a
simple linear fit to these points, which takes the form
\begin{equation}
\calf_1\big|_{fit}\simeq -0.54 - 1.03\ \ln\a\,.
\eqlabel{univdim22z}
\end{equation}
In this case, the rescaled response, $\tilde{p}_{1,0}\equiv-
{p}_{1,0}/\calf_1(\a)$, is then shown in the right panel of
fig.~\ref{univcdim2b} and again the results suggest a slow convergence
towards  $\tilde{p}_{1,0}=p_{1,0}^l$ as $\a\to 0$. Hence, with both
definitions of the scaling factor in eqs.~\reef{univdim22} and
\reef{deff2}, the results suggest that the same scaling of the `prompt'
response with $p_{1,0}\simeq \log\a\,p_{1,0}^l$ for fast quenches, \ie
with small $\a$.

The reduced response of the present quenches with $\calo_2$ will also
become evident below when we consider the properties of the final
equilibrium state. Finally, we re-iterate that neither $p_{1,0}$ nor
$\t$ is scaled in fig.~\ref{figure4} for the fast quenches. Hence we
observe that the relaxation of the system after a fast quench is almost
'universal', \ie it is essentially independent of $\a$.

We now turn to eqs.~\eqref{ftdim2} and \reef{sfsi2}, which give the
characteristics of the final equilibrium state relative to those of the
initial equilibrium, \ie $T_f/T_i$, $\cale_f/\cale_i$,
$\calp_f/\calp_i$ and $S_f/S_i$. All of these ratios depend on
coefficient $a_{2,4}^\infty$ given in eq.~\reef{aindef2}. In
particular, this coefficient was defined to give a direct measure of
the entropy production during the quench, as shown in eq.~\reef{sfsi2}.
This coefficient is shown for a wide range of $\a$ in
fig.~\ref{figure6}. One simple observation is that in our numerical
simulations, we always found $a_{2,4}^\infty\le0$, which must
intuitively be the case from eq.~\reef{sfsi2} in order that $S_f\ge
S_i$. We might also observe that $|a_{2,4}^\infty|$ is becoming very
small for large $\a$, which agrees with the discussion in section
\ref{result} where we argued that $a_{2,4}^\infty\to0$ for
$\a\to\infty$. Let us add here that according to eq.~\eqref{ftdim2}, we
may describe the quenches going from the CFT to the mass-deformed gauge
theory as always `heating' the system. That is, we always have
$T_f>T_i$, even in the adiabatic system. On the other hand, according
to eq.~\reef{ftdim2cft} for `reverse' quenches going from the
mass-deformed gauge theory to the CFT, the system `cools' if the
transition is sufficiently slow, \ie $a_{2,4}^\infty$ is sufficiently
small. However, as the quenches become faster, $a_{2,4}^\infty$ becomes
large and negative resulting in $T_f>T_i$ again. Specifically, from
\eqref{ftdim2cft}, $T_f>T_i$ requires\footnote{Here we are using
eq.~\reef{a24t2} which states that $\ta_{2,4}^\infty=a_{2,4}^\infty$
for the reverse quenches described by eqs.~\reef{reversedim21} and
\reef{reversedim22}.}
\begin{equation}
-a_{2,4}^\infty> \frac16\,.
\eqlabell{larget2}
\end{equation}
As shown in the right panel of fig.~\ref{figure6}, this threshold
occurs  close to $\a=1$:
\begin{equation}
T_f>T_i\qquad \Longleftrightarrow\qquad \a\lesssim 0.58\,.
\eqlabell{tfgtidim2}
\end{equation}

In certain respects, the results for $a_{2,4}^\infty$ in
fig.~\ref{figure6} for quenches induced by the bosonic mass term are
similar to those in fig.~\ref{figure2t} for quenches induced by the
fermionic mass term. In particular, both figures show a simple
behaviour for both $\a\gg1$ and $\a\ll1$ and a clear transition between
these two around $\a\sim1$. However, there is one striking difference
in that the fast quenches are shown in the left panel of
fig.~\ref{figure6} which shows $-\ha_{2,4}^\infty$ (rather than
$\ln(-\ha_{2,4}^\infty)$) as a function of $\ln \a$. The simple fits
for the fast and slow quenches then take the from
\begin{equation}
\begin{split}
&{\rm slow}:\qquad \ln(-a_{2,4}^\infty)\big|_{fit}\simeq -2.688-1.00\ \ln\a
\,,\qquad \a\gg1 \,,\\
&{\rm fast}:\qquad -a_{2,4}^\infty\big|_{fit}\simeq 0.05-0.17\ \ln\a
\,,\qquad \a\ll1 \,.\\
\end{split}
\eqlabell{sffitsdim2}
\end{equation}

Using eq.~\eqref{ftdim2}, we can translate the asymptotic behaviour in
eq.~\eqref{sffitsdim2} into
\begin{equation}
\frac{\Delta \cale}{\cale_i}\equiv \frac{\cale_f-\cale_f^{adiabatic}}{\cale_i}
\propto
\begin{cases}
&\quad\frac{1}{\a}\ \frac{(m_b^0)^4}{T_i^4}\qquad\ \ {\rm for}\ \a\gg 1\,,\\
&\ln(1/\a)\ \frac{(m_b^0)^4}{T_i^4}\quad{\rm for}\ \a\ll 1\,,
\end{cases}
\eqlabell{redtfitsdim2}
\end{equation}
where
\begin{equation}
\cale_f^{adiabatic}=\ \cale_i\ \left(1-\frac 19 \ln\frac{\pi T_i}{\Lambda_1}
\frac{(m_b^0)^4}{\pi^4 T_i^4}
+\calo\left(\frac{(m_b^0)^8}{T_i^8}\right)\right)\,.
\eqlabell{tfadiabdim2}
\end{equation}
Of course, we can also determine the asymptotic behaviour for the
relative change in the temperature and the pressure using
eq.~\eqref{ftdim2} and the entropy using eq.~\reef{sfsi2}. In
particular then, we note that a discontinuous quench with $\a=0$ would
again seem to produce physical divergences. However, we are again
seeing a weaker response here for the quenches induced by $\calo_2$.
Recall that the analogous divergence in eq.~\reef{guess2} was
proportional to $1/\a^2$ for the quenches with $\calo_3$.

Further, we note that the scaling of $a_{2,4}^\infty$ for $\a\to0$
shown in eq.~\reef{sffitsdim2} is slower than the scaling of the
amplitude of the response for the fast quenches found in our previous
discussion --- this was also found for the analogous results with
$\calo_3$. However, if we assume the `prompt' response takes the form
$p_{1,0}\simeq \log\a\,p_{1,0}^l$ as discussed above, then the
arguments given for the $\calo_3$ case around eq.~\reef{hack1} would in
fact indicate that both $a_{2,4}^\infty$ and $p_{1,0}$ scale in the
same way. Hence the reasons for the discrepancy in the scaling of these
two quantities must be more subtle here than before.

\begin{figure}[t]
\begin{center}
\psfrag{tl}{{$\frac \t\a$}}
\psfrag{a}{{$\frac {\t_{ex}}{\a}$}}
\psfrag{b}{{$\frac {\t_{eq}}{\a}$}}
\psfrag{dp}{{$\dd_{neq}$}}
   \includegraphics[width=3.5in]{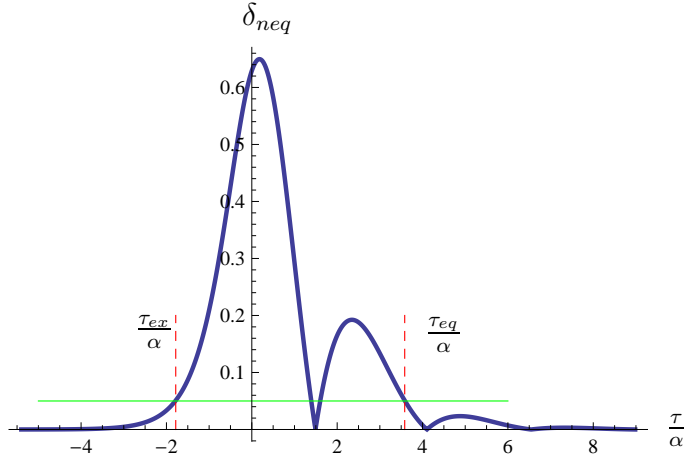}
\end{center}
  \caption{(Colour online)
Extraction of the excitation/equilibration rates for $\a=1$ quench. The
horizontal green line is the threshold for excitation/equilibration
which we define to be $5\%$ away from local equilibrium as determined
by $\delta_{neq}$, as given defined in eq.~\eqref{defddn2}. The dashed
red lines indicate the earliest and latest times of crossing this
threshold, which we denote as $\t_{ex}$ (for excitation time) and
$\t_{eq}$ (for equilibration time), respectively.}\label{relaxdim2p}
\end{figure}
%
%
\begin{figure}[t]
\begin{center}
\psfrag{la}{{$\ln\a$}}
\psfrag{ete}{{$e^{\frac{|\t_{ex}|}{\a}}$}}
\psfrag{lt}{{$\ln \frac{\t_{eq}}{\a}$}}
  \includegraphics[width=2.5in]{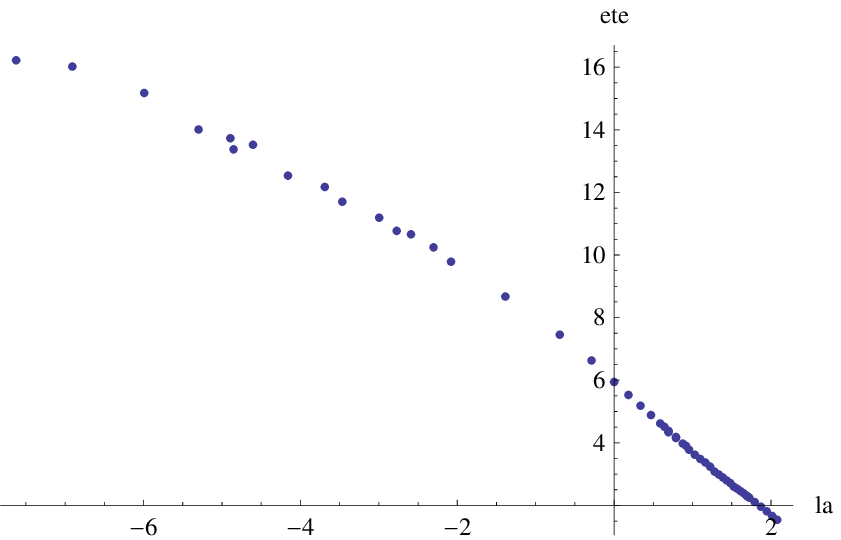}
  \includegraphics[width=2.5in]{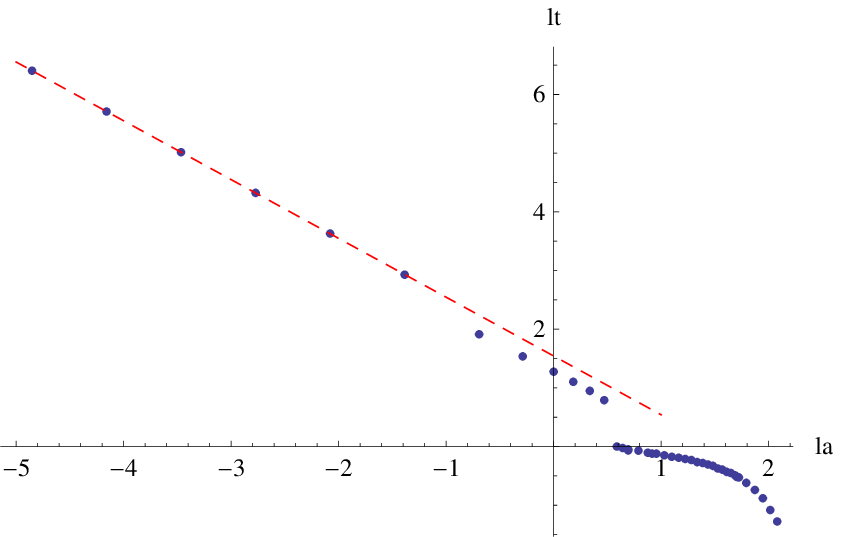}
\end{center}
  \caption{(Colour online)
Excitation time $\t_{ex}$ (left panel) and
equilibration time $\t_{eq}$ (right panel) as a function of $\a$. The
threshold is fixed to be $\e=0.05$ --- see eqs.~\eqref{excitationt} and
\eqref{equilibrationt}. In the left panel, it seems like $|\t_{ex}|/\a\sim \ln(-\ln\a)$
in the limit $\ln\a\to -\infty$. In the right panel, the behaviour
of $\ln(\t_{eq}/\a)$ is linear in $\ln\a$ for small $\a$, as shown with the
red dashed line.
}\label{exeqdim2}
\end{figure}
%
%
We now turn to the excitation and the equilibration time scales in the
evolution of  the normalizable component $p_{1,0}(\t)$. By analogy to
eq.~\eqref{defddn}, we define
\begin{equation}
\dd_{neq}(\t)\equiv \left|
\frac{p_{1,0}(\t)-\left[p_{1,0}(\t)
\right]_{adiabatic}}{p_{1,0}^{equilibrium}} \right|\,,
\eqlabell{defddn2}
\end{equation}
where $\left[p_{1,0}(\t) \right]_{adiabatic}$ is given in
eq.~\reef{adresponsedim2} and
\begin{equation}
p_{1,0}^{equilibrium}=\lim_{\t\to+\infty}p_{1,0}(\t)=-\ln2\,.
\eqlabell{defp12eqiui2}
\end{equation}
A typical behaviour of $\dd_{neq}$ is presented in
fig.~\ref{relaxdim2p}. We define the excitation and  equilibration
times $\t_{ex}$ and  $\t_{eq}$ precisely as before in
eqs.~\eqref{excitationt} and \eqref{equilibrationt}. We indicate these
times in fig.~\ref{relaxdim2p}, as well as the threshold $\e=0.05$. We
extracted  $\t_{ex}$ and $\t_{eq}$ for a wide range of $\a$ and our
results are shown in fig.~\ref{exeqdim2}. As in the previous
discussion, we only expect these time scales to be a useful diagnostic
of fast quenches with $\a<1$. For example, while we did not pinpoint
the precise value of $\a$ such that $\dd_{neq}<\e$ for all
$\t\in(-\infty,+\infty)$, we did establish that such a value of $\a$
occurs prior to $\a=16$.

\begin{figure}[t]
\begin{center}
  \includegraphics[width=3.in,angle=-90]{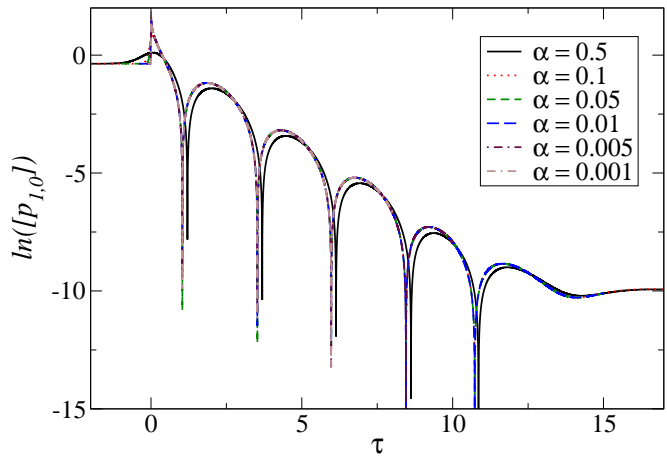}
\end{center}
  \caption{(Colour online) Behaviour of the response coefficients versus time for representative
fast quenches. As is evident in the picture, the same quasinormal mode
governs the dynamics very quickly after the quench.
}\label{figureQNM_2}
\end{figure}

The left panel of fig.~\ref{exeqdim2} suggests that,  in the limit
$\ln\a\to -\infty $,
\begin{equation}
\frac{\t_{ex}}{\a}\sim \ln(-\ln\a)\,,
\eqlabel{oldsatur}
\end{equation}
however, we must say that our results in this regime were not accurate
enough to definitively fix this scaling. According to eq.~\eqref{mq2},
the scaling \eqref{oldsatur} would translate to
\begin{equation}
t_{ex}\sim \calt\ \ln\ln(1/T_i\calt) \qquad{\rm for}\ \ \a\ll 1\,.
\eqlabell{texT2}
\end{equation}
In the right panel of fig.~\ref{exeqdim2}, we see the scaling of the
equilibration time for small $\a$ is
\begin{equation}
\left.\ln\frac{\t_{eq}}{\a}\right|_{fit}\simeq 1.5- 1.0\ \ln\a\,.
\eqlabell{eqfit2}
\end{equation}
Hence for very fast quenches, $\t_{eq}/\a\propto1/\a$ or in terms of
the original boundary time, we have
\begin{equation}
t_{eq}\equiv \calt\ \frac{\t_{eq}}{\a}\ \propto\ \frac{1}{T_i}\,.
\eqlabell{teqT2}
\end{equation}
Much like for quenches of $\calo_3$, here we find that irrespective of
how quickly the system was driven away from the initial equilibrium,
the return to the final equilibrium is determined universally by the
thermal time scale $1/T_i$. As with the previous quenches, here again
the response is essentially controlled by a single quasinormal mode of
the scalar field at a short time after the quench. This behaviour is
well illustrated in fig.~\ref{figureQNM_2}, which displays the
evolution of $[p_{1,0}] \equiv p_{1,0}
-p_{1,0}^{equilibrium}=p_{1,0}+\ln(2)$. A rough fit to the oscillation
periods and slopes indicates that the corresponding quasinormal mode
has $\omega/(2\pi T_i) \simeq (0.64 + i\, 0.4)$, which is consistent
with the expected value of $(0.644 + i\, 0.411)$ found in
\cite{StarinetsNunez}. As in the previous section, the first overtone
can be extracted for about one oscillation period and once again, a
rough fit yields real and imaginary parts of the quasinormal frequency
which agree to within $25\%$ of the expected ones.

\section{Conclusions} \label{kronk}
In this paper, we initiated a program to study quantum quenches in
strongly coupled quantum field theories using holography. We used
gauge-gravity correspondence \cite{m1} to translate the problem of
quenching the coupling $\l_\Delta$ of a relevant operator
$\calo_\Delta$ in planar $\caln=4$ SYM, as given in eq.~\reef{con1}, to
a classical problem in Einstein gravity coupled to a massive scalar
field in asymptotically $AdS_5$ spacetime. Here, we focused on two
cases where $\Delta=2$ and 3 but our analysis is readily extended to
considering general values of $\Delta$ \cite{wip}. Our discussion also
only considered `thermal quenches' where the system begins in a thermal
state of the gauge theory plasma and was limited to a high temperature
limit. That is, during the quench, $\l_{\Delta}$ is always small
compared to the temperature of the initial state $T_i$, \ie
\begin{equation}
|\l_\Delta|\ll T_i^{4-\Delta}\,.
\eqlabell{pertl}
\end{equation}
The latter means that our calculations were {\it perturbative} in the
coupling $\l_\Delta$. However, there was no restriction on the time
scale $\calt$ governing the rate of change of the coupling in the
transition. In particular, we considered profiles of the form given in
eq.~\reef{ratel} and our results were described in terms of the
dimensionless parameter: $\a=\pi T_i\,\calt$. In our analysis, we
considered arbitrary values of $\a$ and in particular, we examined the
limits of adiabatic transitions with $\a\to\infty$ and abrupt quenches
with $\a\to0$. As indicated in eq.~\reef{ratel}, we discussed both the
quenches from the conformal SYM theory to the mass-deformed gauge
theory (with the plus sign for which $\l_\Delta(t\to-\infty)\to0$) and
the {\it reverse} quenches from the mass-deformed gauge theory to the
conformal gauge theory (with the minus sign). In all of these quenches,
we computed, to leading order in the coupling, the response of some
basic gauge invariant observables, namely, the one-point correlators of
the stress tensor $\la T_{ij}\ra$ and the mass operator $\la
\calo_{\Delta}\ra$. A detailed discussion of the results is given in
section \ref{res}.

One of the most interesting results coming from our analysis is that
the `infinitely fast' quenches seem to be ill-defined because of
divergences appearing in the limit $\a\to0$. These divergences appear
most dramatic for quenches of $\calo_3$, the fermionic mass operator,
in expressions such as eqs.~\reef{tfits} or \reef{guess2} which
indicate that the energy of the final state diverges as $1/\a^2$ as
$\a\to0$. This result is of particular note since it is precisely such
`infinitely fast' quenches are studied in the seminal work on the topic
\cite{cc1,cc2}. In their description of these quenches, the evolution
from $t=-\infty$ to $0^-$ is essentially regarded as preparing the
system in the ground state of the initial Hamiltonian and then this
ground state is used as the initial condition at $t=0^+$ as the system
evolves forward with the new `quenched' Hamiltonian. In our more
physical approach to describing fast quenches, the transition from the
initial to the final Hamiltonian is accompanied by an enormous response
in the infinitesimal interval around $t=0$, as can be seen in the
scaling of $p_{2,0}$ for $\calo_3$ and of $p_{1,0}$ for $\calo_2$ as
$\a\to0$. Hence one might worry that our results call into question the
approach of these early studies. Of course, an important difference
between these studies and our work is that the former considered free
or weakly coupled field theories whereas we are studying a strongly
coupled field theory.

However, we must insert a word of caution with respect to our
discussion above. The framework considered here treats the bulk scalar
field perturbatively. In particular, we scaled the amplitude of the
field by a factor $\lambda$, which was then gave a perturbative
expansion of the results for the boundary theory in terms of
$m^0_f/T_i$ or $(m^0_b)^2/T_i^2$, using eq.~\reef{identity} or
\reef{identitydim2}, respectively. However, this approach only properly
accounts for the backreaction of the bulk scalar on the background
spacetime as long as the quench is not too fast. Looking at the
gravitational equations, we find terms containing time-derivatives of
the scalar, \ie terms proportional to $(\partial_v\phi)^2$ in
eq.~\reef{eoms4}, as well as to $\partial_v\phi\partial_r\phi$ in both
eqs.~\reef{eoms2} and \reef{eoms4}. Hence heuristically we should say
that in order to ensure the full backreaction of the bulk scalar
remains small, we must in fact fix $\lambda/\alpha\ll 1$. In other
words, our perturbative calculations only yield reliable results, \eg
in eqs.~\reef{tfits} or \reef{guess2}, as long as
$\frac1{\a^2}\left(\frac{m^0_f}{T_i} \right)^2$ remains small. This
issue is also evident in the observation that with a very high final
temperature (in particular, much larger than the mass scale of the
operator in the quench), $T_f/T_i$, $\cale_f/\cale_i$ and $S_f/S_i$ can
not all diverge in precisely the same way, as would naively be the case
with our perturbative treatment. Rather, the divergences (or scalings)
must arrange themselves such that $\cale_f/\cale_i\propto (T_f/T_i)^4$
and $S_f/S_i\propto (T_f/T_i)^3$. Hence it is clear that our present
analysis does not provide the final story and  to properly understand
the behaviour of the `infinitely fast' quenches, we must turn to a full
nonlinear analysis of the bulk gravitational equations of motion
(\ref{eoms3}--\ref{eoms5}) \cite{wip}. However, to close this
discussion, let us note that the nonlinear analysis of \cite{cy} do
yield divergent results in the limit $\a\to0$.

Further, we wish to stress that the above discussion does not
invalidate the various universal scaling properties and profiles that
were found in section \ref{res} for $\a\to0$. Rather these should be
interpreted in terms of choosing $\lambda\ll \a_{min}$ for some small
value $\a_{min}$. Then for such a sufficiently small amplitude, the
universal behaviours will still correctly describe the physics of
quenches in the regime $\a_{min}<\a\ll1$.

Allowing the couplings of the boundary gauge theory to vary in space
and time introduces a variety of new ultraviolet divergences into the
quantum field theory. The power of holography is that it provides a
well-defined framework for the renormalization of the boundary theory,
even in the presence of these space-time varying couplings. As
discussed in section \ref{holonorm8}, the bulk gravity theory provides
a natural geometric construction which regulates the new divergences
and allows us to identify the appropriate counterterms to renormalize
them in a covariant way. Recall that roughly, the expectation value of
the quenching operator $\calo_\Delta$ and its coupling $\l_\Delta$ are
encoded as normalizable and non-normalizable coefficients
(correspondingly) of the dual bulk scalar,\footnote{We focus here on
relevant operators (with $\Delta<4$). We can also consider operators
with $1\le\Delta<2$ using the `alternate quantization' of the
holographic theory \cite{igor9}. For these, the powers of $\r$
associated with the two coefficients are interchanged in
eq.~\reef{boundarydelta}.} $\phi_\Delta=\phi_{\Delta}(t,\r)$,
\begin{equation}
\begin{split}
\phi_\Delta&\sim {\l_\Delta}\ \r^{4-\Delta}+  \calo_\Delta\ \r^{\Delta}
\qquad{\rm for}\ 2<\Delta< 4\,,\\
\phi_2&\sim {\l_2}\ \r^{2}\ln\r+  \calo_2\ \r^{2}
\qquad{\rm for}\ \Delta= 2\,,
\end{split}
\eqlabell{boundarydelta}
\end{equation}
where $\r$ is the usual Fefferman-Graham coordinate \reef{mfg} in the
asymptotically AdS spacetime. In particular, examining
eq.~\reef{scount} for the fermionic mass operator, we see that there
are additional counterterms proportional to $(\del\l_3)^2$ and $R\,
\l_3^2$, which play a role in backgrounds where $\l_3$ varies in time
(and/or space). We note that this is completely analogous to placing
the boundary theory in a curved background spacetime, which creates the
necessity of including additional counterterms which are functionals of
the background curvatures \cite{surface}. We also note that some of the
new UV divergences, \eg those associated with the two counterterms
above, are logarithmic in the cut-off scale and as a result, new scheme
dependent ambiguities are inevitable in the renormalization procedure
--- we return to this issue below.

Through Einstein's equations \reef{ee}, the bulk scalar backreacts on
the geometry at least at the quadratic order and hence the renormalized
boundary stress-energy tensor must depend on the (non-)normalizable
coefficients of $\phi_\Delta$ at least quadratically. Simple
dimensional arguments imply that for a constant coupling the boundary
stress-energy tensor (having dimension four) can always depend on the
bilinear $\l_\Delta \calo_\Delta$ for any $\Delta$. This combination is
seen to explicitly appear in the energy tensor and the pressure given
in eqs.~\reef{vev1} and \reef{vev12} for for $\Delta=3$ and 2,
respectively. Continuing these dimensional arguments for constant
coupling, the stress tensor can also depend on
\begin{equation}
\begin{split}
\lambda_\Delta^{\ n}&\qquad{\rm for}\ \Delta=4-\frac4n\ {\rm with}
\ n=2,3,4,\cdots\,,\\
\calo_\Delta^{\ m}&\qquad{\rm for}\ \Delta=\frac4m\ {\rm with}
\ m=2,3,4\,.
\end{split}
\eqlabell{wind9}
\end{equation}
For $\Delta=2$ and 3, the dependence as in \eqref{wind9} was indeed
established in \cite{n22,n23} and can again be explicitly seen in
eqs.~\reef{vev1} and \reef{vev12}.

Once $\l_\Delta$ depends on time, additional combinations involving
time-derivatives of $\l_\Delta$ or $\calo_\Delta$ can appear in
boundary stress tensor, as long as the dependence has the correct
mass-dimension four (and is at least quadratic in these coefficients).
In particular, the stress tensor may depend on
\begin{equation}
\begin{split}
\lambda_\Delta^{\ n-2}\,\del_{t}\l_\Delta\,\del_{t}\l_\Delta
\,,\ \lambda_\Delta^{\ n-1}\,\del^2_{tt}\l_\Delta&\qquad{\rm for}\
\Delta=4-\frac2n\ {\rm with}
\ n=2,3,4,\cdots\,,\\
\del_{t}\calo_\Delta\,\del_{t}\calo_\Delta\,,\ \calo_\Delta\,\del^2_{tt}\calo_\Delta
&\qquad{\rm for}\ \Delta=1\,.
\end{split}
\eqlabell{wind8}
\end{equation}
Here we have restricted our attention on relevant operators (with
$\Delta<4$). We also only consider the possibility of an even number of
time-derivatives because otherwise the corresponding operator could not
be extended to a covariant expression. Again, these additional terms
are explicitly seen in eq.~\reef{vev1} for the $\Delta=3$ operator.
Further, as indicated by eq.~\reef{wind8}, no such contributions appear
in eq.~\reef{vev12} for $\Delta=2$.

Returning to the possibility that rapid quenches may produce singular
behaviour, the additional contributions to the boundary stress-energy
tensor involving time derivatives of the couplings $\l_\Delta$, as in
eq.~\eqref{wind8}, should be worrisome. In particular, these terms
suggest that instantaneous changes in the corresponding couplings will
produce a singular response in the stress tensor. Alternatively, if
such a singularity is to be avoided, the response of the normalizable
coefficient $\calo_\Delta$ should be `finely correlated' to cancel the
potential divergence, \eg through the contributions of the $\l_\Delta
\calo_\Delta$ terms. While in section \ref{res}, we saw that for both
operators the specific profile of the leading response was closely
related to that for the coupling, \eg see eq.~\reef{deff}, we must
re-iterate that our perturbative calculations can not fully address the
question of such a cancellation.

Of course, rapid quenches also displayed interesting (potentially
divergent) scalings in the physical properties of the final state, \eg
the energy density as shown in eqs.~\reef{tfits}
and\reef{redtfitsdim2}.\footnote{As noted in section \ref{res}, the
same scalings appear for the temperature, pressure and entropy density
of the final state.} These scalings do not depend on the derivatives of
the couplings as we set $\del_t\l_\Delta(t\to\pm\infty)\to0$ to ensure
that the system begins at equilibrium and eventually settles into a new
equilibrium state. Rather these scalings can be associated with the
behaviour of the expectation value of $\calo_\Delta$ itself, as
follows: Recall the basic observation that when the boundary theory has
a time-varying coupling $\l_\Delta(t)$, the stress-energy tensor is not
longer conserved, \ie see eqs.~\reef{warddiffeo} and
\reef{warddiffeo1}. Instead, the diffeomorphism Ward identity states
that for our quenches:
\begin{equation}
\partial_t\,\cale = \langle\, \calo_\Delta\rangle\ \del_t \l_\Delta\,.
\eqlabell{ldiffeo}
\end{equation}
That is, the time-variation of the coupling is performing work on a
system. Of course, eq.~\reef{ldiffeo} can be written in an integral
form as
 \beq
\cale_f=\cale_i+\int\ \langle\, \calo_\Delta\rangle\ d\l_\Delta\,.
 \labell{new7}
 \eeq
Since we only considered quenches where the change in the coupling was
finite, the (divergent) scaling in this expression for fast quenches
must come from $\langle\, \calo_\Delta\rangle$.

From the above perspective, the integral appearing, \eg
eq.~\reef{aindef} defining $a_{2,4}^\infty$ can be seen to describe the
total work done on the system. In fact, in the subsequent expression
for $\cale_f/\cale_i$ appearing in eq.~\reef{ftdim3}, the extra
constant terms cancel and so we recover precisely eq.~\reef{new7}, \ie
an integral form of eq.~\reef{ldiffeo}. With respect to the comments
above, recall that the leading scaling in the response was slightly
stronger for fast quenches than that found for $a_{2,4}^\infty$. That
is, the integral in eq.~\reef{new7} tempers the scaling of the response
to fast quenches, as was discussed in detail in section \ref{res}. We
have made some preliminary steps towards extending our present work to
quenches with a generic relevant (or marginal) operator with dimension
$\Delta$ \cite{wip}. This work suggests that our previous results
generalize in a simple fashion with
\begin{equation}
a_{2,4}^\infty\propto
\begin{cases}
&\a^{-(2\Delta-4)}\,,\qquad 2<\Delta\le4\,,\\
&\ln(1/\a)\,,\qquad \Delta=2\,,
\end{cases}
\eqlabell{divergentguess}
\end{equation}
for fast quenches with $\a\ll1$. Here we might note that the scaling
above with $\Delta=4$ matches that found in \cite{cy}.

It is well-known that the renormalization procedure in QFT is
scheme-dependent. In the context of gauge/gravity duality, such
scheme-dependence manifests itself with the appearance of finite
counterterms in the holographic renormalization. In this paper, we
carefully enumerated all of the finite counterterms for the situation
where the couplings for $\calo_2$ and $\calo_3$ vary in time (or space)
for the mass-deformed version of $\caln=4$ SYM theory. It is worth
noting that the scheme-dependence in holographic renormalization does
not introduce any ambiguity in the diffeomorphism Ward identity
\eqref{ldiffeo} --- see also eqs.~\reef{warddiffeo} and
\reef{warddiffeo1}. That is, in general, both the stress tensor and the
expectation value $\langle\calo_\Delta\rangle$ are scheme-dependent but
these ambiguities cancel in eq.~\reef{ldiffeo} and this Ward identity
is still true as written in any scheme. This might be contrasted with
the conformal Ward identity, where the scheme-dependent terms can make
an explicit appearance, as shown in eqs.~\reef{wardconf} and
\reef{wardconf1}.

As our analysis showed, the number of scheme-dependent parameters
introduced in renormalization of a general deformation \eqref{con1}
depends on the choice of $\Delta$, the dimension of the operator.
Specifically, we found three ambiguity coefficients in
eq.~\eqref{ambiguities} for $\Delta=3$ and two such coefficients in
eq.~\eqref{ambiguities2} for $\Delta=2$.\footnote{Note that for the
case $\Delta=3$, this means there are two extra coefficients compared
to renormalization of the theory with static couplings, while one extra
coefficient appears for $\Delta=2$.} We would like to emphasize that
these ambiguity coefficients are actually necessary to properly
interpret the bulk gravitational data in the language of the boundary
gauge theory. As a simple example, consider the case $\Delta=2$ with a
constant coupling. As illustrated in eq.~\eqref{equil2}, already to
leading order in $\l_2$ in our high temperature expansion, the energy
density and pressure of $\caln=4$ SYM plasma are modified by a term of
the form $m_b^4\ \ln T$ --- see \cite{n23} for more discussion.
Further, a similar term appears in the expectation value
$\langle\calo_2\rangle$ as shown in eq.~\reef{expO2}. To make sense of
the $\ln T$, one has to introduce an arbitrary scale in the theory, as
shown in eq.~\reef{equil2}. The arbitrariness associated with the scale
$\Lambda$ is encoded in the arbitrariness of the scheme-dependent
coefficient $\dd_1$ in eq.~\eqref{ambiguities2}. We note that the
analogous $\dd_1$ term in eq.~\reef{ambiguities} would produce a
similar effect for $\Delta=3$, \eg the energy density would be modified
by a term proportional to $m_f^4\ \ln T$, but clearly such
contributions would only occur at quartic order in the high temperature
expansion and so were not studied here.

Of course, the above example illustrates that the time variation of the
couplings is not intrinsic to scheme-dependence appearing various
natural `observables'. However, once we consider time-dependent
couplings, one can expect new contributions to the energy density or
pressure proportional to $\ln T$ with coefficients involving
time-derivatives of the coupling $\l_\Delta$. In particular, the latter
must form dimension four operators and so can be read off from the
first line of eq.~\reef{wind8}. Hence there are precisely two such
contributions for $\Delta=3$:
\begin{equation}
(\del_t\l_3)^2\, \ln T\,,\qquad \l_3\ \del^2_{tt}\l_3\ \ln T\,.
\eqlabell{dim3contri}
\end{equation}
As in static case, the proper interpretation of these $\ln T$ terms
requires introduction of two new independent scales, $\Lambda_2$ and
$\Lambda_3$, as appear in eqs.~\reef{vev1qft1} and \reef{vev1qft1cft}.
As shown in eq.~\reef{dddim31}, the ambiguity in the choice of these
scales is then precisely accounted for with the two new
scheme-dependent coefficients, $\dd_2$ and $\dd_3$, appearing in the
finite counterterms in eq.~\reef{ambiguities}.

Of course, our description of quenches with $\calo_2$ contains the same
$\ln \frac{T}{\Lambda}$ ambiguity which appeared in the equilibrium
thermodynamics. However, these quenches also contain a novel power law
ambiguity arising from the finite counterterm proportional to
$R^\gamma\phi$ in eq.~\reef{scountdim2f}. This counterterm contributes
a term proportional to the dimension four operator $(p^l_{1,0})''\sim
\partial_t^2\lambda_2$ in the pressure, as shown in eqs.~\reef{vev12qft1}
and \reef{vev12qft1cft}. However, this term carries a pre-factor
$\Lambda_2^2\,(m_b^0)^2/T_i^4$ here. This makes this contribution novel
in two respects: First of all, the renormalization scale $\Lambda_2$
appears with a power, rather than in a logarithm. Note that we are
therefore free to set this scale to zero, which may be the most natural
choice in this case. Second, this term carries a factor of
$(m_b^0)^2/T_i^2$ and so appears at first order in the high temperature
expansion, in contrast to all of the other perturbative corrections,
\eg in eq.~\reef{vev12qft1} which are second order in this expansion.
Hence (if $\Lambda_2\ne0$), this term formally dominates the behaviour
of the pressure while the mass coupling $\lambda_2$ is varying. Here we
might note that all of the new ambiguities discussed here only play a
role while the couplings are varying. Hence, for example, they are not
important after the coupling has achieved its final value but the
system is still relaxing to its final equilibrium state. Further, they
do not appear in expressions relating the properties of the final and
initial equilibrium states, \eg in eq.~\reef{ftdim3}.

In standard QFT, the scheme-dependence associated with choosing
renormalization scales can typically be eliminated with a judicious
choice of observables or by choosing a physical reference point. For
example, calculating the overall vacuum energy density in a QFT
typically yields a result depending on some cut-off scale and so which
is scheme-dependent. However, any excited stationary state has a
definite energy as measured with respect to this vacuum energy. A
similar strategy might be used in considering the thermal ensemble of
the mass-deformed gauge theory with $\calo_2$. That is, we could choose
a particular reference temperature $T_0\,(\ne0)$ and compare, \eg the
energy density of a given thermal ensemble to that of the reference
ensemble at $T_0$. This prescription would then yield a result for the
equilibrium energy density which is independent of the renormalization
scale $\Lambda$,
 \beq
\tilde \cale(T,T_0)\equiv \cale(T)-\cale(T_0) = \frac 38 \pi^2 N^2
\left(T^4-T_0^4+\frac{m_b^4}{9\pi^4}\ln\frac{T}{T_0}
 \right)\,.
 \labell{grunt4}
 \eeq
Unfortunately, this approach does not extend naturally to quenches
where the mass coupling varies in time. However, another strategy that
is available, at least in this case of the dimension two operator, is
to formulate a new scheme-independent observable by combining $\cale$
and $\langle\calo_2\rangle$. That is, although both of these quantities
are individually scheme-dependent, as shown in eq.~\reef{vev12qft1},
the following combination is independent of both $\Lambda_1$ and
$\Lambda_2$:
\begin{equation}
\tilde{\cale}(\t)\equiv \cale(\t)+\frac{(m_b^0)^2}{\sqrt{6}}\ p_{1,0}^l(\t)
\ \langle\,\calo_2(\t)\rangle\,.
\eqlabell{schem2}
\end{equation}
In particular, this scheme-independence extends to the situation where
the mass coupling varies in time. However, it is not clear if this
definition applies or can be extended to higher orders in the high
temperature expansion. Further, it is not possible to extend this
strategy to define scheme-independent pressure. Similarly, no such
definitions of scheme-independent observables seem possible in the case
of quenches with $\calo_3$.

Hence it seems like that situation here for the holographic QFT's with
time dependent couplings is similar in many respects to QFT in curved
spacetime \cite{birrell}. In a generic background without any special
symmetries, one simply finds that $\langle T_{ij}\rangle$ is ambiguous.
In this context, an `axiomatic' approach has been developed in which
general physically reasonable properties are imposed on the
renormalized stress-energy tensor, without reference to any particular
renormalization techniques \cite{birrell,wald,wald2}. Three of these
conditions would naturally carry over to a general discussion of
quenches, or more generally time-dependent couplings, in QFT. Namely,
the conditions related to causality, finite matrix elements for
orthogonal states and recovering standard results in Minkowski space or
here in the limit of time-independent couplings \cite{wald}. The first
condition, however, was covariant conservation of the stress tensor and
in the present context this would have to be changed to incorporate
work terms induced by the time-varying couplings, as found in
eqs.~\reef{warddiffeo} and \reef{warddiffeo1}. We re-iterate that the
latter were diffeomorphism Ward identities, which naturally appear
without any scheme-dependence in the present holographic
renormalization. We might add that the appearance of scheme-dependent
contributions in the trace anomaly is not resolved by the axiomatic
approach discussed above \cite{wald2} and this feature is also evident
in the present holographic analysis with eqs.~\eqref{wardconf} and
\eqref{wardconf1}.

\begin{figure}[t]
\begin{center}
  \includegraphics[width=3.in,angle=-90]{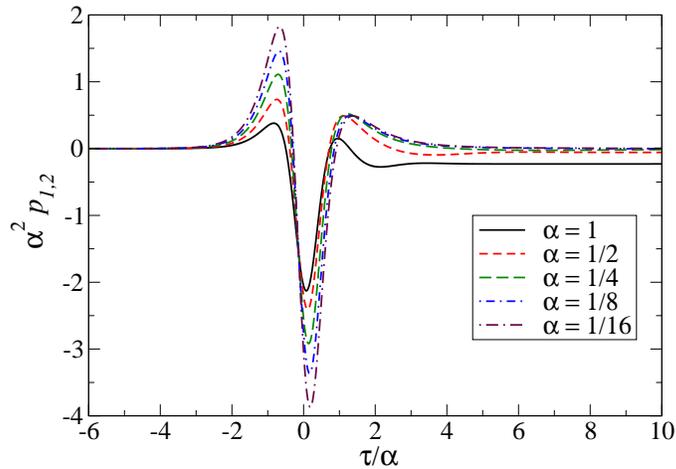}
\end{center}
  \caption{(Colour online)
The evolution of the $\a$-rescaled normalizable
component, $\a^2\ p_{1,2}$, as a function of $\t/\a$
during quenches with the profile given in
eq.~\eqref{p0power} with different values of $\a$.
}\label{ppower}
\end{figure}
Given the ambiguities in the observables, $\cale$, $\calp$ and
$\langle\calo_\Delta\rangle$, we demonstrate that some care must be
taken in defining the excitation and equilibration times. As an
example, we focus on the equilibration time for quenches with $\calo_3$
and consider a profile of the mass coupling which given by
\begin{equation}
p_{1,0}=\biggl(\frac 12 +\frac 12 \tanh\frac\t\a\biggr)
\ \biggl(1+\frac{1}{1+\frac{\t^2}{\a^2}}\biggr)\,.
\eqlabell{p0power}
\end{equation}
With this profile, there is still an exponential turn on of the
coupling are early times but the approach to the final equilibrium
value is now controlled by a power law, \ie $p_{1,0}\simeq 1+
\frac{\alpha^2}{\t^2}+\cdots$ as $\t\to+\infty$. Fig.~\ref{ppower}
shows the response of the normalizable component $p_{1,2}$ in these
quenches for select values of $\a=\{1,\frac 12,\frac 14,\frac
18,\frac{1}{16}\}$. If we use the same definitions as before, \ie
eqs.~\reef{defddn}, \reef{defp12eqiui} and \reef{equilibrationt}, the
results for the equilibration time scale are essentially unchanged.
That is, $\t_{eq}\sim 1$ or $t_{eq}\sim 1/T_i$ and for $\t\gtrsim
\t_{eq}$, $p_{1,2}$ follows the adiabatic response
\eqref{adresponse}.\footnote{We also explicitly checked that the
essentially same QNM behaviour occurs for these quenches
\eqref{p0power} as shown in fig.~\ref{figureQNM_3}.} Now we may also
consider defining the equilibration time scale $\t_{eq}^\calo$ directly
in terms of a more physical quantity such as the expectation value of
$\calo_3$ --- similar remarks would apply using $\cale$ or $\calp$
below. That is, we would replace our previous definitions with
\begin{equation}
\dd_{neq}^\calo(\t)\equiv \left|\frac{\calo_3(\t)-
\calo_3^{adiabatic}(\t)}{\calo_3^{equilibrium}}\right|\,,
\labell{dodo2}
\end{equation}
where
\begin{equation}
\calo_3^{equilibrium}=\lim_{\t\to+\infty} \calo_3(\t)=
\frac{\Gamma(3/4)^4}{2\sqrt{2}\pi^2}N^2 T_i^2 m_f^0\,.
\labell{dodo3}
\end{equation}
Then the equilibration time $\t_{eq}^\calo$ would be defined as the
latest time at which $\dd_{neq}^\calo$ exceeds some threshold, \ie
$\dd_{neq}^\calo(\t>\t_{eq}^\calo)<\epsilon$. Unfortunately, this
approach would lead to ambiguous results because of the
scheme-dependent contribution in the expression for $\calo_3$ given in
eq.~\eqref{vev1qft1}. In particular, we find
\begin{equation}
\dd_{neq}^\calo(\t)\Big|_{ambiguity}\ \propto\ p_{1,0}''\ \ln\frac{\pi T_i}{\Lambda_2}
\ \sim\ \frac{\a^2}{\t^4}\ \ln\frac{\pi T_i}{\Lambda_2}\sim \left(\frac{\t_{eq}}{\t}\right)^4
\ \a^2\ln\frac{\pi T_i}{\Lambda_2}\,,
\eqlabell{ambiguityddpo}
\end{equation}
where $\t_{eq}$ is the previously defined equilibration time. This
ambiguity could substantially affect the computation of the relaxation
time $\t_{eq}^\calo$ (\ie making it longer than the thermal time scale)
if the renormalization scheme is such that
\begin{equation}
\a^2\ln\frac{\pi T_i}{\Lambda_2}\gg 1\,.
\eqlabell{longrelax}
\end{equation}
Of course, one may regard the real source of the issue here to be the
power law behaviour of the profile \reef{p0power} at late times. The
latter was introduced precisely to enhance the importance of the
scheme-dependent terms in eq.~\eqref{vev1qft1}, as these issues would
not have occurred with the exponential fall-off of the original profile
\reef{mq3}. Still this example stands as a cautionary note in studying
the excitation and equilibration times in generic time-dependent
settings.

Further it is interesting that, in some cases, the renormalization
scheme ambiguities can be used to remove the divergent scaling
behaviour appears in the one-point correlators because of the rapid
variation of the couplings in these observables.\footnote{These
divergences should be distinguished from that in the expectation value
of the corresponding operator $\calo_\Delta$, as discussed above.}
Consider a quench of $\calo_3$ where the coupling $\l_3$ exhibits a
power law growth in the vicinity of $t\ge0$ with
\begin{equation}
\l_3(t)=t^\b\,,\qquad 0<\b<1\,.
\eqlabell{singquench}
\end{equation}
From eq.~\eqref{vev1}, we can see that for generic values of the
scheme-dependent coefficients $\dd_2$ and $\dd_3$, the observables of
the theory would diverge as $t\to 0$. The divergence introduced by the
time variation of the coupling $p_0\propto \l_3$ is given by
\begin{equation}
\begin{split}
8\pi G_5\ \cale\big|_{coupling-divergence}=&-\frac{1}{12} (p_0')^2
+\frac13 p_0 p_0''+\frac 12 \dd_2 (p_0')^2\\
&\propto t^{2(\b-1)} (3\b-4+6\b \dd_2)\,,\\
8\pi G_5 \calp\big|_{coupling-divergence}=&-\frac{1}{36} (p_0')^2-\frac{1}{18} p_0
p_0''-2 \delta_3 (p_0')^2-2 \delta_3 p_0 (p_0'')+\frac12 \delta_2 (p_0')^2\\
&\propto t^{2(\b-1)}(-144\delta_3\beta+72\delta_3-3\beta+2+18\delta_2\beta)\,,\\
16\pi G_5\ \la\calo_3\ra\big|_{coupling-divergence}=&\frac12 p_0''+2\dd_2 p_0''\propto
t^{\b-2}(1+4\dd_2)\,.
\end{split}
\eqlabell{divobs}
\end{equation}
Here again, without loss of generality, we have set $a_1=0$. Hence by
choosing
\begin{equation}
\dd_2=-\frac 12 +\frac{2}{3\b}\,,\qquad \dd_3=\frac{7-6\b}{36(2\b-1)}\,,
\eqlabell{schemechoice}
\end{equation}
we would remove the divergences in the energy density and the pressure.
However, the expectation value of the operator $\calo_3$ would still be
divergent. Alternatively, one can remove these coupling-induced
divergences in the pressure  and $\la\calo_3\ra$, with the energy
density remaining divergent as $t\to 0$. Presumably, this example
illustrates that it is more useful to consider the scaling behaviour of
observables characterizing the final equilibrium state.

Our analysis revealed that for rapid quenches with $\calt T_i\ll 1$,
the system relaxes, almost universally, on thermal scales $\sim
{1}/{T_i}$ --- see eqs.~\eqref{teqT} and \eqref{teqT2}. Such a response
might be an artifact of our treatment of the quenches perturbatively in
the amplitude of the coupling, effectively about its zero value. It
would be interesting to study the full nonlinear problem where not only
the rate change of the coupling but also the amplitude of the coupling
(and its variation over the history of a quench) can be arbitrary. From
figs.~\ref{figure2} and \ref{figure4}, it is clear that even though the
relaxation time scale for fast quenches of $\l_3$ and $\l_2$ couplings
is roughly the same, the amplitude of the response due to quenches of
$\l_3$ is quadratically enhanced (in the rate of change) compared to
that for quenches of $\l_2$. In this sense, the quenches due to
$\calo_3$ operator more closely follow the intrinsic time-scale of its
coupling. Given our perturbative analysis of the quenches here, we
could treat quenches due to $\l_2$ and $\l_3$ separately. It would be
interesting to study ``supersymmetric'' quenches with
\begin{equation}
\l_2(t)=\l_3(t)^{\,2}\,,
\eqlabell{susyquench}
\end{equation}
which would necessarily require a full nonlinear analysis.

Another regime of the quenches which exhibited interesting universal
behaviour was $\a\gg1$, corresponding to slow nearly adiabatic
quenches. Recall that for fast quenches ($\a\ll 1$), the normalizable
mode of the bulk scalar relaxes to its final equilibrium with a
quasinormal mode behaviour, \eg see figs.~\ref{figureQNM_3} and
\ref{figureQNM_2}. In contrast, the relaxation in the nearly adiabatic
regime occurs without excitation of any quasinormal modes. Rather for
$\a\gtrsim 1$, the deviation from the purely adiabatic evolution in
eq.~\reef{adresponse} is governed entirely by derivatives of the source
profile, \eg $\a\ \del_\t p_{1,0}(\t/\a)$ for $\calo_3$ quenches or by
$\a\ \del_\t p_{1,0}^l(\t/\a)$ for $\calo_2$ quenches. Indeed, in
either case with an arbitrary source $p_0=p_0({\t}/{\a})$ and $\a\gg1$,
we can solve the bulk scalar wave equation \eqref{dec2} perturbatively
in $1/\a$:
\begin{equation}
\phi_1= p_0({\t}/{\a} )\ \phi_1^e(\r)\ +\ \sum_{n=1}^\infty\ \a^{-n}\,
\ \phi_1^{(n)}({\t}/{\a},\,\r)\,,
\eqlabell{slow1}
\end{equation}
where $\phi_1^e(\r)$ is the equilibrium solution normalized such that
the leading coefficient in the asymptotic expansion is one. For
example, for the $m^2=-3$ scalar, the latter would be the solution
given in eq.~\reef{solve3} divided by a factor of $\lambda\mu$. Now for
$n=1$, we find
\begin{equation}
\begin{split}
&\phi_1^{(1)}({\t}/{\a},\,\r)=\a\, \del_\t p_0(\t/\a)\ \ g^{(1)}(\r)\,,
\end{split}
\eqlabell{slow2}
\end{equation}
where $g^{(1)}$ satisfies the inhomogeneous equation
\begin{equation}
-\frac 12 (1-\r^4)\ \del^2_{\r\r}g^{(1)}+\frac{3+\r^4}{2\r}\
\del_\r g^{(1)}+\frac{m^2}{2\r^2}\, g^{(1)}=
\frac{3}{2\,\r}\, \phi_1^{e}-\del_\r\phi_1^{e}\,.
\eqlabell{slow22}
\end{equation}
Similarly, successively higher derivatives of $p_0$ appear at higher
orders in this expansion. Hence the entire dynamics of the linearized
solution is simply governed by the nearly adiabatic driving source.

Quantum quenches from a thermal initial state were discussed previously
in \cite{cc2}. The major difference between this seminal study and our
present analysis is that we are considering a strongly coupled field
theory while the authors of \cite{cc2} investigated quantum quenches of
a free scalar field. Effectively, the momentum modes in the latter
theory reduce to a collection of decoupled simple harmonic oscillators
(SHO's). In \cite{cc2}, it was shown that a(n instantaneous) quench of
an individual SHO from an initial frequency $\w_i$ and temperature
$T_i$ to a final frequency $\w_f$ results in an effective thermal final
state with temperature $T_f$
\begin{equation}
T_f=\frac12\ \biggl(1+\frac{\w_f^2}{\w_i^2}\biggr)
\ T_i\,,\qquad {\rm for}\qquad T_i\gg \w_i\,.
\eqlabell{cc2t}
\end{equation}
A striking result here is that if we also choose $\w_f\ll\w_i$, then
the final temperature is one-half the initial temperature. Quite
generally, $\w_f<\w_i$ results in $T_f<T_i$, \ie the system cools! A
similar cooling effect was found for quenches of the mass in the field
theory where $T_i\gg m_i\gg m_{fin}$, using an effective temperature
determined by averaging over all of the momentum modes. In fact,
$T_f=T_i/2$ was found for such quenches in $d=2$ and 3. However, in
$d=4$, the effective temperature was dominated by high momentum modes
and they found $T_f=T_i$.

Recall that our analysis also applies to the same high temperature
regime, \eg $T_i\gg m^0_{f,b}$. Hence it is interesting to compare the
above results to those for our quenches from the massive gauge theory
to the CFT, in which case we also satisfy $m_i\gg m_{fin}$. The
expressions for $T_f/T_i$ are given in eqs.~\eqref{ftdim3cft} and
\eqref{ftdim2cft} for these quenches with $\calo_3$ and $\calo_2$,
respectively. Recall that in section \ref{gravityX}, we argued that
$a^\infty_{2,4}\sim0$ for $\a\gg1$ --- see the discussion around
eqs.~\reef{ad3} and \reef{ad3x2} --- and hence it is clear that these
quenches are indeed cooling the system provided the transition was
sufficiently slow. However, as the characteristic time scale of a
quench becomes shorter than a typical thermal scale $1/ T_i$, we also
saw that $a_{2,4}^\infty$ grows large and so the corresponding quenches
always result in an increase of the temperature. The precise thresholds
between the cooling and heating regimes are given in
eqs.~\eqref{thresholds} and \reef{tfgtidim2}. Of course, since our
analysis applies for the high temperature regime, any changes in the
temperature are perturbatively small in $m^0_{b,f}/{T_i}$ and so much
like the result of \cite{cc2} for $d=4$, we have $T_f\simeq T_i$ to
leading order. Hence we may conclude that the thermal quenches studied
in \cite{cc2} for a free field and those studied here for a strongly
interacting QFT share certain similar features. Of course, we pointed
out above that for fast quenches, the behaviour of these two systems is
qualitatively dissimilar.

We would like to contrast our framework with previous analytic
investigations using the AdS-Vaidya solution \cite{vaidya}. These
studies investigated issues related to thermalization of a holographic
plasma by considering the gravitational collapse of a thin shell of
`null dust'. Of course, the latter scenarios might also be regarded as
a holographic quantum quenches. However, we would note that by design,
the null dust does not excite any local gravitational degrees of
freedom, in particular, on the horizon. For example, the spacetime
becomes that of a static black hole immediately beyond the boundary
where the bulk stress tensor goes to zero. In the dual description
then, this scenario injects energy into the gauge theory with an
`exotic' probe which does not excite any of the local degrees of
freedom in the holographic plasma. Hence this approach may be useful to
study questions related to the formation and growth of the horizon,
however, it can not address many aspects of the holographic quench. For
example, it does not reveal the scaling behaviour of the response or
the quasinormal response of the horizon. Our analysis indicates that
both of these features play an important role in the equilibration of
the system. It is interesting that the Vaidya solutions seem to be a
good approximation to a quench induced with a massless bulk scalar in a
small amplitude expansion \cite{shiraz}, similar to that used in our
analysis. The analytic approach of \cite{shiraz} also allows one to
systematically improve the gravitational description within this
expansion.

There are numerous aspects of the holographic quenches described here
that are left for future analysis. First and foremost, we would like to
extend our perturbative treatment to a full nonlinear analysis of the
gravitational equations of motion. However, let us outline a few other
potentially interesting extensions:
\nxt In this paper, we only considered the response as measured by a
few simple one-point correlators. These are representatives of {\it
local} physical observables in the holographic theory. An open question
remains to analyze the behaviour, in particular the relaxation, of {\it
non-local} physical observables, such as higher-point correlation
functions, Wilson lines or entanglement entropy. Previous studies
\cite{vaidya} indicate that these non-local observables equilibriate
more slowly than the local observables  and so it will be interesting
to see if this effect persists in the present framework which accounts
the full quasinormal response of the horizon.
\nxt Our present analysis was restricted to quenches which are
spatially isotropic and homogeneous. As a result, we
eliminated\footnote{We would like to thank Rob Leigh for raising this
issue.} some of the potentially important relaxation channels of the
system, namely, the sound and the shear modes in off-equilibrium gauge
theory plasma. In a more general anisotropic or inhomogeneous quench
where such modes are excited, one may expect that the hydrodynamic
modes will have much slower decay rates than the quasinormal modes of
the bulk scalar found in the present work. Inevitably then, these new
modes would dominate the late-time behaviour of the system, however,
their influence on the equilibration time would still depend on how
efficiently they are produced during the quench. We defer the
consideration such inhomogeneous quenches to future work.
\nxt We focused on the properties of the energy density and the
pressure of gauge theory plasma undergoing a quench. We pointed out
that both of these observables are renormalization scheme dependent. It
is interesting to address the question of ambiguities in the entropy
density away from equilibrium \cite{booth}. In particular then, it
would be interesting to explore how a choice of a renormalization
scheme affects the definition of the entropy density.

\section*{Acknowledgments}
We would like to thank Colin Denniston, Carlos Hoyos, Romuald Janik,
Pavel Kovtun, Sung-Sik Lee, Rob Leigh, Jorma Louko, 
David Mateos, Anton van Niekerk, Jo\~ao Penedones, Eric Poisson, Misha
Smolkin and Toby Wiseman for useful discussions. Research at Perimeter
Institute is supported by the Government of Canada through Industry
Canada and by the Province of Ontario through the Ministry of Research
\& Innovation. AB, LL and RCM gratefully acknowledge support from NSERC
Discovery grants. Research by LL and RCM is further supported by
funding from the Canadian Institute for Advanced Research.

\appendix

\section{Discretization of the evolution equation}\label{appd}

We describe here the basic strategy employed to discretize the evolution equation.
The schematic structure of such equation, for a field $f(\t,\r)$, takes the form:
\begin{equation}
0=\del_{\t\r}f- A_{\r\r}\ \del^2_{\r\r} f-A_\t\ \del_\t f +A_{\r}\ \del_\r f+A_0\ f+J_0\,,
\eqlabell{app1}
\end{equation}
for suitably defined multiplicative factors $A_{\cdots}$,  and a source
term $J_0$, all explicit functions of $(\t,\r)$. Notice the principal
part of the system, \ie
\begin{equation}
0=\left( \del_{\t\r}- A_{\r\r}\ \del^2_{\r\r} \right) f\,,
\end{equation}
is a second order partial differential equation admitting two
propagating modes. One of them has a characteristic speed given by
$A_{\r\r}$ while the other propagates instantly inwards from the
boundary (a consequence of having chosen incoming characteristics to
foliate the spacetime). Further, notice that for a scenario with a
black hole present from the start, as is the case for our studies here,
$A_{\r\r} > 0 ~(<0) $ outside (inside) the horizon. Thus {\em only one
boundary condition} can be prescribed at the AdS boundary.

A natural strategy to discretize the system, taking advantage of the
characteristic structure explicitly displayed by equation (\ref{app1}),
has been presented in~\cite{Isaac} (and further extended in~\cite{diss}
to include dissipation and in~\cite{adapt} to allow for adaptive
meshes). In our implementation, we follow this approach to obtain a
second order accurate implementation in the following way. We introduce
a uniform discrete spatial grid covering the domain $\rho \in [0,
L_{\rho}]$ with $N_{\rho} + 1$ points  located at $\rho_i = (i-1)
d\rho$ ($d\rho \equiv L_{\rho}/(N_{\rho})$. Discrete, evenly spaced,
time levels are introduced as $\tau_n = \tau_o + (n-1) d\tau$ (with
$d\tau = \alpha d\rho$). The continuous function $f(\tau,\rho)$ is then
represented by values on this grid structure as $f(\tau_n,\rho_i)\equiv
f^n_i$. Only two time levels are required for discretizing our target
equation as follows. Consider $f$ is known at level $n$, \ie all values
$f^{n}_{i}$ are known, as well as those at level $n+1$ for $\rho <
\rho_i$, to obtain the value of $f^{n+1}_i$ a centered discretization
at the virtual point $(n+1/2,i-1/2)$ is achieved by discretizing the
terms involved as,
\begin{eqnarray}
\partial_{\tau\rho} f |^{n+1/2}_i &=& \frac{f^{n+1}_i-f^{n+1}_{i-1} - f^n_i + f^n_{i-1}}{d\rho d\tau} \,,\\
\partial^2_{\rho} f |^{n+1/2}_i &=& \frac{f^{n+1}_i-f^{n+1}_{i-1} + f^{n+1}_{i-1} +
f^n_{i+1} -2 f^n_i + f^n_{i-1}}{2 d\rho^2} \,,\\
\partial_{\tau} f |^{n+1/2}_i &=& \frac{f^{n+1}_i + f^{n+1}_{i-1} - f^n_i - f^n_{i-1}}{2 d\tau }\,,\\
\partial_{\rho} f |^{n+1/2}_i &=& \frac{f^{n+1}_i - f^{n+1}_{i-1} + f^n_i - f^n_{i-1}}{2 d\rho }\,,\\
f |^{n+1/2}_i &=& \frac{f^{n+1}_{i-1} + f^n_i } {2}\,.
\end{eqnarray}
The coefficients $A_{\cdots}$ and the source $J_0$ are evaluated
analytically at the virtual point. Plugging these expressions into the
equation and solving for $f^{n+1}_i$ defines the scheme employed.
Notice the resulting scheme is thus an inwards ``marching'' algorithm
that provides the value of $f^{n+1}_i$. Further the ``slanted'' nature
of this scheme (\ie the forward offset of the second derivative at
level $n$) is simply a reflection of the characteristic structure of
the system. Since initial data is given at an initial hypersurface, and
boundary conditions at $\rho_1 = 0$, this scheme suffices to integrate
all points with the exception of $\rho_2, \rho_{N_{\rho}+1}$ where the
simple change,
\begin{eqnarray}
\partial^2_{\rho} f |^{n+1/2}_2 &=&  \frac{f^n_{3} -2 f^n_2 + f^n_{1} } {d\rho^2} \,,\\
\partial^2_{\rho} f |^{n+1/2}_{N_{\rho+1}} &=& \frac{f^{n+1}_{N_{\rho+1}}-f^{n+1}_{N_{\rho}} + f^{n+1}_{N_{\rho-1}} +
f^n_{N_{\rho+1}} -2 f^n_{N_{\rho}} + f^n_{N_{\rho-1}}}{2 d\rho^2} \,,
\end{eqnarray}
suffices to provide a first accurate order approximation which does not affect the overall second order convergence of the solution.

We specify the characteristic initial data $f^{initial}(\r)$ at initial time $\t=\t_1$:
\begin{equation}
f(\t=\t_1,\r)=f^{initial}(\r)\,,
\eqlabell{initial}
\end{equation}
and the boundary data $f^{boundary}$ at $\r_1=0$,
\begin{equation}
f(\t,\r=\r_1)=f^{boundary}(\t)\,.
\eqlabell{boundary}
\end{equation}

\begin{figure}[t]
\begin{center}
  \includegraphics[width=2.6in]{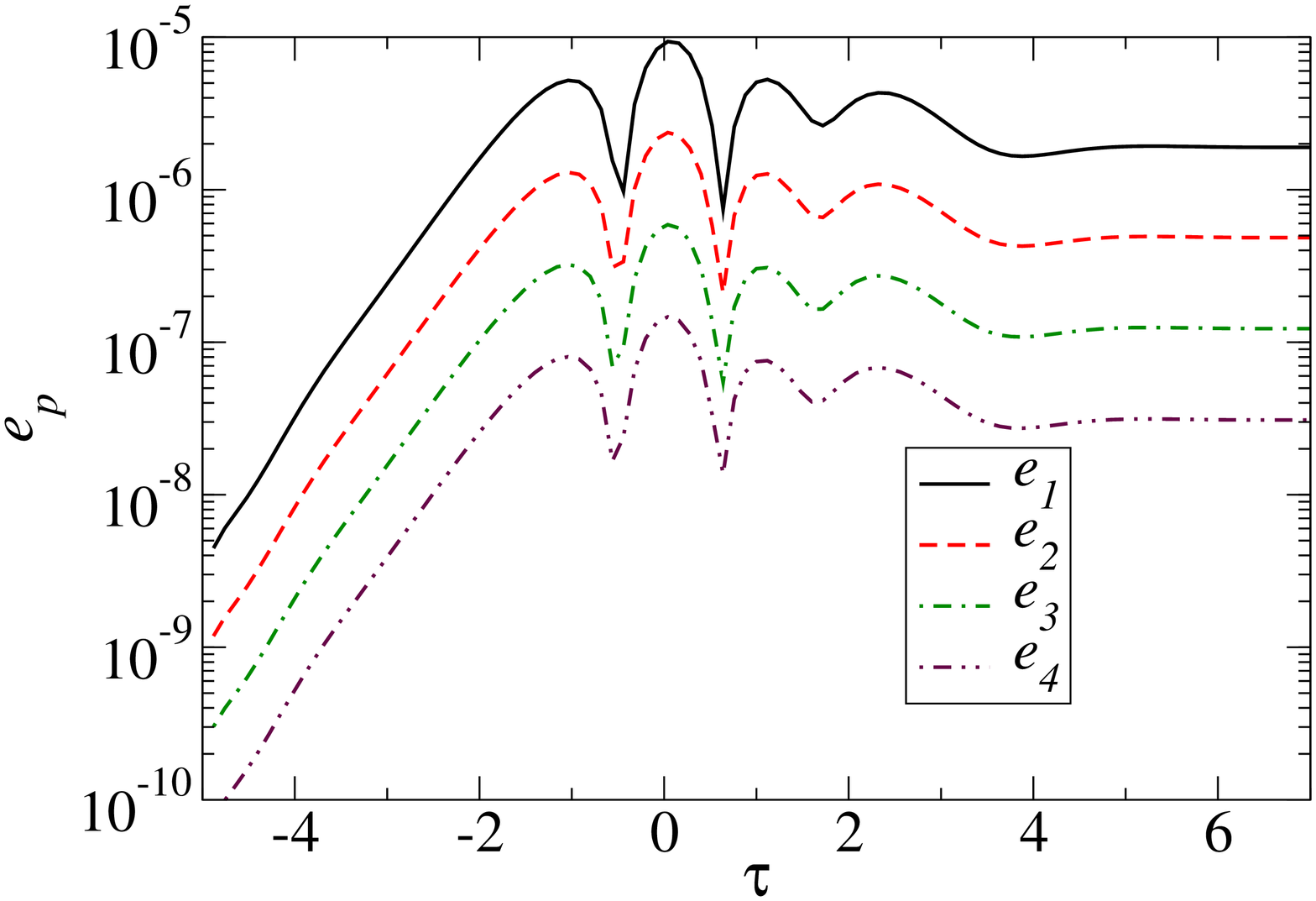}
  \includegraphics[width=2.4in]{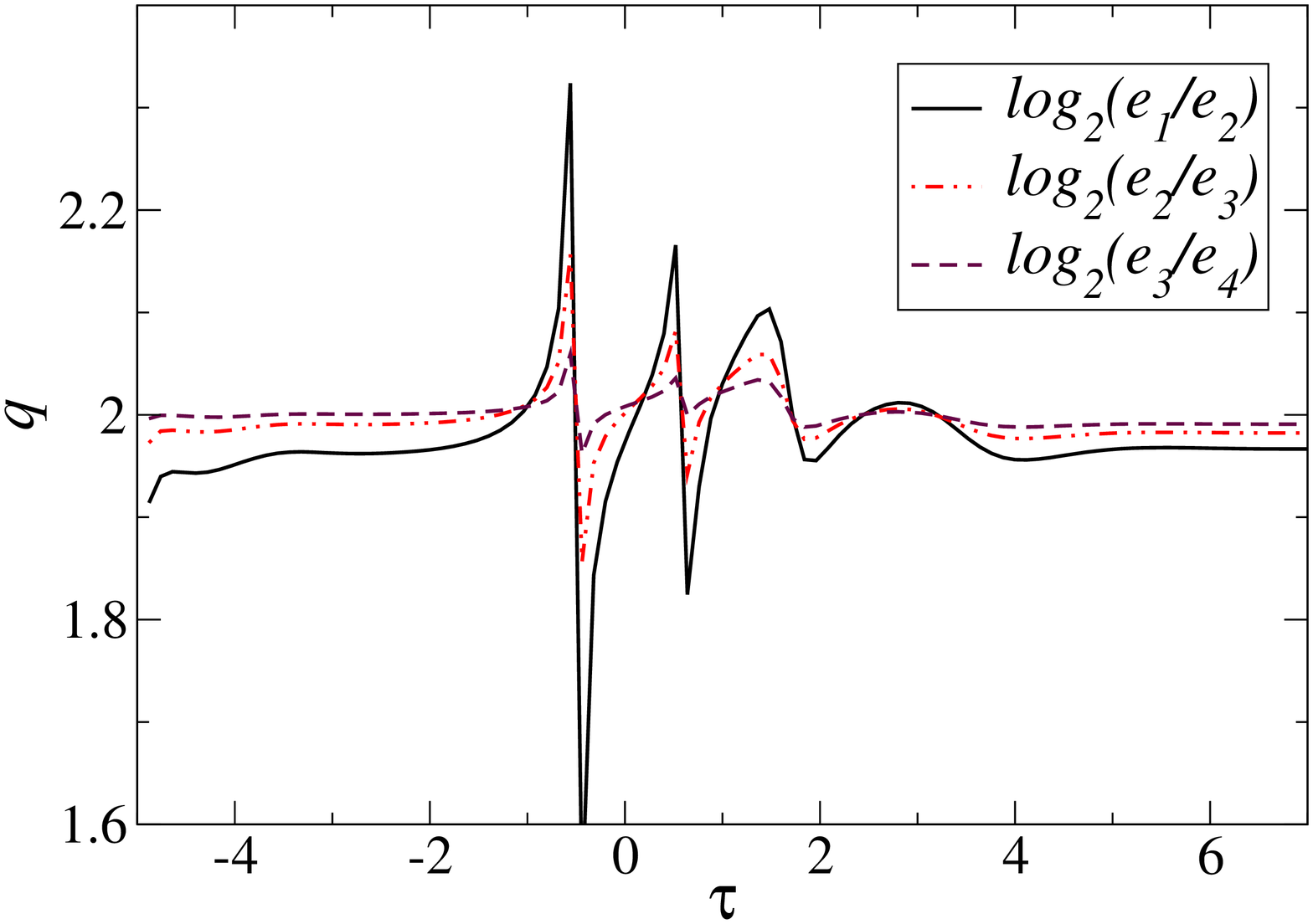}
\end{center}
  \caption{(Colour online) Convergence tests. (Left panel) $L_2$ norm of the difference between solutions
obtained halving the discretization length. (Right panel) Estimated convergence rate from the numerical
solution, indicating the numerical solution indeed converges to second order
as expected.
}\label{fig:convergence}
\end{figure}

As a representative test of the code's convergence behaviour, we
consider the bosonic case and obtain solutions for different grid sizes
$N_{\rho} = 2^p \times 100$  ($p=0,\ldots,4$) and simulate the system
until equilibration. We monitor the ($L_2$ norm of the) difference of
solutions with successive values of p (defining $e_{p}\equiv
||\phi_{p+1}-\phi_{p}||_2$) as well as the convergence rate $q$
calculated as,
\begin{equation}
2^q = \frac{||\phi_{p+1}-\phi_{p}||_2}{||\phi_{p+2}-\phi_{p+1}||_2} = \frac{e_{p}}{e_{p+1}} \,.
\end{equation}
The results are shown in figure \ref{fig:convergence}, the left panel
illustrates how $e_{p}$ decreases as the resolution is improved while
the right one the convergence rate $q$ illustrating second order
convergence behaviour.

As a last remark, we stress that in the general case metric variables
governed by an evolution equation (\eg $\Sigma$ determined by
eq.~\eqref{eoms1}) can be integrated using the same strategy described
above. Thus, this algorithm together with straightforward integration
along the radial direction (\eg for $A$ determined by
eq.~\eqref{eoms2}) suffice for implementing the complete system
provided consistent boundary conditions are provided at the boundary.
For other strategies see ~\cite{cy} (employing also a characteristic
formulation of the equations) and
~\cite{Bantilan:2012vu,Heller:2012je}) (for works implementing a Cauchy
formulation).

\end{document}